\documentclass[10pt,prd,english]{revtex4}
\usepackage{amssymb}
\usepackage{amsmath}
\usepackage{bm}
\usepackage{mathptmx}
\usepackage{amsthm}
\usepackage{mathtools}
\usepackage{emerald}
\usepackage{slantsc}
\usepackage{array}
\usepackage{url}
\usepackage{tikz}
\usepackage{graphics}

\def\be{\begin{equation}}
\def\ee{\end{equation}}
\def\ba{\begin{eqnarray}}
\def\ea{\end{eqnarray}}
\def\bit{\begin{itemize}}
\def\eit{\end{itemize}}
\def\eq{\equiv}
\def\bsu{\begin{subequations}}
\def\esu{\end{subequations}}

\def\a{\alpha}
\def\a {\alpha}
\def\b{\beta}
\def\g{\gamma}     
\def\d{\delta}

\def\e{\epsilon}

\def\l{\lambda}
\def\m{\mu}
\def\n{\nu}
   
\def\p{\phi}

\def\s{\sigma}
\def\t{\tau}

\def\la{\label}

\def\ri{\right}

\def\aj{{\it The Astronomical Journal}}
\def\apj{{\it The Astrophysical Journal}}

\def\apjl{{\it The Astrophysical Journal Letters}}
\def\mnras{{\it Monthly Notices of the Royal Astronomical Society}}
\def\prd{{\it Physical Review D}}

\def\aap{{\it Astronomy \& Astrophysics}}

\usepackage{color}
\expandafter\ifx\csname package@font\endcsname\relax\else
 \expandafter\expandafter
 \expandafter\usepackage
 \expandafter\expandafter
 \expandafter{\csname package@font\endcsname}%
\fi

\newcommand{\il}{\mathcal{I}_{A_l}}

\newcommand{\ilrr}{\frac{\mathcal{I}_{A_l}(t-r)}{r}}
\newcommand{\ilrrp}{\frac{{\mathcal{I}}^{(p)}_{A_l}(t-r)}{r}}
\newcommand{\eislrr}{\frac{\epsilon_{iba}\mathcal{S}_{bA_{l-1}}(t-r)}{r}}
\newcommand{\eislrrp}{\frac{\epsilon_{iba}{\mathcal{S}}^{(p)}_{bA_{l-1}}(t-r)}{r}}
\newcommand{\ejslrr}{\frac{\epsilon_{jba}\mathcal{S}_{bA_{l-1}}(t-r)}{r}}
\newcommand{\ejslrrp}{\frac{\epsilon_{jba}{\mathcal{S}}^{(p)}_{bA_{l-1}}(t-r)}{r}}

\newcommand{\eisbr}{\frac{\epsilon_{iba}\mathcal{S}_b}{r}}
\newcommand{\ejsbr}{\frac{\epsilon_{jba}\mathcal{S}_b}{r}}

\newcommand{\Mc}{\mathcal M}
\newcommand{\Ic}{\mathcal I}
\newcommand{\Sc}{{\mathcal S}}
\newcommand{\ffrac}[2]{\frac{{\displaystyle #1}}{{\displaystyle #2}}}

\newcommand{\dksi}{{\hat\partial}}
\newcommand{\dtau}{\hat\partial_\tau}
\newcommand{\dtautau}{\hat\partial_{\tau\tau}}
\newcommand{\dtz}{\hat\partial_{t^*}}
\newcommand{\wm}{w_{\scriptscriptstyle(M)}}
\newcommand{\ws}{w_{\scriptscriptstyle(S)}}
\newcommand{\phim}{\phi_{\scriptscriptstyle(M)}}
\newcommand{\phis}{\phi_{\scriptscriptstyle(S)}}
\newcommand{\chim}{\chi_{\scriptscriptstyle(M)}}
\newcommand{\chis}{\chi_{\scriptscriptstyle(S)}}
\newcommand{\fim}{\Phi_{\scriptscriptstyle(M)}}
\newcommand{\fis}{\Phi_{\scriptscriptstyle(S)}}
\newcommand{\fig}{\Phi_{\scriptscriptstyle(G)}}
\newcommand{\ddxim}{\ddot x^i_{\scriptscriptstyle(M)}}
\newcommand{\ddxis}{\ddot x^i_{\scriptscriptstyle(S)}}
\newcommand{\aGi}{\ai\limits_{\scriptscriptstyle (G)}{}\left(\tau,{\bm\xi}\right)}
\newcommand{\aMi}{\ai\limits_{\scriptscriptstyle (M)}{}\left(\tau,{\bm\xi}\right)}
\newcommand{\aSi}{\ai\limits_{\scriptscriptstyle (S)}{}\left(\tau,{\bm\xi}\right)}
\newcommand{\DXGi}{\DXii\limits_{\scriptscriptstyle (G)}{}\left(\tau,{\bm \xi}\right)}

\newcommand{\XG}{\Xii\limits_{\scriptscriptstyle (G)}{}\left(\tau,{\bm \xi}\right)}
\newcommand{\XGo}{\Xii\limits_{\scriptscriptstyle (G)}{}\left(\tau_0,{\bm \xi}\right)}
\newcommand{\DXiim}{\DXii\limits_{\scriptscriptstyle (M)}{}\left(\tau,{\bm \xi}\right)}
\newcommand{\DXiis}{\DXii\limits_{\scriptscriptstyle (S)}{}\left(\tau,{\bm \xi}\right)}
\newcommand{\Xiim}{\Xii\limits_{\scriptscriptstyle (M)}{}\left(\tau,{\bm \xi}\right)}
\newcommand{\Xiis}{\Xii\limits_{\scriptscriptstyle (S)}{}\left(\tau,{\bm \xi}\right)}

\newcommand{\Xiimo}{\Xii\limits_{\scriptscriptstyle (M)}{}\left(\tau_0,{\bm \xi}\right)}
\newcommand{\Xiiso}{\Xii\limits_{\scriptscriptstyle (S)}{}\left(\tau_0,{\bm \xi}\right)}

\newcommand{\DelG}{\Del\limits_{\scriptscriptstyle (G)}{}\left(\tau,\tau_0\right)}
\newcommand{\Delm}{\Del\limits_{\scriptscriptstyle (M)}{}\left(\tau,\tau_0\right)}
\newcommand{\Dels}{\Del\limits_{\scriptscriptstyle (S)}{}\left(\tau,\tau_0\right)}
\newcommand{\Delt}{\Del\left(\tau,\tau_0\right)}

\def\ai{\operatornamewithlimits{\alpha^i}}
\def\h{\operatornamewithlimits{h}}
\def\q{\operatornamewithlimits{q}}
\def\D^i{\operatornamewithlimits{D^{\it i}}}
\def\Del{\operatornamewithlimits{\Delta}}
\def\DXii{\operatornamewithlimits{\dot\Xi^{\it i}}}
\def\Xii{\operatornamewithlimits{\Xi^{\it i}}}

\def\Xij{\operatornamewithlimits{\Xi^{\it j}}}
\makeatletter \@addtoreset{equation}{section} \makeatother
\def\Quadrat#1#2{{\vcenter{\hrule height #2
  \hbox{\vrule width #2 height #1 \kern#1
    \vrule width #2}
  \hrule height #2}}}

\begin{document}
\author{Pavel Korobkov}
\affiliation{Solovetsky Monastery, Arkhangelsk Region 164070, Russia}
\author{Sergei M. Kopeikin}\email{kopeikins@missouri.edu}\thanks{Corresponding author}
\affiliation{Department of Physics \& Astronomy, University of Missouri, 322 Physics Bldg., Columbia, Missouri 65211, USA}
\title{General Relativistic Theory of Light Propagation in Multipolar Gravitational Fields}
\date{\today}
\maketitle

\section{Introduction}\label{intr1}
\subsection{Statement of the problem}\la{intr2}

Direct experimental detection of gravitational waves \index{gravitational wave} is a fascinating but yet unsolved problem of modern fundamental physics  Enormous efforts have been undertaken to make progress in its solution both by theorists and experimentalists \cite{brown,tama,astone,ligo,cerd}. The main theoretical efforts are presently focused on calculation of templates of the gravitational waves emitted by coalescing binary systems \index{binary system!coalescing} comprised of neutron stars and/or black holes \cite{luc,dmrbala,buo1,buo,dmr1,dmr} as well as creation of improved filtering technique for gravitational wave detectors \cite{owen,unga} which will enable the extraction of the gravitational wave signal from all kind of interferences present in the noisy data collected by the gravitational wave observatories \index{gravitational wave observatory}. Direct experimental efforts have led to the construction of several ground-based optical interferometers 
with the length of arms reaching a few miles \cite{rile,sigg,acer,fuj,geo600}. Certain work is under way to build super-sensitive cryogenic-bar gravitational-wave detectors of Weber's type \cite{weber,coc,bar1,agu}. Spaceborne laser interferometric detectors like LISA \cite{lisa}, NGO \citep{e-lisa} or ASTROD \cite{astrod, astrod-gw} may significantly increase the sensitivity of the gravitational-wave detectors and revolutionize the field of gravitational physics. \index{gravitational physics} 

There is no doubt, the detection of gravitational waves by the specialized gravitational wave antennas would provide the most direct evidence of the existence of these elusive ripples in the fabric of spacetime. On the other hand, there exist a number of astronomical phenomena which might be used for an indirect detection of gravitational waves and understanding gravitational physics of astrophysical systems emitting gravitational waves. It is worth emphasizing that the present-day ground-based gravitational wave \index{gravitational wave} detectors are sensitive to a rather narrow band of the gravitational wave spectrum ranging between 1000$\div$1 Hz \cite{thrn,sch}. Spaceborne gravitational wave detectors may bring the sensitivity band down to 1 mHz \citep{e-lisa,astrod-gw}. In order to explore the ultra-long gravitational wave phenomena much below 1 mHz we have to rely upon other astronomical techniques for example, pulsar timing \citep{riles_2013,kramer_2010} or radiometry of polarization of cosmic microwave background radiation \citep{polnarev_1995,polnarev_2008}. The short gravitational waves can be generated by coalescing binary systems of compact astrophysical objects like neutron stars and/or black holes \citep{sathya_2013,war,hi,raj}. The ultra-long gravitational waves can be generated by localized gravitational sources and/or topological defects in the early universe \cite{gris,gris1,allen,allen+,marvil}.  

Astronomical observations are conducted with electromagnetic waves (photons) of different frequencies over the spectrum spreading from very long radio waves to gamma rays. Therefore, in order to detect the effects of the gravitational waves by astronomical technique one has to solve the problem of propagation of electromagnetic waves through the field generated by a localized gravitationally-bounded distribution of masses which motion is determined by general relativity. Notice that gravitational wave detectors are optical interferometers making use of light propagation. Therefore, correct understanding of the process of physical interaction of the laser beam of the interferometer with incoming gravitational waves is important for their unambiguous recognition and detection. Equations of propagation of electromagnetic signals must be derived in the framework of the same theory of gravity in order to keep description of gravitational and electromagnetic phenomena on the same theoretical ground. We draw attention of the reader that the parametrized post-Newtonian (PPN) formalism \cite{will} does not comply \index{PPN formalism} with this requirement. The PPN formalism is constructed on the ground of plausible physical hypothesis and assumptions about alternatives to general relativity but it does not demand their overall conformity. Therefore, straightforward application of PPN formalism to discuss gravitational physics may eventually lead to incorrect results. As an example, we point out that PPN formalism does not produce the consistent post-Newtonian equations of motion of extended bodies even in a simplified case of two PPN parameters, $\beta$ and $\gamma$, when more subtle effects of body's gravitational multipoles are taken into account \cite{kopvlas}. PPN formalism applied to interpret gravitational light-ray deflection experiments by major planets of the solar system \citep{kop_2001,fomkop_2003,fom_2010IAUS} created a notable ``speed-of-light versus speed-of-gravity'' controversy \citep{willapj} which originates in the inability of PPN formalism to distinguish between physical effects of gravity and electromagnetism due to the non-covariant nature of PPN parametrization limited merely by the metric manifolds \citep{kopfom_2006,kopeikin_book}. 

In the present chapter we rely upon the Einstein theory of general relativity \index{general relativity} and assume that light propagates in vacuum that is the interstellar medium has no impact on the speed of light propagation. General relativity is a geometrized theory of gravity and it assumes that both gravity and electromagnetic field propagate locally in vacuum with the same speed which is equal to the fundamental speed $c$ in special theory of relativity \cite{LL}. Gravitational field is found as a solution of the Einstein field equations. The electromagnetic field is obtained by solving the Maxwell equations on the curved spacetime manifold derived at previous step from Einstein's equations. In the approximation of geometric optics the electromagnetic signals propagate along null geodesics of the metric tensor \cite{LL,mtw,wald}.

The problem of finding solutions of the null ray equations in curved spacetime attracted many researchers since the time of discovery of general relativity. Exact solutions of this problem were found in some, particularly simple cases of symmetric spacetimes like Schwarzschild's or Kerr's black hole, homogeneous and isotropic Friedman-Leme\^itre-Robertson-Walker (FLRW)\index{FLRW} cosmology, plane gravitational wave \index{gravitational wave!plane}, etc. \cite{LL,mtw,wald,mac}. For quite a long time these exact description of null geodesics was sufficient for the purposes of experimental gravitational physics. However, real physical spacetime has no symmetries and the solution of such current problems as direct detection of gravitational waves, interpretation of an anisotropy and polarization of cosmic microwave background radiation (CMBR)\index{CMBR}, exploration of inflationary models of the early universe, finding new experimental evidences in support of general relativity and quantum gravity\index{quantum gravity}, development of higher-precision relativistic algorithms for space missions, and many others, can not fully rely upon the mathematical techniques developed mainly in the symmetric spacetimes. Especially important is to find out a method of integration of equations of light geodesics in spacetimes having time-dependent gravitational perturbations of arbitrary multipole order.

In the present chapter we consider a case of an isolated astronomical system embedded to an asymptotically-flat spacetime. This excludes Friedmann-Leme\^itre-Robertson-Walker spacetime which is not asymptotically-flat \index{spacetime!asymptotically-flat}(see chapter \ref{chap7}). We assume that gravitational field of this system is characterized by an infinite set of gravitational multipoles emitting gravitational waves. Precise and coordinate-free mathematical definition of the asymptotic flatness is based on the concept of conformal infinity \index{conformal infinity}that was worked out in a series of papers \cite{bondi,sachs,penr,geroch,adm,ash} and we recommend the textbook of R. Wald \cite{wald} for a concise but comprehensive mathematical introduction to this subject. The asymptotic flatness implies the existence of a comformal compactification \index{conformal compactification} of the manifold but it may not exist for a particular case of an astrophysical system. There is also coordinate-free definition of gravitational multipoles of an isolated system given by Hansen \index{multipoles!Hansen}\cite{hansen}. However, they make sense only in stationary spacetimes but are not applicable for time-dependent gravitational fields which make them useless for practical analysis of real astronomical observations and for interpretation of relativistic effects in propagation of electromagnetic signals. We use neither Hansen's multipoles in the present chapter nor the comformal compactification technique. Moreover, because observable effects of gravitational waves are weak we shall consider only linear effects of gravitational multipoles so that relativistic effects produced by non-linearity of the gravitational field will be ignored. The linearized approximation of general theory of relativity represents a straightforward and practically useful approach to description of the multipole structure of the gravitational field of a localized astronomical system. The multipolar gravitational formalism was developed by several scientists, most notably by Kip Thorne \index{multipoles!Thorne}\cite{thorne} and Blanchet \& Damour \index{multipoles!Blanchet-Damour}\cite{bld1,bld2,bld3,bl,di}, and we use their results in the present chapter.

\begin{figure}
\includegraphics[width=\textwidth]{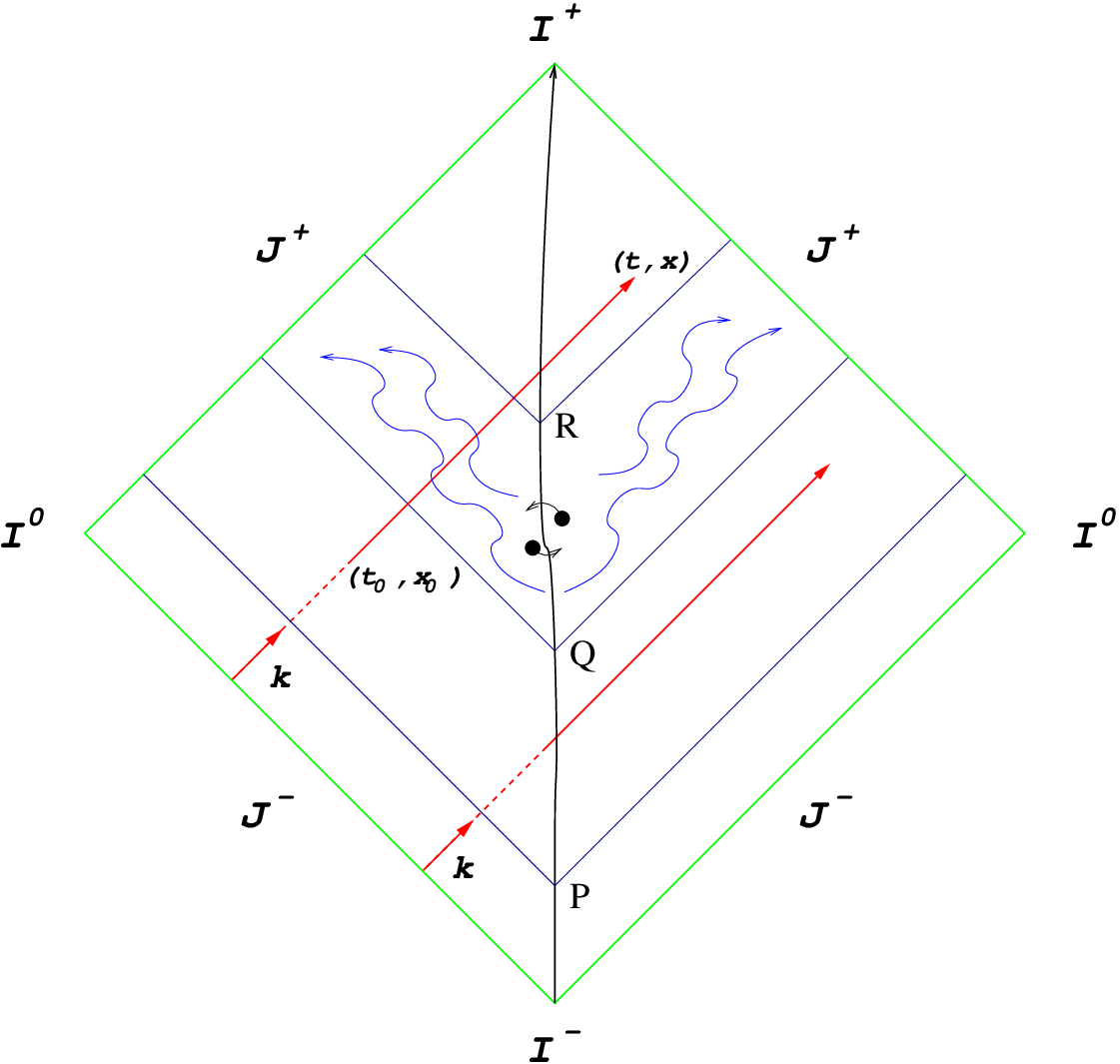}
\caption[The Penrose Diagram]{The Penrose diagram \index{Penrose diagarm}shows a worldline of the isolated system (binary star) originating at past timelike infinity\index{infinity!timelike} $I^-$ and ending at the future timelike infinity $I^+$. Both light and gravity propagate along null geodesics going from the past null infinity $J^-$ to the future null infinity\index{infinity!null} $J^+$. A particular light geodesic is emanating from the event $(t_0,{\bm x}_0)$ and ending at the event $(t,{\bm x})$. Extrapolation of this light geodesic\index{light geodesics} to the past null infinity defines the null wave vector $k^\alpha=(k^0,{\bm k})$ of the electromagnetic wave \index{electromagnetic wave}under consideration.}
\label{gugu}
\end{figure}

Exact formulation of the problem under discussion in this chapter is as follows (see Fig \ref{gugu}).
We assume (in a less restrictive sense than the comformal compactification technique does) that spacetime \index{spacetime}under consideration is asymptotically flat \index{spacetime!asymptotically flat} and
covered globally by a single coordinate chart\index{coordinate chart}, $x^\alpha=(ct, {\bm x})$, where $t$ is coordinate time \index{coordinate time} and ${\bm x}$ are spatial coordinates. An electromagnetic wave (photon) \index{photon}is emitted by a source of light at time $t_0$ at point ${\bm x}_0$ towards an observer which receives the photon at the instant of time $t$ and at the point ${\bm x}$. Photon propagates through the time-dependent gravitational field of the isolated astronomical system\index{astronomical system!isolated} emitting gravitational waves. The structure of the gravitational field is described by a set of Blanchet-Damour multipole moments of the system which are functions of the retarded time $t-r/c$, where $r$ is the distance from the isolated system to the field point and $c$ is the fundamental speed. The retardation is physically due to the finite speed of gravity which coincides in general relativity with the speed of light\index{speed of light} \index{light!speed}in vacuum. Some confusion may arise in the interpretation of the observational effects because both gravity and light propagates in general relativity with the same speed. The discrimination between various effects is still possible because gravity and light propagate to observer from different sources and along different spatial directions (characteristics of the null cone in spacetime)  \cite{kopeikin_book}.

Our task is to find out the relations connecting various physical parameters (direction of propagation\index{light!direction of propagation}, frequency\index{light!frequancy}, polarization\index{light!polarization}, intensity\index{light!intensity}, etc.) of the electromagnetic signal at the point of emission with those measured by the observer. Gravitational field affects propagation of the electromagnetic signal and changes its parameters along the light ray. Observing these changes allow us to study various properties of the gravitational field of the astronomical system and to detect gravitational waves with astronomical technique. The rest of the chapter is devoted to the mathematical solution of this problem.

\subsection{Historical background}

Propagation of electromagnetic signals through stationary gravitational field\index{gravitational field!stationary} is a rather well-known subject having been originally discussed in classic textbooks by \citep{60,31}. Presently, almost any standard textbook on relativity describes solution of the null geodesic equations in the field of a spherically-symmetric and rotating body or black hole. This solution is practical since it is used, for example, for interpretation of gravitational measurements and light-propagation experiments in the solar system \citep{kopeikin_book}. Another application is gravitational lensing in our galaxy and in cosmology \cite{gl1,gl2,gl3}. Continually growing accuracy of astronomical observations demands much better treatment of secondary effects in the propagation of light produced by the perturbations of the gravitational field associated with the higher-order gravitational multipoles \index{multipoles!gravitational}of planets and Sun\citep{32}. Time-dependent multipoles \index{multipoles!time dependent}emit gravitational waves which perturb propagation of light with the characteristic period of the gravitational wave. Influence of these periodic perturbations on light propagation parameters is important for developing correct strategy for understanding the principles of operation of gravitational wave detectors\index{gravitational wave detector} as well as for searching gravitational waves by astronomical techniques.

Among the most interesting sources of periodic gravitational waves
with a well-predicted behaviour are binary systems\index{binary system} comprised of two
stars orbiting each other around a common barycenter \index{barycenter} (center of mass). Indirect
evidence in support of existence of gravitational waves emitted by
binary pulsars was given by Joe Taylor with collaborators
\cite{joe,weist}. However, direct observation of gravitational
waves remains a challenging problem for experimental gravitational
physics. The expected spectrum of gravitational waves \index{gravitational waves!spectrum}
extends from $\sim 10^4$~Hz to $10^{-18}$~Hz \cite{thrn,sch}.
Within that range, the spectrum of periodic waves from known
binary systems extends from about $10^{-3}$~Hz -- the frequency of
gravitational radiation from a contact white-dwarf binary\index{binary system!white dwarf}
\cite{war}, through the $10^{-4}$ to $10^{-6}$~Hz -- the range of
radiation from the main-sequence binaries \index{binary system!main sequence stars}\cite{hi}, to the
$10^{-7}$ to $10^{-9}$~Hz -- the frequencies emitted by binary
supermassive black holes\index{black hole!supermassive} presumably lurking in active galactic
nuclei (AGN)\index{active galatic nuclei}\index{AGN} \cite{raj}. The dimensionless strain of these waves at the
Earth, $h$, may be as great as $10^{-21}$ at the highest
frequencies, and as great as $3\times 10^{-15}$ at the lowest
frequencies in the spectrum of gravitational waves
\cite{thrn,sch}.

Sazhin \cite{sazhin} was the first who suggested the method of detection of gravitational waves emitted
from a binary system by using timing observations of a background pulsar, more distant than the binary lying on the line of sight \index{line of sight} which 
passes sufficiently close in the sky to the binary. He had shown that
the integrated time delay for the
propagation of an electromagnetic pulse near the binary is proportional to $1/d^2$
where $d$ is the impact parameter \index{light ray!impact parameter}of the unperturbed trajectory of the signal. Similar idea was independently proposed  by Detweiler \cite{det} who has focused on discussing an application of pulsar timing for detection of a stochastic cosmological background of gravitational waves.
More recently,
Sazhin \& Saphonova \cite{ssap} have made estimates of the probability of
observation of such an effect for pulsars in globular clusters and
showed that the probability can be high, reaching 97\%. Sazhin-Detweiler idea is currently used in pulsar timing arrays to detect gravitational waves in nano-Hertz frequency band \citep{tinto_2011,liwex_2011,yardley_2011}.

Wahlquist \cite{9} proposed another approach to the detection of periodic
gravitational waves based on Doppler tracking \index{Doppel tracking}of spacecraft travelling
in deep space. His approach is restricted by the plane gravitational wave\index{gravitational waves!plane}
approximation developed earlier by Estabrook \& Wahlquist \cite{10}.
Tinto (\cite{11}, and references therein) made the most recent theoretical
contribution in this research area. The Doppler tracking technique is
used in deep space missions for detection of gravitational waves by seeking for the characteristic triple
signature in the continuously recorded phase of radio waves in the radio link between the ground station and spacecraft. The presence of this specific signature would indicate to the influence of the Doppler signal by a
gravitational wave crossing the line of sight from the spacecraft to
observer \cite{12,armstrong_2006LRR}.

Braginsky {\it et al.} \cite{13,14} raised a question about a possibility of using Very-Long Baseline Interferometry (VLBI)\index{VLBI}\index{Very-Long Baseline Interferometry}
as a detector of stochastic gravitational waves \index{gravitational waves!stochastic} produced in the early universe.  This idea
had been also investigated by Kaiser \& Jaffe \cite{15} and, the most prominently, by Pyne {\it et al.} \cite{16} and Gwinn {\it et al.} \cite{17}
who
showed that the overall effect in the time delay of VLBI signal is proportional to the strain of the metric
perturbation, $h$, caused by the plane gravitational wave. They also calculated the pattern of proper motions\index{proper motion} of quasars over all the sky as an indicator of the presence of quadrupole and high-order harmonics of ultra-long gravitational wave and
set an observational limit on the energy
density of such gravitational waves\index{gravitational waves!ultra-long} present in the early universe. Montanari
\cite{18} studied the perturbations of polarization\index{light!polarization} of electromagnetic radiation\index{radiation!electromagnetic} propagating in
the field of a plane gravitational wave\index{gravitational waves!plane} and found that the effects are
exceedingly small, unlikely to be observable.

Fakir (\cite{19,19+}, and references therein) has suggested
to use astrometry \index{astrometry}to detect the periodic variations in
apparent angular separations of appropriate light sources, caused by
gravitational waves emitted by isolated sources of gravitational
radiation. He was not able to develop a self-consistent approach to tackle the
problem with a necessary completeness and rigour. For this reason, his estimate of
the effect of the deflection of light caused by gravitational wave perturbation, is too optimistic. Another attempt to work out a more consistent approach
to the calculation of the light deflection angle by the radiation field of an arbitrary source of
gravitational waves has been undertaken by Durrer \cite{20}. However, the
calculations have been done only for the plane wave approximation. Nonetheless, the result
obtained was extrapolated to the case of
the localized source of gravitational waves without convincing justification. For this
reason the magnitude of the periodic changes of the light deflection angle was largely overestimated. The same
misinterpretation of
the effect of gravitational waves from localized sources can be found in the paper by Labeyrie \cite{21} who studied a
photometric modulation of background sources of light (stars) by gravitational waves
emitted by fast-orbiting binary stars. Because of the erroneous predictions, the expected detection of 
gravitational waves from VLBI
observations of a radio source GPS QSO 2022+171 undertaken by Pogrebenko
{\it et al.} \cite{22} was not based on firm theoretical ground and did not lead to success.

Damour \& Esposito-Far\`ese \cite{23} have studied the deflection of light and
integrated
time delay caused by the time-dependent gravitational field generated by a
localized astrophysical source lying in the sky close to the line of sight to a background source of light. They worked in a quadrupole approximation and explicitly calculated the effects of the retarded gravitational field of the astrophysical source in its
near, intermediate, and wave zones by making use of the Fourier-decomposition technique.
Contrary to the claims of Fakir \cite{19,19+} and Durrer
\cite{20} and in agreement with Sazhin's \cite{sazhin}
calculations, they found that the contribution of the wave-zone and intermediate-zone fields to the deflection angle vanish exactly due to some remarkable mutual cancellations of different components of the gravitational field.
The leading, total
time-dependent deflection of light is created only by the quasi-static, near-zone
quadrupolar part of the gravitational field.

Damour and Esposito-Farese \cite{23} analyzed propagation of light under a simplifying condition that the impact parameter of the light ray is small with respect to the distances from observer and the source of light to the isolated system. We have found \cite{smk1,ksh1} another way around to solve the problem of propagation of electromagnetic waves in the quadrupolar field of the gravitational waves emitted by the system without making any assumptions on mutual disposition of observer, source of light, and the system, thus, significantly improving and extending the result of paper \cite{23}.
At the same time the paper \cite{ksh1} did not answer the question about the impact of the other, higher-order gravitational multipoles of the isolated system on the process of propagation of electromagnetic signals. This might be important if the effective gravitational wave emission of an octupole and/or higher-order multipoles is equal or even exceeds that of the quadrupole as it may be in case of, for example, highly-asymmetric stellar collapse \cite{msmk}, nearly head-on collision of two stars, or break-up of a binary system caused by a recoil of two black holes \cite{bek}.

In the present chapter we work out a systematic approach to the problem of propagation of light rays in the field of arbitrary gravitational multipole.
While the most papers on light propagation consider both a light source and
an observer as being located at infinity we do not need these assumptions. For this reason, our approach is generic and applicable for any mutual configuration of the source of light and
observer with respect to the source of gravitational radiation. The
integration technique which we use for finding solution of the equations of
propagation of light rays was worked out in series of our papers \cite{smk1,ksh1,kokopol,kopmak_2007}.

The metric tensor and coordinate systems involved in our
calculations are described in section \ref{mtgc78} along with gauge
conditions imposed on the metric tensor. The equations of
propagation of electromagnetic waves in the geometric optics
approximation are discussed in section \ref{bn3e5} and the method of their
integration is given in section \ref{acv5d}. Exact solution of
the equations of light propagation and the exact form of relativistic perturbations
of the light trajectory and the coordinate speed of light are obtained in section \ref{dyb5k}. Section \ref{mq6yh} is devoted to
the derivation of the primary observable relativistic effects - the integrated time delay, the deflection angle, the frequency shift, and the rotation of the plane of polarization of an electromagnetic wave. We discuss in sections \ref{byq6f} and \ref{dg8j4} two
limiting cases of the most interesting
relative configurations of the source of light, the observer, and the source of
gravitational waves -- the gravitational-lens configuration (section \ref{byq6f}) and the case of a plane gravitational wave (section \ref{dg8j4}). 

\subsection{Notations and Conventions}\label{nac9h}

We consider a spacetime manifold \index{manifold} which is asymptotically flat at infinity \cite{tetrad}. Metric tensor \index{metric tensor} of the spacetime manifold is denoted by $g_{\alpha\beta}$
and its perturbation 
\be\la{aqz1}
h_{\alpha\beta}=g_{\alpha\beta}-\eta_{\alpha\beta}\;.
\ee
The determinant of the metric tensor is negative, and is denoted as $g={\rm det}[g_{\alpha\beta}]$. A four-dimensional, fully antisymmetric Levi-Civita symbol \index{Levi-Civita symbol} $\epsilon_{\alpha\beta\gamma\delta}$ is defined in accordance with the convention $\epsilon_{0123}=+1$.

In the present chapter we use a geometrodynamic system of units \index{system of units--geometrized} \cite {mtw} such that the fundamental speed, $c$, and the universal gravitational constant, $G$, are equal to
unity, that is $c=G=1$. spacetime is assumed to be globally covered by a Cartesian-like coordinate system $
(x^\a)\eq(x^0,x^1,x^2,x^3)\eq(t,x,y,z)\;, $ where $t$ and $(x,y,z)$ are time
and space coordinates respectively. This coordinate system is
reduced at infinity to the inertial Lorentz coordinates defined up to a
global Lorentz-Poincare transformation \index{Lorentz-Poincare transformation}\cite{fock}. Sometimes we shall
use spherical coordinates \index{coordinates!spherical} $ (r,\theta,\phi) $
related to $(x,y,z)$ by a standard transformation
\begin{equation}\la{aqz2}
x=r\sin\theta\cos\phi\;,\quad y=r\sin\theta\sin\phi\;,\quad
z=r\cos\theta\;.
\end{equation}
Spatial coordinates $(x^i)\eq(x,y,z)$ in some equations will be denoted with a boldface font, ${\bm x}\eq (x^i)$.

We shall operate with various geometric objects which have tensor indices. We agree that Greek (spacetime) indices $\a,\b,\g,\ldots$ range from 0 to 3, and Latin (space) indices $a,b,c,\ldots$ run from
1 to 3. If not specifically stated the opposite, the Greek indices are raised and lowered by means of the Minkowski metric \index{Minkowski metric}
$
\eta_{\alpha\beta}\equiv{\rm diag}(-1,1,1,1)
$, for example, $A^\a=\e^{\a\b}A_\b$, $B_{\a\b}=\e_{\a\m}\e_{\b\n}B^{\m\n}$, and so on. The spatial indices are raised and lowered with the help of the Kronecker symbol (a unit matrix), $\d_{ij}\equiv{\rm diag}(1,1,1)$. Regarding this rule the following conventions for the Cartesian coordinates \index{coordinates!Cartesian} hold: $x^i=x_i$ and $ x^0=-x_0$.

Repeated indices are summed over in accordance with Einstein's rule \cite{mtw}, for example,
\be\la{axz1}
A^{\a\b\g}B_{\a\m}\eq A^{0\b\g}B_{0\m}+A^{1\b\g}B_{1\m}+A^{2\b\g}B_{2\m}+A^{3\b\g}B_{3\m}\;.
\ee 
In the linearized (with respect to $G$) approximation of general relativity used in the present chapter, there is no difference between spatial vectors and co-vectors nor between upper and lower space indices. Therefore, we do not distinguish between the superscript and subscript spatial indeses. For example, for a dot (scalar) product of two space vectors we have
\begin{equation}
A^iB_i=A_iB_i\equiv A_1B_1+A_2B_2+A_3B_3\;.
\end{equation}
In what follows, we shall commonly use the spatial multi-index notations for three-dimensional, Cartesian tensors \cite{thorne} like this
\begin{equation}
 \il\equiv \mathcal{I}_{a_1\ldots a_l}\;.
\end{equation}
A tensor product of $l$ identical spatial vectors $k^i$ will be denoted as a three-dimensional tensor having $l$ indices
\begin{equation}
 k_{a_1}k_{a_2}\ldots k_{a_l}\equiv k_{a_1\ldots a_l}\;. \end{equation}
Full symmetrization with respect to a group of spatial indices of a Cartesian tensor \index{tensor!Cartesian} will be denoted with round brackets embracing the indices
\begin{equation}
Q_{(a_1\ldots a_l)}\equiv \frac{1}{l!}\sum_\sigma Q_{\sigma(1)\ldots\sigma(l)}\;,
\end{equation}
where $\sigma$ is the set of all permutations of $(1,2,...,l)$ which makes $Q_{a_1\ldots a_l}$ fully symmetric in $a_1\ldots a_l$.

It is convenient to introduce a special notation for symmetric trace-free (STF)\index{tensor!STF}\index{tensor!symmetric trace-free} Cartesian tensors by making use of angular brackets around STF indices. The explicit expression of the STF part of a tensor $Q_{a_1\ldots a_l}$ is \cite{thorne,bld1}
\begin{equation}
Q_{<a_1\ldots a_l>}\equiv\sum_{k=0}^{[l/2]}a^l_k\delta_{(a_1a_2}\cdot\cdot\cdot\delta_{a_{2k-1}a_{2k}}S_{a_{2k+1}\ldots a_l)b_1b_1\ldots b_kb_k}\;,
\end{equation}
where $[l/2]$ is the integer part of the number $l/2$, 
\begin{equation}
S_{a_{1}\ldots a_l}\equiv Q_{(a_1\ldots a_l)}\;, 
\end{equation}
and the numerical coefficients
\begin{equation}
a^l_k=\frac{(-1)^k}{(2k)!!}\frac{l!}{(2l-1)!!}\frac{(2l-2k-1)!!}{(l-2k)!}\;.
\end{equation}
We also assume that for any integer $l\ge 0$
\begin{equation}
 l!\equiv l(l-1)\ldots 2\cdot 1\;,\qquad 0!\equiv 1\;,
\end{equation}
and
\begin{equation}
 l!!\equiv l(l-2)(l-4)\ldots (2\mbox{ or} 1)\;,\qquad 0!!\equiv 1\;.
\end{equation}
One has, for example,
\ba\la{azq3} 
T_{<ab>}&=&T_{(ab)}-\frac13\d_{ab}T_{cc}\;,\\\la{azq4}
T_{<abc>}&=& T_{(abc)}
                    -\frac{1}5\delta_{ab}T_{(cjj)}
                      -\frac{1}5\delta_{bc}T_{(ajj)}
                        -\frac{1}5\delta_{ac}T_{(bjj)}\;,
\ea
and so on.

Cartesian tensors of the mass-type (mass) multipoles\index{multipole!mass-type} $\mathcal {I}_{<A_l>}$
and spin-type (spin) multipoles \index{multipoles!spin-type} $\mathcal {S}_{<A_l>}$ entirely
describing gravitational field outside of an isolated astronomical system \index{astronomical system!isolated}
 are always STF objects that can be checked by inspection of the definition following from the multipolar
 decomposition of the metric tensor perturbation $h_{\alpha\beta}$ \cite{thorne,bld1}. For this reason,
  to avoid the appearance of overcomplicated index notations we shall omit the angular
   brackets around the spatial indices of these (and only these) Cartesian tensors, that is we adopt: $\il\equiv\mathcal {I}_{<A_l>}$
    and $\mathcal {S}_{A_l}\equiv\mathcal {S}_{<A_l>}$.

We shall also use transverse (T) and transverse-traceless (TT) \index{tensor!Cartesian!transverse}\index{tensor!Cartesian!transverse-traceless}
Cartesian tensors in our calculations \cite{mtw,thorne,bld1}. These objects are defined by making
use of the operator of projection \index{operator of projection}
\be\la{azq5}P_{jk}\equiv
\delta_{jk}-k_{jk}\;,\ee 
onto the plane orthogonal to a unit vector $k_j$. This operator plays a role of a Kroneker symbol \index{Kroneker symbol!two-dimensional} in the two dimensional space in the sense that $P_{ij}P_{jk}=P_{ik}$, and $P_{ii}=2$. Definitions of the transverse and transverse-traceless tensors is
\cite{thorne,smk1}
\begin{eqnarray}
Q_{a_1\ldots a_l}^{\rm T}&\equiv& P_{a_1b_1}P_{a_2b_2}...P_{a_lb_l}Q_{b_1\ldots b_l}\;,
\\
Q_{a_1\ldots a_l}^{\rm TT}&\equiv&\sum_{k=0}^{[l/2]}b^l_k P_{(a_1a_2}\cdot\cdot\cdot P_{a_{2k-1}a_{2k}}W_{a_{2k+1}\ldots a_l)b_1b_1\ldots b_kb_k}\;,
\end{eqnarray}
where again $[l/2]$ is the integer part of $l/2$, $W_{a_{1}\ldots a_l}\equiv Q_{(a_1\ldots a_l)}^{\rm T}$,
and the numerical coefficients
\begin{equation}
b^l_k=\frac{(-1)^k}{4^{k}}\frac{l(l-k-1)!!}{k!(l-2k)!}\;.
\end{equation}
For instance,
\begin{equation}
Q_{ab}^{\rm TT}\equiv
P_{i(a}P_{b)j}Q_{ij}-\frac{1}2P_{ab}P_{jk}Q_{jk}\;.
\end{equation}

We shall also use the polynomial coefficients $C_l(p_1,\ldots ,p_n)$ in some of our equations \index{polynomial coefficients}. They are defined by
\begin{equation}\label{pco}
C_l(p_1,\ldots ,p_n)\equiv \frac{l!}{p_1!\ldots p_n!}\;,
\end{equation}
where $l$ and $p_i$ are positive integers such that
$\sum_{i=1}^{n}p_i=l$.
We introduce a Heaviside step function \index{Heviside step function},  $H(p-q)$, such that on the set of whole numbers
\begin{equation}
\label{H}
  H(p-q)=\begin{cases}0,& \text{if $p\leq q$\;,}\\
                  1,&\text{if $p > q$\;.}
                 \end{cases}
\end{equation}

Partial derivatives of any differentiable function, $f=f(t,{\bm x})$, are denoted as follows: $f_{,0}=\partial f/\partial t$ and $f_{,i}=\partial f/\partial x^i$. In general, comma standing after a function denotes a partial derivative with respect to a corresponding coordinate: $f,_\alpha\equiv \partial f(x)/\partial x^\alpha $. A dot above function denotes a total derivative of the function with respect to time
\be
\dot f\equiv df/dt =\partial f/\partial t+\dot x^i\partial f/\partial x^i\;,\ee 
where $\dot x^i$ denotes velocity along the integral curve $x^i=x^i(t)$ parametrized with coordinate time $t$. In this chapter the integral curves are light rays, and the derivatives $\dot x^i$ are taken along the light ray trajectory $(x^i)={\bm x}(t)$.
Sometimes the partial derivatives with respect to space coordinate $x^i$ will be also denoted as $\partial_i\equiv \partial /\partial x^i$, and the partial time derivative will be denoted as $\partial_t\equiv \partial /\partial t$. A covariant derivative with respect to the coordinate $x^\alpha$ will be denoted as $\nabla_\alpha$.

We shall introduce and distinguish notations for integrals taken with respect to time at a fixed spatial point, from those taken along a light-ray trajectory. Specifically,
the time integrals from a function $F(t, {\bm x})$, where ${\bm x}$ is a fixed point in space, are denoted as
\begin{equation}
\label{i1}
         F^{(-1)}(t, {\bm x})\equiv \int\limits_{-\infty}^t
                                                   F(\tau, {\bm x}) d\tau\;,
             \qquad\qquad
         F^{(-2)}(t, {\bm x})\equiv \int\limits_{-\infty}^t
                             F^{(-1)} (\tau, {\bm x}) d\tau\;.
\end{equation}
The time integrals from a function $F(t,{\bm x})$ taken on a light ray suggest that the spatial coordinate ${\bm x}$ is a function of time ${\bm x}\equiv{\bm x}(t)$, taken along the light ray. These integrals are denoted as
\begin{equation}
\label{i2}
          F^{[-1]}(t,{\bm x})\equiv
                    \int\limits_{-\infty}^{t}
                       F(\tau,{\bm x}(\tau))d\tau,
          \qquad\qquad
       F^{[-2]}(t,{\bm x})\equiv
                    \int\limits_{-\infty}^{t}
                         F^{[-1]}(\tau,{\bm x}(\tau))d\tau\;,
\end{equation}
where ${\bm x}\equiv{\bm x}(t)$ in the right side of these definitions.
The integrals in (\ref{i1}) represent functions of time, $t$, and spatial, ${\bm x}$, coordinates. The integrals in (\ref{i2}) are functions of time, $t$, only.

Partial time derivative of the order $p$ from a function $F(t,{\bm x})$ is denoted by
\begin{equation}
\label{tder2}
F^{(p)}(t,{\bm x})=\frac{\partial^p F(t,{\bm x})}{\partial t^p}\;,
\end{equation}
so that its action on the time integrals eliminates integration in the sense that
\begin{equation}
\label{tder3}
F^{(p)}(t,{\bm x})=\frac{\partial^{p+1}F^{(-1)}(t,{\bm x})}{\partial t^{p+1}}=\frac{\partial^{p+2}F^{(-2)}(t,{\bm x})}{\partial t^{p+2}}\;.
\end{equation}
Total time derivative of the order $p$ from a function $F(t,{\bm x})$ is denoted by
\be\label{tde2n}
F^{[p]}(t,{\bm x})=\frac{d^p F(t,{\bm x})}{dt^p}\;.
\end{equation}
The reader can easily confirm that
\begin{equation}
\label{tder4}
F^{[p]}(t,{\bm x})=\frac{d^{p+1}F^{[-1]}(t,{\bm x})}{dt^{p+1}}=\frac{d^{p+2}F^{[-2]}(t,{\bm x})}{dt^{p+2}}\;.
\end{equation}

In what follows, we shall denote spatial vectors by the bold italic letters, for instance, $A^i\equiv{\bm A}$, $k^i\equiv {\bm k}$, etc. The Euclidean dot product \index{dot product} between two spatial vectors, for example ${\bm a}$ and ${\bm b}$, is denoted with a dot between them: $a^ib_i={\bm a}\cdot{\bm b}$. The Euclidean wedge (cross) product \index{wedge product}\index{cross product} between two spatial vectors is denoted with a symbol $\times$, that is
$\epsilon_{ijk}a^jb^k=({\bm a}\times{\bm b})^i$.
Other particular notations will be introduced as soon as they appear in text.

\section{The Metric Tensor, Gauges and Coordinates}\label{mtgc78}

\subsection{The canonical form of the metric tensor perturbation}

We consider an isolated astronomical system emitting gravitational waves and assume that gravitational field is weak everywhere so that the metric tensor can be expanded in a Taylor series with respect to the powers of gravitational constant $G$ which labels the order of products of the metric tensor perturbations that are kept in the solution of the Einstein equations. We shall consider only a linearised post-Minkowskian approximation\index{approximation!post-Minkowskian} \index{post-Minkowskian approximation} of general relativity and discard all terms of the order of $G^2$ and higher. The  metric tensor is a linear combination of the Minkowski metric,\index{Minkowski metric} $\eta_{\alpha\beta}$, and a small perturbation $h_{\alpha\beta}$
\begin{eqnarray}
\label{m} g_{\alpha\beta}=\eta_{\alpha\beta}+Gh_{\alpha\beta}+O(G^2)\;,\\\label{m+}
g^{\alpha\beta}=\eta_{\alpha\beta}-Gh^{\alpha\beta}+O(G^2)\;,
\end{eqnarray}
where $h_{\alpha\beta}\ll 1$ and we use $\eta_{\alpha\beta}$ to rise and lower indices so that, for example, $h^{\alpha\beta}=\eta^{\alpha\mu}\eta^{\beta\nu}h_{\mu\nu}$.
In many cases the origin of the coordinates is placed to the center of mass of the astrophysical system. It eliminates the dipole component of the gravitational field which is associated with a coordinate degree of freedom. In some cases, however, it is necessary to keep the dipole component unrestricted in order to determine position of the center of mass of the system under consideration with respect to another coordinate chart which is introduced independently for solving some other astronomical problems. This is important for unambiguous interpretation of gravitational experiments done with astrometric instruments \citep{kopmak_2007,kopeikin_book}. Fact of the matter is that the displacement of the center of mass of an astrophysical system from the origin of the coordinates induces translational deformations of the higher-order multipole moments of the gravitational field which introduce a bias to the physical values of the multipoles. Therefore, physical interpretation of the observed values of the multipoles requires identification of the dipole moment and subtraction of the coordinate deformations caused by it. We shall discard the dipole component of the gravitational field in our solution.

The most general expression for the linearised perturbation of the metric tensor outside of the astronomical system emitting gravitational radiation was derived by Blanchet and Damour \cite{bld1} by solving Einstein's equations. The perturbation is given in terms of the symmetric and trace-free (STF) mass and spin multipole moments \index{multipole moment!symmetric trace-free}\index{multipole moment!mass-type}\index{multipole moment!spin-type}(similar formulas were derived by Thorne \citep{thorne}) and is described by the following expression
\begin{equation}
\label{gau1}
 h_{\alpha\beta}=h_{\alpha\beta}^{\rm can.}+
         w_{\alpha,\beta}+
              w_{\beta,\alpha}\;,
\end{equation}
where $w_\alpha$ are, the so-called, gauge functions describing the freedom in the choice of coordinates covering the manifold. The {\it canonical} perturbation, $h_{\alpha\beta}^{\rm can.}$, obeys the homogeneous wave equation in vacuum
\begin{equation}
\label{apo}
\Box h_{\alpha\beta}^{\rm can.}=0\;,
\end{equation}
which solution is chosen as
\begin{align}
\label{bm}
h_{00}^{\rm can.}(t,{\bm x})=&\frac{2{\Mc}}{r}+
      2\sum_{l=2}^{\infty}\frac{(-1)^l}{l!}
       \left[\frac{{\Ic}_{A_l}(t-r)}{r}\right]_{,A_l}\; ,
\\
h_{0i}^{\rm can.}(t,{\bm x})=& -\frac{2\epsilon_{ipq}{\Sc}_p(t-r)N_q}{r^2}-
                 4\sum_{l=2}^{\infty}\frac{(-1)^ll}{(l+1)!}
          \left[\frac{\epsilon_{ipq}
                   {\Sc}_{pA_{l-1}}(t-r)}{r}\right]_{,qA_{l-1}}+
 \\
               &4\sum_{l=2}^{\infty}\frac{(-1)^l}{l!}
          \left[\frac{\dot{\mathcal
          I}_{iA_{l-1}}(t-r)}{r}\right]_{,A_{l-1}}\;,\notag
\\
h_{ij}^{\rm can.}(t,{\bm x})=&\delta_{ij}h_{00}^{\rm can.}(t,{\bm x})+q_{ij}^{\rm can.}(t,{\bm x})\;,
\\
\label{em}
q_{ij}^{\rm can.}(t,{\bm x})=&4\!\sum_{l=2}^{\infty}\!\!\frac{(-1)^l}{l!}\!\!
       \left[\frac{\ddot{\Ic}_{ijA_{l-2}}(t-r)}{r}
           \right]_{,A_{l-2}}\!\!\!\!-
        \!8\!\sum_{l=2}^{\infty}\!\frac{(-1)^ll}{(l+1)!}\!\!
         \left[\frac{\epsilon_{pq(i}\dot
          {\Sc}_{j)pA_{l-2}}(t-r)}{r}\right]_{,qA_{l-2}}\;.
\end{align}
Here $\Mc$ and $\Sc_i$ are the total mass and spin (angular momentum) of the system, and
$\il $ and $\Sc_{A_l}$ are two independent sets of mass-type and
spin-type multipole moments, $N^i=x^i/r$ is a unit vector directed
from the origin of the coordinate system to the field point.
Because the origin of the coordinate system has been chosen at the
center of mass, the expansions \eqref{bm} -- \eqref{em} do not
depend on the mass-type dipole moment, $\Ic_i$, which is equal to zero by definition.
We emphasize that in the linearised approximation the total mass $\Mc$ and spin $\Sc_i$ of the
astronomical system are constant while all other multipoles are functions of
time which temporal behaviour obeys the equations of motion derived from the law of conservation of the stress-energy tensor of the system \cite{LL,mtw}. Gravitational waves emitted by
the system reduce its energy, linear and angular momenta. This effect does not appear in the linearised general relativity but
in higher order approximations we would obtain where the mass, spin, and linear momentum of the system must be
considered as functions of time like any other multipole. Higher-order gravitational perturbations in the metric going beyond (\ref{bm})--(\ref{em}) are shown in a review paper by Blanchet \citep{lrb}. They are not of concern in the present chapter.  

The {\it canonical} metric tensor (\ref{bm})--(\ref{em}) depends on the multipole moments
$\Ic_{A_l}(t-r)$ and $\Sc_{A_l}(t-r)$ taken at the retarded instant of time. The retardation is
explained by the finite speed of propagation of gravity (light propagation will be considered below). In the near zone of the isolated system
the retardation due to the propagation of gravity is small and all functions of time in the metric
tensor can be expanded in Taylor series around the present time $t$ \cite{thorne,bld1}. This near-zone expansion of
the metric tensor is called the post-Newtonian expansion leading to the post-Newtonian successive approximations \citep{kopeikin_book}. The post-Newtonian expansion can be smoothly matched to the solution of the linearized Einstein equations in the
domain of space being occupied by matter of the isolated system. The matching allows us to express the multipole moments in terms
of matter variables \cite{bld2}
\begin{equation}
\label{mjk}
\sigma\equiv T^{00}+T^{kk}\;,\qquad\qquad \sigma^i\equiv T^{0i}\;,
\end{equation}
where $T^{\a\b}$ is the stress-energy tensor \index{stress-energy tensor}of matter bounded in space.
In the first post-Newtonian approximation \index{approximation!post-Newtonian} the multipole moments have a matter-compact support \cite{bld2}
\begin{eqnarray}
\label{zqm}
\Ic^{\rm 1PN}_{A_l}&=&\int_V d^3{\bm x}\left\{x_{<A_l>}\sigma+\frac{|{\bm x}|^2x_{<A_l>}}{2(2l+3)}\partial^2_t\sigma-\frac{4(2l+1)x_{<iA_l>}}{(l+1)(2l+3)}\partial_t\sigma_i\right\}+O(c^{-4})\;,\\\nonumber\\\label{zqm1}
\Sc^{\rm 1PN}_{A_l}&=&\int_V d^3{\bm x}\epsilon_{pq<a_l}x_{A_{l-1}p>}\sigma_q+O(c^{-2})\;,
\end{eqnarray}
where notations have been explained in section \ref{nac9h}.
In the higher post-Newtonian approximations the multipole moments have contributions coming directly from the stress-energy tensor of gravitational field (Landau-Lifshitz pseudotensor) which have non-compact support. Therefore, the multipole moments are
expressed by more complicated functionals \cite{lrb}. Radiative approximation of the {\it canonical}
metric tensor reveals that contribution of the tails of gravitational waves \index{gravitational wave!tail} must be added to the definitions of the multipole
moments \eqref{zqm}, \eqref{zqm1} so that the multipole moments in the radiative zone of the isolated
system read \citep{bld3,bl,bks}
\begin{eqnarray}
\label{zqm2}
\Ic_{A_l}&=&\Ic^{\rm 1PN}_{A_l}+2\Mc\int\limits_0^{+\infty}d\zeta\ddot{\Ic}^{\rm 1PN}_{A_l}(t-r-\zeta)\left[\ln\left(\frac{\zeta}{2b}\right)+\frac{2l^2+5l+4}{l(l+1)(l+2)}+\sum_{k=1}^{l-2}\frac{1}{k}\right]\;,
\\\nonumber\\
\label{zqm3}
\Sc_{A_l}&=&\Sc^{\rm 1PN}_{A_l}+2\Mc\int\limits_0^{+\infty}d\zeta\ddot{\Sc}^{\rm 1PN}_{A_l}(t-r-\zeta)\left[\ln\left(\frac{\zeta}{2b}\right)+\frac{l-1}{l(l+1)}+\sum_{k=1}^{l-1}\frac{1}{k}\right]\;,
\end{eqnarray}
where $b$ is a normalization constant which value is supposed to be absorbed to the definition of the origin of
time scale in the radiative zone but this statement has not been checked so far.

\subsection{The harmonic coordinates}\index{coordinates!harmonic}

Equation (\ref{gau1}) holds in an arbitrary gauge imposed on the metric tensor.
The harmonic gauge is defined by the condition \citep{kopeikin_book}
\begin{equation}
\label{gai}
2h^{\alpha\beta}_{\;\;\,,\beta}-h^{,\alpha}=0\;,
\end{equation}
where $h\equiv h^\m_{\;\m}$. The gauge condition \eqref{gai} reduces the Einstein vacuum field equations to
the wave equation \eqref{apo} for the gravitational potentials $h^{\alpha\beta}$. Harmonic
coordinates $x^\alpha$ are defined as solutions of the homogeneous wave equation $\Box x^\alpha=0$ up to the
gauge functions $w^\alpha$. In particular, the harmonic {\it canonical} coordinates are defined by the condition
that all gauge functions $w_\alpha=0$. The {\it canonical} metric tensor \eqref{bm} -- \eqref{em} depends on
two sets of multipole moments \cite{thorne,bld1} which reflects the existence of only two degrees of freedom of a free (detached from matter) gravitational field in general relativity \cite{LL,mtw,wald}. At the same time one can obtain a generic
expression for the harmonic metric tensor by making use of infinitesimal
coordinate transformation 
\begin{equation}\label{coort}x'^\alpha=x^\alpha-w^\alpha\end{equation} from
the \textit{canonical harmonic} coordinates $x^\alpha $ to arbitrary\index{gauge!harmonic}\index{harmonic gauge}
harmonic coordinates $x'^\alpha $ with the harmonic gauge
functions $w^\alpha$ which satisfy to a homogeneous wave equation
\begin{equation}
\label{sor}
\Box w^\alpha=0\;.
\end{equation}
The most general solution of this vector equation contains four sets of STF multipoles \cite{thorne,bld1}
\begin{eqnarray}\label{poh}
w^0&=&\displaystyle{\sum_{l=0}^{\infty}}
\left[\frac{{\cal W}_{A_l}(t-r)}{r}\right]_{,A_l}\;,\\
\nonumber\\\nonumber\\\label{boh}
w^i&=&\displaystyle{\sum_{l=0}^{\infty}}
\left[\frac{{\cal X}_{A_l}(t-r)}{r}\right]_{,iA_l}+
\displaystyle{\sum_{l=1}^{\infty}}
\left[\frac{{\cal Y}_{iA_{l-1}}(t-r)}{r}\right]_{,A_{l-1}}+\\\nonumber&&
\displaystyle{\sum_{l=1}^{\infty}}\left[\epsilon_{ipq}\frac{{\cal
Z}_{qA_{l-1}}(t-r)}{r}\right]_{,pA_{l-1}}\;,
\end{eqnarray}
where ${\cal W}_{A_l}$, ${\cal X}_{A_l}$, ${\cal Y}_{iA_{l-1}}$, and ${\cal
Z}_{qA_{l-1}}$ are Cartesian tensors depending on the retarded time. Their specific form is a matter of computational convenience (or the boundary conditions) for derivation and interpretation of observable effects but it does not affect the invariant quantities like the phase of electromagnetic wave propagating through the field of the multipoles.

The most convenient choice simplifying the structure of the metric tensor perturbations, is given by the following gauge functions
\begin{align}
\label{w0}
       w^0=&\sum_{l=2}^{\infty}\ffrac{(-1)^l}{l!}
            \left[\frac{
            {\Ic}^{(-1)}_{A_l}(t-r)}{r}\right]_{,A_l},
\\
               w^i=& \sum_{l=2}^{\infty}\ffrac{(-1)^l}{l!}
\label{wi}
            \left[\frac{
            {\Ic}^{(-2)}_{A_l}(t-r)}{r}\right]_{,iA_l}-
\\
        &4\sum_{l=2}^{\infty}\ffrac{(-1)^l}{l!}
                 \left[\ffrac{{\Ic}_{iA_{l-1}}(t-r)}{r}
                                    \right]_{,A_{l-1}}+\notag
\\
&4\sum_{l=2}^{\infty}\ffrac{(-1)^ll}{(l+1)!}
              \left[\ffrac{\epsilon_{iba}
                             {\mathcal
                             S}^{(-1)}_{bA_{l-1}}(t-r)}{r}\right]_{,aA_{l-1}}.\notag
\end{align}

These functions, after they are substituted to equation \eqref{gau1}, transform the {\it canonical} metric tensor perturbation to a remarkably simple form
\begin{eqnarray}\label{adm1}
h_{00}&=&\frac{2{\cal M}}{r}\;,\\
\nonumber\\\label{adm2}
h_{0i}&=&-\frac{2\epsilon_{ipq}{\cal S}_p N_q}{r^2}\;,\\
\nonumber\\\label{adm3}
h_{ij}&=&\delta_{ij}h_{00}+h^{TT}_{ij}\;,\\
\nonumber\\\label{adm4}
h^{TT}_{ij}&=& P_{ijkl}q_{kl}^{\rm can.}\;,
\end{eqnarray}
where the TT-projection differential operator $P_{ijkl}$,
applied to the symmetric tensors depending on both time and spatial coordinates,
is given by
\begin{eqnarray}
P_{ijkl}&=&(\delta_{ik} - \Delta^{-1} \partial_i  \partial_k)
              (\delta_{jl} - \Delta^{-1} \partial_j  \partial_l) -
\frac{1}{2} (\delta_{ij} - \Delta^{-1} \partial_i  \partial_j)
              (\delta_{kl} - \Delta^{-1} \partial_k  \partial_l)\;,
\end{eqnarray}
and $\Delta$ and $\Delta^{-1}$  denote the Laplacian and the inverse Laplacian \index{Laplacian} respectively.

When comparing the {\it canonical} metric tensor with that given by equations \eqref{adm1}--\eqref{adm4} it is instructive to keep in mind that $^{(-2)}\ddot{\cal{I}}_{A_l}(t-r) = {\cal{I}}_{A_l}(t-r)$ and
$\Delta ({\cal{I}}_{A_l}(t-r)/r) =
\ddot{\cal{I}}_{A_l}(t-r)/r$ for $r \ne 0$. This is a consequence of the fact that function
$^{(-2)}\ddot{\cal{I}}_{A_l}(t-r)$ is a solution of
the homogeneous d'Lambert's
equation, that is, $\Box \left[^{(-2)}{\cal{I}}_{A_l}(t-r)/r\right] =0$
for $r \ne 0$. We also notice that ${\cal{I}}_{A_l}(t-r)/r=\Delta^{-1}\left[\ddot{\cal{I}}_{A_l}(t-r)/r\right]$ and ${^{(-2)}\cal{I}}_{A_l}(t-r)/r=\Delta^{-1}\left[{\cal{I}}_{A_l}(t-r)/r\right]$. The metric tensor harmonic perturbation (\ref{adm1})--(\ref{adm4}) is similar to the Coulomb gauge in electrodynamics \cite{jackson,LLE}.\index{gauge!Coulomb}

\subsection{The ADM coordinates}\index{coordinates!Arnowitt-Deser-Misner}

The  Arnowitt-Deser-Misner (ADM) gauge condition in the linear approximation is given by two equations \cite{adm}
\begin{eqnarray}\label{admgauge}
2h_{0i,i} - h_{ii,0} = 0\;, \quad \quad
3h_{ij,j} - h_{jj,i} = 0\;,
\end{eqnarray}
where the second equation holds exactly, and for any function $f=f(t,{\bm x})$ we use notations: $f_{,0}=\partial f/\partial t$ and $f_{,i}=\partial f/\partial x^i$.
For comparison, the harmonic gauge condition \eqref{gai} in the linear approximation
reads:
\begin{eqnarray}\label{harmon}
2h_{0i,i} - h_{ii,0} = h_{00,0}\;, \quad \quad
2h_{ij,j} - h_{jj,i} = - h_{00,i}\;.
\end{eqnarray}
The ADM gauge condition (\ref{admgauge}) brings the space-space component of the
metric to the following form
\begin{eqnarray}
\label{xxx}
g_{ij}&=& \delta_{ij} (1 + \frac{1}{3} h_{kk}) +
h^{TT}_{ij}\;,
\end{eqnarray}
where $h^{TT}_{ij}$ denotes the transverse-traceless part of $h_{ij}$ and $h_{kk}=3h_{00}$. The ADM and harmonic gauge conditions are not compatible inside the regions occupied by matter. However, outside of matter they can co-exist simultaneously. Indeed, it is straightforward to check out that the metric tensor \eqref{adm1}--\eqref{adm4} satisfies both the harmonic and the ADM gauge conditions in the linear approximation along with the assumption that $\dot{\cal M}=0$. This was first noticed in \cite{ksh1}. We call the coordinates in which the metric tensor is given by equations \eqref{adm1}--\eqref{adm4} as the ADM-harmonic coordinates.\index{coordinates!ADM-harmonic}\index{ADM-harmonic coordinates}

The experimental problem of detection of gravitational waves is reduced to the observation of motion of test particles in the field of the incident or incoming gravitational wave. These test particles are photons in the electromagnetic wave used in observations and mirrors in ground-based gravitational-wave detectors or pulsars and Earth in case of using a pulsar timing array. The gravitational wave affects propagation of photons and perturbs motion of the mirrors or pulsars and Earth. These perturbations must be explicitly calculated and clearly separated from noise to avoid possible misinterpretation of observable effects due to the gravitational wave.
It turns out that the \textit{canonical} form of the
metric tensor \eqref{bm} -- \eqref{em} is well-adapted for
performing an analytic integration of equations of light rays.
At the same time, freely-falling mirrors (or pulsars and Earth) experience
influence of gravitational waves emitted by the isolated
astronomical system and move with respect
to the coordinate grid of the \textit{canonical harmonic} coordinates in a complicated way. For this reason, the perturbations produced
by the gravitational waves on the light propagation get
mixed up with the motion of massive test particles in these coordinates.

Arnowitt, Deser and Misner
\cite{adm} showed that there exist {\it canonical ADM} coordinates which have a special property such that freely-falling massive particles are not moving with respect to this coordinates despite that they are
perturbed by the gravitational waves. This means that the ADM coordinates themselves are not inertial \index{coordinates!inettial} and, although have an advantage in treating motion of massive test particles, should be used with care in the interpretation of gravitational wave experiments. Making use of the {\it canonical ADM} coordinates simplifies analysis of the gravitational wave effects observed at gravitational wave observatories (LIGO, LISA, NGO, etc.) or by astronomical technique because the motion of observer (proof mass) is excluded from the equations. However, the mathematical structure of the metric tensor in the
\textit{canonical ADM} coordinates does not allow us to
directly integrate equations for light rays analytically because it contains terms that are instantaneous functions of time. Integrals from these instantaneous functions of time cannot be performed explicitly  \cite{ksh1}.

The ADM-harmonic coordinates have the advantages of both harmonic and ADM coordinates. Thus, the
ADM-harmonic coordinates allow us to get a full analytic solution of the
light-ray equations and to eliminate the effects produced by the
motion of observers with respect to the coordinate grid caused by the influence of gravitational waves. In other words, {\it all} physical effects produced by gravitational waves are contained merely in the solution of the equations of light propagation. This conclusion is, of course, valid in the linear approximation of general relativity and is not extended to the second approximation where gravitational-wave effects on light and motion of observers can not be disentangled and have to be analysed together.

Similar ideology based on the introduction of TT coordinates \index{coordinates!TT}, has been earlier applied for analysis of the output signal of the gravitational-wave detectors with freely-suspended masses \cite{weber,LL,mtw} placed to the field of a plane gravitational wave, that is at the distance far away from the localized astronomical system emitting gravitational waves where the curvature of the gravitational-wave front is negligible. Our ADM-harmonic coordinates are an essential generalization of the standard TT coordinates because they can be constructed at an arbitrary distance from the astronomical system, thus, covering the near, intermediate and radiative zones.

\section{Equations of Propagation of Electromagnetic Signals}\label{bn3e5}
\subsection{Maxwell equations in curved spacetime}

In this section we treat gravitational field exactly without approximation. Therefore, all indices are raised and lowered by means of the
metric tensor $g_{\alpha\beta}$ with $g^{\alpha\beta}$ defined in
accordance with the standard rule
$g^{\alpha\beta}g_{\beta\gamma}=\delta^\alpha_\beta$. The general
formalism describing the behavior of electromagnetic radiation in
an arbitrary gravitational field is well known and can be found, for example, in textbooks \cite{LL,mtw,penrose} or in reviews \cite{ehlers,frolov}. Electromagnetic field is defined in terms of the (complex)
electromagnetic tensor $F_{\alpha\beta}$ as a solution of the
Maxwell equations. In the high-frequency limit one can approximate
the electromagnetic tensor $F_{\alpha\beta}$ as \cite{LL,mtw}
\begin{equation}
\label{ff1}
F_{\alpha\beta}=\{A_{\alpha\beta}\exp(i\varphi)\}\;,
\end{equation}
where $A_{\alpha\beta}$ is a slowly varying (complex) amplitude
and $\varphi$ is a rapidly varying phase of the electromagnetic
wave which is called {\it eikonal} \cite{LL,synge}, and $i$ is the imaginary unit, $i^2=-1$. In the most general case of propagation of light in a transparent medium the eikonal is a complex function which real and imaginary parts are connected by the Kramers-Kr\''onig dispersion relations \cite{jackson}. We shall consider propagation of light in vacuum and neglect the imaginary part of the eikonal that is associated with absorption. Of course, the amplitude, $A_{\a\b}$, and phase, $\varphi$,  are functions of both time and spatial coordinates. 

The
source-free (vacuum) Maxwell equations are given by \cite{LL,mtw}
\begin{equation}
\label{m1a}
\nabla_\alpha F_{\beta\gamma}+\nabla_\beta F_{\gamma\alpha}+\nabla_\gamma
F_{\alpha\beta}=0\;,
\end{equation}
\begin{equation}
\label{m1b}
\nabla_\beta F^{\alpha\beta}=0\;,
\end{equation}
where $\nabla_\alpha$ denotes covariant differentiation. Taking a
covariant divergence from equation (\ref{m1a}), using
equation (\ref{m1b}) and applying the rule of commutation of covariant derivatives of a tensor field of a second rank, we obtain the covariant wave equation for the
electromagnetic field tensor
\begin{equation}
\label{m2}
{\Box}_g F_{\alpha\beta}+R_{\alpha\beta\mu\nu}F^{\mu\nu}+R_{\mu\alpha}F^{\;\;\mu}_{\beta}-R_{\mu\beta}F^{\;\;\mu}_{\alpha}=0\;,
\end{equation}
where ${\Box}_g\equiv g^{\a\b}\nabla_\alpha\nabla_\beta$, $R_{\alpha\beta\mu\nu}$ is the Riemann curvature tensor,\index{Riemann tensor}\index{tensor!Riemann} and $R_{\alpha\beta}=R^\gamma_{\;\;\alpha\gamma\beta}$ is the Ricci tensor\index{Ricci tensor}\index{tensor!Ricci} (definitions of the Riemann and Ricci tensors in this chapter are the same as in the textbook \cite{wald}). We consider the case of propagation of light in vacuum where the stress-energy tensor of matter, $T^{\a\b}$, is absent. Due to the Einstein equations it yields $R_{\alpha\beta}=0$. Hence, in our case \eqref{m2} is reduced to a more simple form
\begin{equation}
\label{m2+}
{\Box}_g F_{\alpha\beta}+R_{\alpha\beta\mu\nu}F^{\mu\nu}=0\;.
\end{equation}
Differential operator ${\Box}_g$ in \eqref{m2} taken along with the Riemann and Ricci tensors is called de Rham's operator \index{de Rham's operator} for the electromagnetic field \cite{mtw,rahm}.

\subsection{Maxwell equations in the geometric optics approximation}\label{goap}\index{approximation!geometric optics}\index{geometric optics}

Let us now assume that the electromagnetic tensor
$F_{\alpha\beta}$ shown in (\ref{ff1}) can be expanded
with respect to a small dimensionless parameter
$\varepsilon=\l_{\rm em}/L$ where ${\l_{\rm em}}$ is a characteristic wavelength of
the electromagnetic wave and $L$ is a characteristic radius of
spacetime curvature. The parameter $\varepsilon$ is a bookkeeping parameter of the high-frequency approximation in expansion of the electromagnetic field beyond the limit of the geometric optics.  
More specifically, we assume that the
expansion of the electromagnetic field given by equation
\eqref{ff1} has the following form \cite{mtw}
\begin{equation}
\label{m3}
F_{\alpha\beta}=\left(a_{\alpha\beta}+ \varepsilon
b_{\alpha\beta}+\varepsilon^2
c_{\alpha\beta}+...\right)\exp\left(\frac{i\varphi}{\varepsilon}\right)\;,
\end{equation}
where $a_{\a\b}$, $b_{\a\b}$, $c_{\a\b}$, etc. are functions of time and spatial coordinates.

Substituting expansion (\ref{m3}) into equation (\ref{m1a}), taking into account
a definition of the electromagnetic wave vector, $l_\alpha\equiv\partial\varphi/\partial x^\alpha$, and arranging
the terms with similar powers of $\varepsilon$, lead to the chain of equations
\begin{eqnarray}
\label{m4a}
l_\alpha\; a_{\beta\gamma}+l_\beta\; a_{\gamma\alpha}+l_\gamma\;
a_{\alpha\beta}&=&0\;,\\
\label{m4b}
\nabla_\alpha\; a_{\beta\gamma}+\nabla_\beta\; a_{\gamma\alpha}+\nabla_\gamma\;
a_{\alpha\beta}&=&-i\left(
l_\alpha\; b_{\beta\gamma}+l_\beta\; b_{\gamma\alpha}+l_\gamma\;
b_{\alpha\beta}\right)\;,
\end{eqnarray}
where we have neglected the effects of spacetime curvature which are of the order of $O(\varepsilon^2)$ that are too small to measure.

Similarly, equation (\ref{m1b}) gives a chain of equations
\begin{eqnarray}
\label{m5a}
l_\beta a^{\alpha\beta}&=&0\;,\\
\label{m5b}
\nabla_\beta a^{\alpha\beta}+il_\beta\; b^{\alpha\beta}&=&0\;,
\end{eqnarray}
where we again neglected the effects of spacetime curvature.

Equation (\ref{m5a}) implies that the amplitude, $a_{\a\b}$, of the electromagnetic field tensor is
orthogonal in the four-dimensional sense to a wave vector $l_\alpha$, at least, in the first approximation.
Contracting equation (\ref{m4a}) with $l_\alpha$ and accounting for
(\ref{m5a}), we find that the wave vector $l_\alpha$ is null, that is
\begin{equation}
\label{1q}
l_\alpha l^\alpha=0\;.
\end{equation}
Taking a covariant derivative from this equation and using the fact that 
\be\label{bh8v3}
\nabla_{[\beta}\;l_{\alpha]}=0\;,
\ee
because $l_\alpha=\nabla_\alpha\varphi$, one can show that the vector
$l_\alpha$ obeys the null geodesic equation\index{null geodesic}
\begin{equation}
\label{p0}
{l}^\beta\nabla_\beta{l}^\alpha=0\;.
\end{equation}
It means that the null vector ${l}^\alpha$ \index{null vector} is parallel
transported along itself in the curved spacetime. Equation \eqref{p0} can be expressed more explicitly
as
\begin{equation}
\label{ff2}
\frac{d{l}^\alpha}{d\sigma}+\Gamma^\alpha_{\beta\gamma}{l}^\beta
{l}^\gamma=0\;,
\end{equation}
where $\sigma$ is an affine parameter \index{affine parameter} along the light-ray trajectory, and
\begin{equation}
\label{bb1}
\Gamma^\alpha_{\beta\gamma}=\frac{1}{2}g^{\alpha\mu}\left(
\partial_\gamma g_{\mu\beta}+\partial_\beta g_{\mu\gamma}-
\partial_\mu g_{\beta\gamma}\right)\;,
\end{equation}
are the Christoffel symbols. \index{Christoffel symbols}

Finally, contracting equation (\ref{m4b}) with $l^\g$, and using (\ref{m4a}), (\ref{m5a}), (\ref{m5b}) along with (\ref{bh8v3}) we can show that in the first approximation
\begin{equation}
\label{m6}
l^\g\nabla_\g a_{\alpha\beta}+\vartheta a_{\alpha\beta}=0\;,
\end{equation}
where 
\begin{equation}\label{dzp}
\vartheta\equiv(1/2)\nabla_\alpha{l}^\alpha\;,
\end{equation}
is the expansion of the light-ray congruence \index{congruence} defined at each point of spacetime by the derivative of the wave vector ${l}^\alpha$.

Equation (\ref{m6}) represents the law of propagation of the tensor amplitude of electromagnetic wave along the light ray. In the most general general case, when the expansion $\vartheta\not=0$, the tensor amplitude of the electromagnetic wave is not parallel-transported along the light rays. It can be shown that the expansion $\vartheta$ of the light-ray congruence is defined only by the stationary components of the gravitational field of the isolated astronomical system determined by its mass ${\cal M}$, and spin ${\cal S}^i$, but it does not depend on the higher-order multipole moments. It means that gravitational waves do not contribute to the expansion of the light-ray congruence in the linearised approximation of general relativity and their impact on $\vartheta$ is postponed to the terms of the second order of magnitude with respect to the universal gravitational constant $G$.

\subsection{Electromagnetic eikonal and light-ray geodesics}\label{eeg}

\subsubsection*{The unperturbed congruence of light rays}

We have assumed that geometric optics approximation is valid and electromagnetic waves propagate in vacuum. We also assume that each electromagnetic wave has a wavelength $\l_{\rm em}$ much smaller than the characteristic wavelength $\l_{\rm gw}$ of gravitational waves emitted by the isolated astronomical system. Physical speed of light in vacuum, measured locally, is equal to the speed of propagation of gravitational waves, and is equal to the fundamantal speed $c$ in tangent Minkowski spacetime. We have neglected all relativistic effects associated with the curvature tensor of spacetime in equations of light propagation. In accordance with the consideration given in previous section \ref{goap}, a kinematic description of propagation of each electromagnetic wave can be given by tracking position of its phase $\varphi$, which is a null hypersurface in spacetime, as a function of time or by following the congruence of light rays that are orthogonal to the phase. Quantum electrodynamics tells us that the light rays are tracks of massless particles of the quantized electromagnetic field (photons) which are moving along light-ray geodesics defined by equation (\ref{ff2}).

Particular solution of these equations can be found after imposing the initial-boundary conditions
 \begin{equation}\label{ibc}
      {\bm x}(t_0)= {\bm x}_0 ,\qquad
     \frac {d {\bm x}(-\infty)}{dt}= {\bm k}\;.
 \end{equation}
These conditions determine the spatial position, $ {\bm x}_0$, of an electromagnetic signal (a photon\index{photon}) at the time of its emission, $t_0$, and the initial direction of its propagation given by the unit vector, $ {\bm k}$, at the past null infinity \index{past null infinity}, that is at the infinite spatial distance and at the infinite past \cite{mtw,wald} where the spacetime is assumed to be flat (see Fig. \ref{gugu}). We imply that vector $ {\bm k}$ is directed towards observer. Notice that the initial-boundary conditions \eqref{ibc} have been chosen as a matter of convenience only. Instead of them, we could chose two boundary conditions when both the point of emission and that of observation of the electromagnetic signal are fixed in time and space. It is always possible to convert solution of equations of the null geodesics given in terms of the initial-boundary conditions to that given in terms of the boundary conditions. We discuss it in section \ref{h6tv7}.

In the next sections we will derive an explicit form of equations of null geodesics and solve them by iterations with the initial boundary conditions (\ref{ibc}). At the first iteration we can neglect relativistic perturbation of photon's motion and approximate it by a straight line
\begin{equation}
\label{nt}x^i= x^i_N(t)\equiv x^i_0+k^i(t-t_0)\;,
\end{equation}
where $t_0$ is the time of emission of electromagnetic signal, $x_0^i$ are spatial coordinates of the source of the electromagnetic signal taken at the time $t_0$, and $k^i={\bm k}$ is the unit vector along the trajectory of photon's motion defined in \eqref{ibc}.

\begin{figure}
\includegraphics[width=\textwidth]{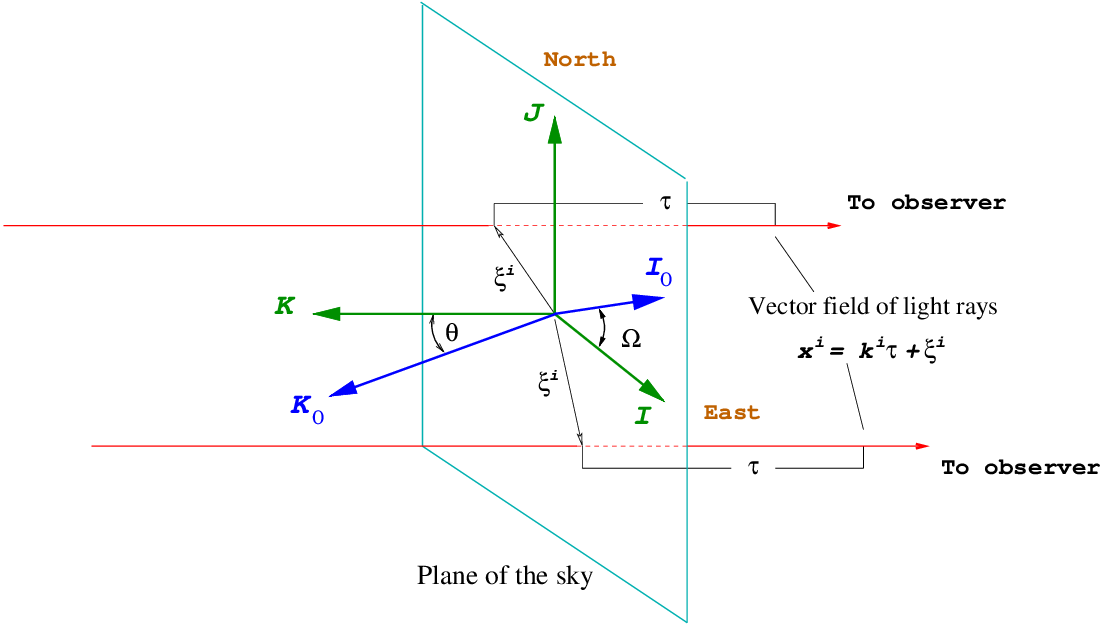}
\caption[Astronomical Coordinates]{Astronomical coordinates \index{coordinates!astronomical} used in calculation of light propagation. The origin of
the coordinates is at the center-of-mass of the source of gravitational
waves. The bundle of light rays is defined by a vector field $k^i$. Vector $K^i=-k^i+O(c^{-2})$ is directed from observer towards the source
of light. Vector $K_0^i$ is directed from the observer towards the source of gravitational
waves. We define $K_0^i=-N^i=-x^i/r$, where $x^i$ are the
coordinates of observer with respect to the source of gravitational waves, and
$r=| {\bm x}|$. The picture shows the plane of the sky\index{plane of the sky} being orthogonal to vector $K^i$.}
\label{bundle}
\end{figure}

The bundle of light rays makes 2+1 split of space by projecting any point in space onto the plane being orthogonal to the bundle (see Figure \ref{bundle}). 
This allows to make a transformation to new independent variables, $\tau$ and $\xi^i$, defined as follows
\begin{equation}
\label{nv} \tau=k_ix^i_N \qquad ,\qquad \xi^i=P^i_{\;j}x^j_N\;,
\end{equation}
where 
\be\label{pr8c2}
P^i_{\;j}=\delta^i_j-k^i k_j\;,
\ee
is the operator of projection on the plane being orthogonal to ${\bm k}$.
It is easy to see that the parameter $\tau$ is equivalent to time 
\begin{equation}
\label{tt*}
      \tau\equiv {\bm k}\cdot {\bm x}=t-t^*\;,
\end{equation}
where 
\be\label{z5d1a}
t^*\equiv {\bm k}\cdot{\bm x}_0-t_0\;,
\ee 
is the time of the closest approach of the electromagnetic signal to the origin of the spatial coordinates which is taken, in our case, coinciding with the center of mass of the isolated astronomical system. Because for each light ray the time $t^*$ is fixed, we conclude that the time differential $d\tau=dt$ on the light ray. The reader may expect that the results of our calculation of observable quantities are to depend on the parameters $\xi^i$ and $t^*$. This is, however, not true since $\xi^i$ and $t^*$ depend on the choice of the origin of the coordinates and direction of its spatial axes that is they are coordinate-dependent. The observable quantities have nothing to do with the choice of coordinates and, thus, $\xi^i$ and $t^*$ cannot enter the expressions for observable quantities. Inspection of the resulting equations in the sections which follow, shows that parameters  $\xi^i$ and $t^*$ do vanish from the observed quantities.

The unperturbed light-ray trajectory \eqref{nt} written in terms
of the new variables \eqref{nv} reads
\begin{equation}
\label{nt2}
x^i_N(\tau)=k^i\tau+\xi^i\;,
\end{equation}
so that the new variable $\xi^i\equiv{\bm \xi}= {\bm k}\times( {\bm x}\times {\bm k})$ should be understood as a vector drawn from the origin of the coordinate system towards the point of the closest approach of the ray to the origin. For vectors $k^i$ and $\xi^i$ are orthogonal, the unperturbed distance $r_N=\sqrt{x_i x^i}$ between the photon and the origin of the coordinate system
\begin{equation}\label{r_N}
r_N=\sqrt{\tau^2+d^2}\;,
\end{equation}
where $d=|{\bm \xi}|$ is the impact parameter of the unperturbed light-ray trajectory with respect to the coordinate origin.

We introduce two other operators of partial derivatives with respect to $\tau$ and $\xi^i$ determined for any smooth function taken on the congruence of light rays. These operators will be denoted with a hat above them and are defined as
\begin{equation}
\label{do}
\hat\partial_\tau\equiv\frac{\partial}{\partial\tau}\qquad,\qquad
\hat\partial_i\equiv P_i^{\;j}\frac{\partial}{\partial\xi^j}\;,\qquad,\qquad\hat\partial_{t^*}\equiv\frac{\partial}{\partial t^*}
\end{equation}
so that, for example,
\begin{equation}
\label{ort}
k^i\hat\partial_i =0\;.
\end{equation}

An important consequence of the projective structure of the bundle of the light rays is that for any smooth function $F(t, {\bm x})$ defined on the light-ray trajectories, one has
\begin{align}
\label{rf1}
\left[\left(\frac{\partial}{\partial x^i}+
       k_i\frac{\partial}{\partial t}\right)F(t\,,\,{\bm x})
        \right]_{{\bm x}= {\bm x}_0+{\bm k}(t-t_0)}      =&
    \left(\frac{\partial}{\partial\xi^i}+
       k_i\frac{\partial}{\partial\tau}\right)
        F(t^*+\tau\,,\,{\bm \xi}+{\bm k}\tau)\;,
\\\label{q23}
\left[\left(\frac{\partial}{\partial t}+k^i\frac{\partial}{\partial x^i}
       \right)F(t\,,\,{\bm x})
        \right]_{{\bm x}={\bm x}_0+{\bm k}(t-t_0)}      =&
       \frac{d}{d\tau}
        F(t^*+\tau\,,\,{\bm \xi}+{\bm k}\tau)\;,\\
 \label{rf2}
\left[\frac{\partial}{\partial t}F(t\,,\,{\bm x})\right]_{{\bm
x}={\bm x}_0+ {\bm k}(t-t_0)}=&\frac{\partial}{\partial t^*}
F(t^*+\tau\,,\,{\bm \xi}+{\bm k}\tau).
\end{align}
Here, in the left sides of Eqs. \eqref{rf1}--\eqref{rf2} one must, first, calculate the partial derivatives and only after that substitute the unperturbed trajectory of the light ray, $ {\bm x}={\bm
x}_0+ {\bm k}(t-t_0)$, while in the right side of these equations one, first, substitute the unperturbed trajectory parameterized by the variables $\tau$ and $\xi^i$ and, then, differentiate. Equations \eqref{rf1}--\eqref{rf2} define the commutation rule of interchanging the operations of substitution of the light ray trajectory to a function defined on spacetime manifold and the calculation of the partial derivatives from the function. It turns out to be very effective for analytic integration of the light-ray geodesic equations. 

It is worth noticing that \eqref{rf2} allows us to re-write \eqref{rf1} as follows
\begin{equation}
\label{za34}
\left[\frac{\partial F(t,{\bm x})}{\partial x^i}\right]_{{\bm x}= {\bm x}_0+{\bm k}(t-t_0)}
       =
    \left(\frac{\partial}{\partial\xi^i}+
       k_i\frac{\partial}{\partial\tau}-k_i\frac{\partial}{\partial t^*}\right)
        F(t^*+\tau\,,\,{\bm \xi}+{\bm k}\tau)\;.
\end{equation}
This equation will be used later for decomposing STF spatial derivatives from the potentials of gravitational field depending on retarded time.

\subsubsection*{The eikonal equation}

The eikonal $\varphi$ is related to the wave vector $l^\a$ of the electromagnetic wave as $l_\alpha=\partial_\alpha\varphi$. This definition along with equation \eqref{1q} immediately gives us a differential Hamilton-Jacobi equation for the eikonal propagation \citep{LL}
\begin{equation}
\label{2n}
g^{\alpha\beta}\frac{\partial\varphi}{\partial x^\alpha}\frac{\partial\varphi}{\partial x^\beta}=0\;.
\end{equation}
The unperturbed solution of this equation is a plane electromagnetic wave
\begin{equation}
\label{3s}
\varphi_N=\varphi_0+\omega k_\alpha x^\alpha\;,
\end{equation}
where $\varphi_0$ is a constant, $\omega=2\pi\nu_\infty$, $\nu_\infty$ is a constant frequency of the electromagnetic wave at infinity, and $k_\alpha$ is the unperturbed direction of the co-vector $l_\alpha$. Equation (\ref{2n}) assumes that $k_\alpha$ is a null co-vector with respect to the Minkowski metric in the sense that
\begin{equation}
\label{2z}
\eta^{\alpha\beta}k_\alpha k_\beta=0\;.
\end{equation}
We postulate that the co-vector $k_\alpha=(-1,{\bm k})$ where the unit Euclidean vector ${\bm k}$ is defined at past null infinity by equation \eqref{ibc}.

In the linearized approximation of general relativity the eikonal can be decomposed in a linear combination of unperturbed, $\varphi_N$, and perturbed, $\psi$, parts
\begin{equation}
\label{7z}
\varphi=\varphi_N+\omega\psi\;,
\end{equation}
so that the wave co-vector
\begin{equation}
\label{7ez}
l_\alpha=\omega\left(k_\alpha+\frac{\partial\psi}{\partial x^\alpha}\right)\;,
\end{equation}
Making use of equations \eqref{m+} and \eqref{2n}--\eqref{7ez} yield a partial differential equation of the first order for the perturbed part of the eikonal
\begin{equation}
\label{7b}
k^\alpha\frac{\partial\psi}{\partial x^\alpha}=\frac{1}{2}h_{\alpha\beta}k^\alpha k^\beta\;.
\end{equation}
This equation can be solved in all space by the method of characteristics \cite{mchar} which, in the case under consideration, are  the unperturbed light-ray geodesics given by Eq. (\ref{nt2}). Hence, after making use of relation \eqref{q23} one gets an ordinary differential equation for finding the eikonal perturbation
\begin{equation}
\label{7b+}
\frac{d\psi}{d\tau}=\frac{1}{2}h_{\alpha\beta}^{\rm can.}(\tau,{\bm\xi})k^\alpha k^\beta+\hat\partial_\tau\left(k^iw^i-w^0\right)\;,
\end{equation}
where both the gauge functions $w^\alpha$ and the canonical metric tensor perturbation $h_{\alpha\beta}^{\rm can.}$ are taken on the unperturbed light-ray trajectory. In particular, the components of the metric tensor perturbation have the following form
\begin{align}
\label{nm}
 h_{00}^{\rm can.}=&\frac {2\Mc}r
                        +2
                        \sum_{l=2}^\infty\sum_{p=0}^l\sum_{q=0}^p
                        \h\limits_{{\scriptscriptstyle (M)}}{}\!\!_{00}^{lpq}(t^*,\tau,{\bm\xi})\;,
\\
    h_{0i}^{\rm can.}=&-\frac{2\epsilon_{iba}\Sc^bN^a}{r^2}
                    +4\sum_{l=2}^\infty\sum_{p=0}^{l-1}\sum_{q=0}^p\left[
                                   \h\limits_{{\scriptscriptstyle (S)}}{}\!\!_{0i}^{lpq}(t^*,\tau,{\bm\xi})
                    +\h\limits_{{\scriptscriptstyle (M)}}{}\!\!_{0i}^{lpq}(t^*,\tau,{\bm\xi})\right]\;,
\\
\label{nhij}
h_{ij}^{\rm can.}=&\delta_{ij}h_{00}^{\rm can.}+q_{ij}^{\rm can.}\;,
\\
\label{nqij}
q_{ij}^{\rm can.}=&4\sum_{l=2}^\infty\sum_{p=0}^{l-2}\sum_{q=0}^p
                                     \q\limits_{{\scriptscriptstyle (M)}}{}\!\!_{ij}^{lpq}(t^*,\tau,{\bm\xi})
                   -8\sum_{l=2}^\infty\sum_{p=0}^{l-2}\sum_{q=0}^p
                                 \q\limits_{{\scriptscriptstyle
                                 (S)}}{}\!\!_{ij}^{lpq}(t^*,\tau,{\bm\xi})\;,
\end{align}
where
\begin{align}
 \h\limits_{{\scriptscriptstyle (M)}}{}\!\!_{00}^{lpq}(t^*,\tau,{\bm\xi})=&
\label{hoom}
        \frac{(-1)^{l+p-q}}{l!}C_l(l-p,p-q,q)
k_{<a_1\ldots a_p}\dksi_{a_{p+1}\ldots a_l>}
                  \dtau^{q}
                     \left[\frac{\Ic^{(p-q)}_{A_{l}}(t-r)}r\right]\;,
                     \\
\label{hoim}
 \h\limits_{{\scriptscriptstyle(M)}}{}\!\!_{0i}^{lpq}(t^*,\tau,{\bm\xi})=&
        \frac{(-1)^{l+p-q}}{l!}C_{l-1}(l-p-1,p-q,q)\times\\\nonumber&
      k_{<a_1\ldots a_p}\dksi_{a_{p+1}\ldots a_{l-1}>}
                              \dtau^{q}
                              \left[\frac{\Ic^{(p-q+1)}_{iA_{l-1}}(t-r)}r\right],
                                             \\
\label{hois}
\h\limits_{{\scriptscriptstyle(S)}}{}\!\!_{0i}^{lpq}(t^*,\tau,{\bm\xi})=&
  \frac{(-1)^{l+p-q}\;l}{(l+1)!}C_{l-1}(l-p-1,p-q,q)\times\\\nonumber&
                              (\dksi_a+k_a\dtau-k_a\dtz)\dtau^{q}
                            \left[\frac{\epsilon_{iab}\Sc^{(p-q)}_{bA_{l-1}}(t-r)}r   \right]\;,
       \\
\label{qijm}
 \q\limits_{{\scriptscriptstyle(M)}}{}\!\!_{ij}^{lpq}(t^*,\tau,{\bm\xi})=&
\frac{(-1)^{l+p-q}}{l!}C_{l-2}(l-p-2,p-q,q)\times\\\nonumber&
      k_{<a_1\ldots a_p}\dksi_{a_{p+1}\ldots a_{l-2}>}
                              \dtau^{q}
                              \left[\frac{\Ic^{(p-q+2)}_{ijA_{l-2}}(t-r)}r\right]\;,
         \\
\label{qijs}
\q\limits_{{\scriptscriptstyle(S)}}{}\!\!_{ij}^{lpq}(t^*,\tau,{\bm\xi})=&
        \frac{(-1)^{l+p-q}l}{(l+1)!}C_{l-2}(l-p-2,p-q,q)(\dksi_a+k_a\dtau-k_a\dtz)\times\\\nonumber&
      k_{<a_1\ldots a_p}\dksi_{a_{p+1}\ldots a_{l-2}>}
                              \dtau^{q}
                              \left[\frac{\epsilon_{ba(i}\Sc^{(p-q+1)}_{j)bA_{l-2}}(t-r)}r\right]\;.
\end{align}
All quantities in the right side of \eqref{hoom}--\eqref{qijs}, which are explicitly shown as functions of $x^i$, $r=|{\bm x}|$ and $t$, must be understood as taken on the unperturbed light-ray trajectory and expressed in
terms of $\xi^i$, $d=|{\bm\xi}|$, $\tau$ and $t^*$ in accordance with equations \eqref{tt*}, \eqref{r_N}. For example, the ratio $\Ic^{(p-q)}_{A_{l}}(t-r)/r$ in equation \eqref{hoom} must be understood as
\begin{equation}
\label{kob}
\frac{\Ic^{(p-q)}_{A_{l}}(t-r)}r\equiv\frac{\Ic^{(p-q)}_{A_{l}}(t^*+\tau-\sqrt{\tau^2+d^2})}{\sqrt{\tau^2+d^2}}\;,
\end{equation}
and the same replacement rule is applied to the other equations. After accounting for (\ref{nm})--(\ref{qijs}), equation \eqref{7b+} can be solved analytically with the mathematical technique shown in section \ref{int}.

\subsubsection*{Light-ray geodesics}

The geodesic equation for light rays is given in \eqref{ff2}. It is reduced to a more explicit form after making use of the linearized post-Minkowskian expressions for the Christoffel symbols 
\begin{eqnarray}
\label{bb2}
\Gamma^0_{00}&=&-\frac{1}{2}\partial_t h_{00}(t, {\bm x})\;,\\
\label{bb3}
\Gamma^0_{0i}&=&-\frac{1}{2}\partial_i h_{00}(t, {\bm x})\;,\\
\label{bb4}
\Gamma^0_{ij}&=&-\frac{1}{2}\left[\partial_i h_{0j}(t, {\bm x})+
\partial_j h_{0i}(t, {\bm x})-\partial_t h_{ij}(t, {\bm x})\right]\;,\\
\label{bb5}
\Gamma^i_{00}&=&\partial_t h_{0i}(t, {\bm x})-\frac{1}{2}
\partial_i h_{00}(t, {\bm x})\;,\\
\label{bb6}
\Gamma^i_{0j}&=&\frac{1}{2}\left[\partial_j h_{0i}(t, {\bm x})-
\partial_i h_{0j}(t, {\bm x})+\partial_t h_{ij}(t, {\bm x})\right]\;,\\
\label{bb7}
\Gamma^i_{jp}&=&\frac{1}{2}\left[\partial_j h_{ip}(t, {\bm x})+
\partial_p h_{ij}(t, {\bm x})-\partial_i h_{jp}(t, {\bm x})\right]\;.
\end{eqnarray}

The affine parameter $\sigma$ in this equation is an implicit function of the coordinate time $t$. Relation between $\sigma$ and $t$ is derived from the time component of \eqref{ff2}
\be\label{z3x1a}
\frac{d^2t}{d\sigma^2}+\left(\Gamma^0_{00}+\Gamma^0_{0p}\dot x^p+\Gamma^0_{pq}\dot x^p\dot x^q\ri)\left(\frac{dt}{d\s}\ri)^2=0\;,
\ee
where the dot above coordinates denote a derivative with respect to the coordinate time, $\dot x^p\equiv dx^p/dt$.
Effectively, there is no need to solve (\ref{z3x1a}) explicitly as we are not interested in the parameter $\s$ because the coordinate time $t$ is more practical parameter which can be measured with clocks. Therefore, we express the spatial components of the geodesic equation \eqref{ff2} in terms of the Christoffel symbols \eqref{bb2}--\eqref{bb7}, and replace differentiation with respect to the canonical parameter $\s$ by differentiation with respect to the coordinate time $t$.  With the help of equation (\ref{z3x1a}) the spatial components of the geodesic equation for light ray propagation becomes
\begin{align}
\label{eq1a}
 \ddot{x}^i(t)=&\frac{1}2 h_{00,i}-
                          h_{0i,0}-
                  \frac{1}2 h_{00,0}\dot{x}^i-
                             h_{ik,0}\dot{x}^k-
                            ( h_{0i,k}-
                               h_{0k,i})\dot{x}^k-\\
\nonumber
              &  h_{00,k} \dot{x}^k\dot{x}^i-
            \left( h_{ik,j}-
           \frac{1}2 h_{kj,i}\right)\dot{x}^k\dot{x}^j+
       \left(\frac{1}2h_{kj,0}-
                        h_{0k,j}\right)\dot{x}^k\dot{x}^j\dot{x}^i ,
\end{align}
where $h_{00}$, $h_{0i}$, $h_{ij} $ are components of the metric tensor taken on the unperturbed light-ray trajectory as shown in equations \eqref{nm}--\eqref{qijs}, that is $h_{\alpha\beta}\equiv
  h_{\alpha\beta}(t, {\bm x}_N(t))$.

Equation \eqref{eq1a} can be further simplified after substituting the unperturbed light-ray trajectory
\eqref{nt2} to the right side of Eq. \eqref{eq1a} and making use of equation \eqref{rf1}. Working in arbitrary coordinates one obtains
\begin{equation}
\label{eq2+}
\frac{d^2 x^i(\tau)}{d\tau^2}=
         \ffrac{1}{2}k^\alpha k^\beta\,\dksi_i h_{\alpha\beta}^{\rm can.}-
      \dtau
    \left( k^\alpha h_{i\alpha}^{\rm can.}-
       \ffrac{1}{2}\,k^ik^jk^p q^{\rm can.}_{jp}\right)
    -\dtautau(w^i-k^iw^0).
\end{equation}
where all functions in equation \eqref{eq2+} are
taken (before any differentiation) on the unperturbed light-ray
trajectory given by equation \eqref{nt2} and the gauge functions $w^\alpha$ (they are explained in (\ref{gau1})) have not yet been specified which means that equations \eqref{eq2+} are gauge-invariant. We discuss the choice of the gauge functions later on in next subsection.

The main advantage of the form \eqref{eq2+} to which we have reduced the light ray propagation equation (\ref{ff2}) is the convenience of its analytic integration. Indeed, when we integrate along the light-ray path the following rules, applied to any
smooth function $F(\tau,{\bm \xi})$, can be used
\begin{align}
\label{ip1kk}
             \int\frac{\partial}{\partial\tau}F(\tau,{\bm \xi})\,d\tau
                                             =&F(\tau,{\bm \xi})
                                +C({\bm\xi})\;,\\\nonumber\\
\label{ip2kk}
\int\frac{\partial}{\partial\xi^i}F(\tau,{\bm \xi})\,d\tau
  =&\frac{\partial}{\partial\xi^i}\int F(\tau,{\bm \xi})\,d\tau\;.
\end{align}
This means that terms which are represented as partial derivatives
with respect to the time parameter $\tau$ can be immediately integrated out
by making use of \eqref{ip1kk}. At the same time \eqref{ip2kk}
shows that one can change the order of integration
and differentiation with respect to the parameter $\xi^i$. It allows us to calculate the integral along the light ray from a more simple scalar expression  instead of integrating a vector function.
This technique will be demonstrated explicitly in next sections.

Equation \eqref{eq2+} is linear
with respect to the perturbation of the metric tensor,
$h_{\alpha\beta}$. Hence, it can be linearly decomposed in the 
equation for perturbations of the light-ray trajectory caused
separately by mass and spin multipole moments. Substitution of the
metric tensor \eqref{bm} -- \eqref{em} to Eq. \eqref{eq2+} and
replacement of spatial derivatives with respect to $x^i$ with
those with respect to parameters $\xi^i$ and $\tau$ by making use
of \eqref{rf2} yield the following linear superposition 
\begin{equation}
\label{ddx} \ddot x^i=\ddot{x}^i_{\scriptscriptstyle(G)}+\ddxim+\ddxis\;,
\end{equation}
where $\ddxim$ and $\ddxis$ are the components of photon's
coordinate acceleration caused by mass and spin multipoles of the
metric tensor respectively, and $\ddot{x}^i_{\scriptscriptstyle(G)}$ is the gauge-dependent acceleration. These components read
\begin{equation}
\label{ddxg}
\ddot{x}^i_{\scriptscriptstyle(G)}=\dtautau\left[(k^i\phi^0-\phi^i)+(k^i w^0-w^i)\right]\;,
\end{equation}
\begin{align}
\label{ddxm}
 \ddxim=&2(\dksi_i-k_i\dtau)\frac{\Mc}r+\\\notag&
2\dksi_i
   \sum_{l=2}^\infty\sum_{p=0}^l\sum_{q=0}^p
     \frac{(-1)^{l+p-q}}{l!}
    C_l(l-p,p-q,q)H(2-q) \times
\\
&\left(
    1-\frac{p-q}l
   \right)
       \left(
               1-\frac{p-q}{l-1}
         \right)
k_{<a_1\hdots a_p}\dksi_{a_{p+1}\hdots a_l>}
    \dtz^{p-q}\dtau^q
     \left[\ilrr\right]-
    \notag
\\\notag
            &2\dtau\sum_{l=2}^\infty\sum_{p=0}^l
            \frac{(-1)^{l+p}}{l!}
                                  C_l(l-p,p)
\left(
    1-\frac{p}l
  \right)\times\\\notag
&\left\{
    \left(
        1+\frac{p}{l-1}
      \right)
\right.
k_{i<a_1\hdots a_p}\dksi_{a_{p+1}\hdots a_l>}\dtz^p
       \left[\ilrr\right]-
    \notag
\\
    &\frac{2p}{l-1}
\left.
    k_{<a_1\hdots a_{p-1}}\dksi_{a_p\hdots a_{l-1}>}\dtz^p
        \left[\frac{{\Ic}_{iA_{l-1}}(t-r)}{r}\right]
             \right\}\;,\notag
\end{align}
and
\begin{align}
\label{ddxs} \ddxis=&
2\left(k_j\dksi_{ia}-\delta_{ij}\dksi_{a\tau}\right)\ejsbr-\\\notag&
4k_j\dksi_{ia}
    \sum_{l=2}^{\infty}\sum_{p=0}^{l-1}\sum_{q=0}^{p}
        \frac{(-1)^{l+p-q}l}{(l+1)!}
C_{l-1}\left(l-p-1,p-q,q\right)H(2-q)\times
\\&
    \left(1-\frac{p-q}{l-1}\right)
k_{<a_1\hdots a_p}\dksi_{a_{p+1}\hdots a_{l-1}>}
        \dtz^{p-q}\dtau^q\left[\ejslrr\right]+
    \notag
\\&
4\left(\dksi_a-k_a\dtz\right)\dtau
\sum_{l=2}^{\infty}\sum_{p=0}^{l-1}
        \frac{(-1)^{l+p}l}{(l+1)!}
C_{l-1}\left(l-p-1,p\right)\times
    \notag
\\ &\left(1-\frac{p}{l-1}\right)
k_{<a_1\hdots a_p}\dksi_{a_{p+1}\hdots a_{l-1}>}
        \dtz^p\left[\eislrr\right]+
    \notag
\\&
4k_j\dtau
\sum_{l=2}^{\infty}\sum_{p=0}^{l-1}
        \frac{(-1)^{l+p}l}{(l+1)!}
C_{l-1}\left(l-p-1,p\right)\\\notag&
k_{<a_1\hdots a_p}\dksi_{a_{p+1}\hdots a_{l-1}>}
        \dtz^p\left[ \frac {\epsilon_{jba_{l-1}}\mathcal {\dot S}_{\hat ibA_{l-2}}(t-r)}{r}\right]&&\;,\notag
\end{align}
where a dot above any function denotes a partial time derivative with respect to the parameter $t^*$ \cite{dsd},
$w^\alpha$ are the gauge functions coming from (\ref{gau1}), and $\phi^\alpha$ are the gauge
functions which appear as a result of integration of the light ray propagation equation.
These terms enter the equations of motion in the form of combination $k^i\phi^0-\phi^i$
given in next section. It is worth emphasizing that we do not intend to separate
$k^i\phi^0-\phi^i$ in two functions, $\phi^i$ and
$\phi^0 $ because such a separation is not unique while the linear combination
$k^i\phi^0-\phi^i$ is unambiguous. We have not combined functions
$\phi^\alpha$ with the gauge functions $w^\alpha$ for two reasons:
\begin{enumerate}
\item to indicate that the solution of
equations of light-ray geodesic, performed in one specific
coordinate system, leads to generation of terms which can be
eliminated by gauge transformation,
\item to simplify the final form
of the result of the integration as all terms with second and higher
order time derivatives are immediately integrated in accordance with \eqref{ip1kk}.
\end{enumerate}

Gauge functions $w^\alpha$ are still arbitrary which makes our equations gauge-invariant.
However, for the sake of physical interpretation of the
result of integration of equations of light-ray geodesic, we shall
choose a specific form of functions $w^\alpha$ to make our
coordinate system both harmonic and ADM which makes the coordinate description of motion of free-falling 
particles in these coordinates simple. Specific form of the gauge
functions $w^\alpha$ at arbitrary field point is shown in equations \eqref{poh}, \eqref{boh} and their form at any point on the light-ray trajectory is given in equations \eqref{nw0},\eqref{nwi}.

It is important to notice that all terms depending on mass-type multipoles of the order
$l$ in the right sides of \eqref{ddxm} have a numerical factor $1-p/l$ where $p$ is the summation index. Such terms vanish when $p=l$. It means that \eqref{ddxm} does not contain terms with the time derivatives of the order $l$ from the multipoles of the $l$-th order which, actually, describe gravitational wave emission from the astrophysical system because they decay slowly as $1/r$. The same is true with regard to the spin-type multipoles of the order $l-1$ in \eqref{ddxs} -- the time derivatives of the order of $l-1$ from the spin-type multipoles (which decay as $1/r$) vanish in the right side of \eqref{ddxs}. Explicit analytic integration of such terms would be impossible but they simply do not present in the solution of general relativistic equations of light propagation for the reason mentioned above. It is this property of the null geodesic in general relativity which prevents the amplification of the gravitational wave perturbation for a light ray propagating closely to an astrophysical system emitting multipolar gravitational radiation (binary star, etc.). This fact was established in \cite{ksh1} in a quadrupole
approximation and extended to any multipole in \citep{kokopol}. Present chapter confirms this result. The reader should notice that the cancellation of these terms occurs only in general relativity. 

\subsubsection*{Gauge freedom of equations of propagation of light}\label{gfa35}
Gauge functions ${w}^\alpha$, generating the
coordinate transformation from the canonical harmonic coordinates
to the ADM-harmonic ones, are given by equations \eqref{w0}, \eqref{wi}. They transform the metric tensor as follows
\begin{equation}
\label{amq}
h_{\alpha\beta}^{\rm can.}=h_{\alpha\beta}-\partial_\alpha w_\beta-\partial_\beta w_\alpha\;,
\end{equation}
where $h_{\alpha\beta}^{\rm can.}$ is the canonical form of the metric tensor in harmonic coordinates given by equations \eqref{bm}--\eqref{em} and $h_{\alpha\beta}$ is the metric tensor given in the ADM-harmonic coordinates by equations \eqref{adm1}--\eqref{adm4}.

The gauge functions taken on the light-ray trajectory and expressed in terms of the variables $\xi$ and $\tau$ can be written down in the next form
\begin{eqnarray}
\label{nw0}
{w}^0&=&\sum_{l=2}^\infty\sum_{p=0}^l\sum_{q=0}^p\int^{\tau+t^*}_{-\infty}du
    \h\limits_{{\scriptscriptstyle (M)}}{}\!\!_{00}^{lpq}(u,\tau,{\bm\xi}),
\\
\label{nwi}
{w}^i&=&(\dksi_i+k_i\dtau-k_i\dtz)
    \sum_{l=2}^\infty\sum_{p=0}^l\sum_{q=0}^p
   \int^{\tau+t^*}_{-\infty}dv\int^{v}_{-\infty}du
        \h\limits_{{\scriptscriptstyle (M)}}{}\!\!_{00}^{lpq}(u,\tau,{\bm\xi})\\\nonumber&&
        -4\sum_{l=2}^\infty\sum_{p=0}^{l-1}\sum_{q=0}^p
                    \int^{\tau+t^*}_{-\infty}du\left[
                    \h\limits_{{\scriptscriptstyle
                     (M)}}{}\!\!_{0i}^{lpq}(u,\tau,{\bm\xi})
                       +\h\limits_{{\scriptscriptstyle (S)}}{}\!\!_{0i}^{lpq}(u,\tau,{\bm\xi})\right]\;,
\end{eqnarray}
where $ \h\limits_{{\scriptscriptstyle (M)}}{}\!\!_{00}^{lpq}(u,\tau,{\bm\xi}),
\h\limits_{{\scriptscriptstyle (M)}}{}\!\!_{0i}^{lpq}(u,\tau,{\bm\xi})$ and
  $\h\limits_{{\scriptscriptstyle (S)}}{}\!\!_{0i}^{lpq}(u,\tau,{\bm\xi})$
are defined by the Eqs. \eqref{hoom}, \eqref{hoim} and
\eqref{hois} after making use of the substitution $t^*\rightarrow u$. It is worth noticing  the following relationships
\begin{align}
\label{nw0+}
\frac{\partial{w}^0}{\partial t^*}=&\sum_{l=2}^\infty\sum_{p=0}^l\sum_{q=0}^p
    \h\limits_{{\scriptscriptstyle (M)}}{}\!\!_{00}^{lpq}(t^*,\tau,{\bm\xi}),
\\
\label{nwi+}
\frac{\partial{w}^i}{\partial t^*}=&
    \sum_{l=2}^\infty\sum_{p=0}^l\sum_{q=0}^p
   \int^{\tau+t^*}_{-\infty}du(\dksi_i+k_i\dtau)
        \h\limits_{{\scriptscriptstyle (M)}}{}\!\!_{00}^{lpq}(u,\tau,{\bm\xi})\\\notag&
        -4\sum_{l=2}^\infty\sum_{p=0}^{l-1}\sum_{q=0}^p\left[
                    \h\limits_{{\scriptscriptstyle
                     (M)}}{}\!\!_{0i}^{lpq}(t^*,\tau,{\bm\xi})
                       +\h\limits_{{\scriptscriptstyle (S)}}{}\!\!_{0i}^{lpq}(t^*,\tau,{\bm\xi})\right]\;,
\intertext{and}\label{xox}
k^i\frac{\partial{w}^i}{\partial t^*}-\frac{\partial{w}^0}{\partial t^*}=&
\sum_{l=2}^\infty\sum_{p=0}^l\sum_{q=0}^p\left[
   \int^{\tau+t^*}_{-\infty}du\dtau
        \h\limits_{{\scriptscriptstyle (M)}}{}\!\!_{00}^{lpq}(u,\tau,{\bm\xi})-\h\limits_{{\scriptscriptstyle (M)}}{}\!\!_{00}^{lpq}(t^*,\tau,{\bm\xi})\right]\\\nonumber &
        -4\sum_{l=2}^\infty\sum_{p=0}^{l-1}\sum_{q=0}^p\left[
                    k^i\h\limits_{{\scriptscriptstyle
                     (M)}}{}\!\!_{0i}^{lpq}(t^*,\tau,{\bm\xi})
                       +k^i\h\limits_{{\scriptscriptstyle (S)}}{}\!\!_{0i}^{lpq}(t^*,\tau,{\bm\xi})\right]\;,
\end{align}
which are helpful in calculation of the gravitational shift of the frequency of light.

A linear combination, $k^i\phi^0-\phi^i$, of the gauge-dependent functions $\phi^\alpha$ that appear in \eqref{ddxg}, is given
by the expressions
\begin{equation}
\label{phiphi}
 k^i\phi^0-\phi^i=(k^i\phim^0-\phim^i)+(k^i\phis^0-\phis^i)\;,
\end{equation}
\begin{align}
\label{phim}k^i\phim^0-\phim^i=\;&
2\dksi_i\sum_{l=2}^\infty\sum_{p=2}^l\sum_{q=2}^p
          \frac{(-1)^{l+p-q}}{l!}
                              C_l(l-p,p-q,q)\times
       \\&
            \left(
    1-\frac{p-q}l
   \right)
       \left(
               1-\frac{p-q}{l-1}
         \right)
           k_{<a_1\ldots a_p}\dksi_{a_{p+1}\ldots a_l>}
                 \dtau^{q-2}
         \left[\frac{\Ic^{(p-q)}_{A_l}(t-r)}{r}\right]+
\notag \\&
     2\sum_{l=2}^\infty\sum_{p=1}^l\sum_{q=1}^p
           \frac{(-1)^{l+p-q}}{l!}C_l(l
                                    -p,p-q,q)\times
\notag \\&
      \left(1-\frac{p-q}l\right)\left\{
           \left(1+\frac{p-q}{l-1}\right)
         \right.
           k_{i<a_1\ldots a_p}\dksi_{a_{p+1}\ldots a_l>}
                  \dtau^{q-1}
         \left[\frac{\Ic^{(p-q)}_{A_l}(t-r)}{r}\right]-
\notag \\&
       2\frac{p-q}{l-1}
          \left.
           k_{<a_1\ldots a_{p-1}}\dksi_{a_p\ldots a_{l-1} >}
                   \dtau^{q-1}
            \left[\frac{\Ic^{(p-q)}_{iA_{l-1}}(t-r)}r\right]
          \right\},
\notag
\end{align}
\begin{align}
\label{phis}k^i\phis^0-\phis^i=\;& 2\frac{\epsilon_{iab}k^a\Sc^b}{r}+\\\notag& 4k_j\dksi_{ia}
    \sum_{l=3}^{\infty}\sum_{p=2}^{l-1}\sum_{q=2}^{p}
        \frac{(-1)^{l+p-q}\;l}{(l+1)!}
C_{l-1}\left(l-p-1,p-q,q\right)\times
\\&
    \left(1-\frac{p-q}{l-1}\right)
k_{<a_1\hdots a_p}\dksi_{a_{p+1}\hdots a_{l-1}>}
        \dtau^{q-2}\left[\frac{\epsilon_{jab}\Sc^{(p-q)}_{bA_{l-1}}(t-r)}{r}\right]-
    \notag
\\&
4\left(\dksi_a-k_a\dtz\right)
\sum_{l=2}^{\infty}\sum_{p=1}^{l-1}\sum_{q=1}^{p}
        \frac{(-1)^{l+p-q}l}{(l+1)!}
C_{l-1}\left(l-p-1,p-q,q\right)\times
    \notag
\\& \left(1-\frac{p-q}{l-1}\right)
k_{<a_1\hdots a_p}\dksi_{a_{p+1}\hdots a_{l-1}>}
        \dtau^{q-1}\left[\frac{\epsilon_{iab}\Sc^{(p-q)}_{bA_{l-1}}(t-r)}{r}\right]-
    \notag
\\&4k_a
\sum_{l=2}^{\infty}\sum_{p=0}^{l-1}\sum_{q=0}^{p}
        \frac{(-1)^{l+p-q}l}{(l+1)!}
C_{l-1}\left(l-p-1,p-q,q\right)\times
    \notag
\\& \left(1-\frac{p-q}{l-1}\right)
 k_{<a_1\hdots a_p}\dksi_{a_{p+1}\hdots a_{l-1}>}
        \dtau^q\left[\frac{\epsilon_{iab}\Sc^{(p-q)}_{bA_{l-1}}(t-r)}{r}\right]+
    \notag
\\&4k_j
\sum_{l=2}^{\infty}\sum_{p=1}^{l-1}\sum_{q=1}^{p}
        \frac{(-1)^{l+p-q}l}{(l+1)!}
C_{l-1}\left(l-p-1,p-q,q\right)\times
    \notag
\\&
k_{<a_1\hdots a_p}\dksi_{a_{p+1}\hdots a_{l-1}>}
        \dtau^{q-1}\left[ \frac {\epsilon_{jba_{l-1}}\mathcal {S}^{(p-q+1)}_{\hat ibA_{l-2}}(t-r)}r\right]
    \notag
\end{align}

\subsection{Polarization of light and the Stokes parameters}

\subsubsection*{Reference tetrad}
Propagation of electromagnetic fields in vacuum and evolution of their physical parameters in curved spacetime can be studied with various mathematical techniques. One of the most convenient techniques was worked by Newman and Penrose \cite{npf,tetrad} and is called the Newman-Penrose formalism \index{Newman-Penrose formalism}\cite{frolov}. This formalism introduces at each point of spacetime a null tetrad \index{null tetrad} of four vectors associated with the bundle of light rays defined by the electromagnetic wave vector field $l^\alpha$. The Newman-Penrose tetrad consists of two real and two complex null vectors -- $({l}^\alpha, n^\alpha, m^\alpha, \bar{m}^\alpha)$ -- where the bar above function indicates complex
conjugation. The null tetrad vectors are normalized in such a way that ${n}_\alpha l^\alpha=-1$ and $m_\alpha
\bar{m}^\alpha=+1$ are the only non-vanishing products among the four vectors of the tetrad.

The vectors of the null tetrad are not uniquely determined by specifying $l^\alpha$. Indeed, for a fixed direction $l^\alpha$ the normalization conditions for the tetrad vectors are preserved under the linear transformations (null rotation)\index{null rotation} \cite{npf,frolov}
\begin{eqnarray}
\label{abg5}
l'^\alpha&=&Al^\alpha\;,
\\\label{abg6}
n'^\alpha&=&A^{-1}\left(n^\alpha+\bar{B} m^\alpha+B\bar{m}^\alpha+B\bar{B} l^\alpha\right)\;,
\\\label{abg7}
m'^\alpha&=&e^{-i\Theta}\left(m^\alpha+\bar{B}l^\alpha\right)\;,
\\\label{abg8}
\bar{m}'^\alpha&=&e^{i\Theta}\left(\bar{m}^\alpha+Bl^\alpha\right)\;,
\end{eqnarray}
where $A,\Theta$ are real scalars and $B=B_1+iB_2$ is a complex scalar. These transformations form a four-parameter $(A,B_1,B_2,\Theta)$ subgroup of the homogeneous Lorentz group \index{Lorentz group} which is equivalent to the point-like Lorentz transformations \citep{penrose}.

For doing mathematical analysis of the intensity and polarization \index{polazrization} of electromagnetic waves it is useful to introduce a local orthonormal reference frame of observer moving with a four velocity $u^\alpha$ who is seeing the electromagnetic wave travelling in the positive direction of $z$ axis of the reference frame. It means that at each point of spacetime the observer uses a tetrad frame $e^\alpha_{\;(\beta)}=\left\{e^\alpha_{\;(0)}, e^\alpha_{\;(1)}, e^\alpha_{\;(2)}, e^\alpha_{\;(3)}\right\}$ defined in such a way that
\begin{equation}
\label{p3}
e^\alpha_{\;(0)}=u^\alpha\;\qquad\quad,\qquad\quad
e^\alpha_{\;(3)}=(-{l}_\alpha u^\alpha)^{-1}\left[{l}^\alpha+
({l}_\beta u^\beta)u^\alpha\right]\;,
\end{equation}
and two other vectors of the observer's tetrad, $e^\alpha_{\;(1)}$ and $e^\alpha_{\;(2)}$, are the unit space-like vectors being
orthogonal to each other as well as to $e^\alpha_{\;(0)}$ and
$e^\alpha_{\;(3)}$. In other words, vectors of the observer tetrad are subject to the following normalization conditions
\begin{equation}
\label{norc}
g_{\alpha\beta}e^\alpha_{\;(\mu)}e^\beta_{\;(\nu)}=\eta_{\mu\nu}\qquad\;,\qquad
\eta^{\mu\nu}e^\alpha_{\;(\mu)}e^\beta_{\;(\nu)}=g^{\alpha\beta} \;.
\end{equation}
It is worth noticing that the observer's tetrad $e^\a_{\b}$ has two group of indices. The indices without round brackets run from 0 to 3 and are associated with time and space coordinates. The indices enclosed in the round brackets numerate vectors of the tetrad and also run from 0 to 3. The coordinate-type indices of the tetrad have no relation to the tetrad indices. If one changes spacetime coordinates (passive coordinate transformation\index{passive coordinate transformation}) it does not affect the tetrad indices while the coordinate indices of the tetrad change in accordance with the transformation law for vectors. On the other hand, one can change the tetrad vectors at each point in spacetime by doing the Lorentz transformation (active coordinate transformation\index{active coordinate transformation}) without changing the coordinate chart \cite{tetrad}.

Let us define at each point of spacetime a coordinate basis of static observers
\begin{eqnarray}
\label{dqx}
E^\alpha_{\;(0)}&=&\left(1+\frac{1}{2}h_{00}\;,\;0\;,\;0\;,\;0\right)\;,\\
E^\alpha_{\;(1)}&=&\left(h_{0j}a^j\;,\; a^i-\frac{1}{2}h_{ij}a^j\right)\;,\\
E^\alpha_{\;(2)}&=&\left(h_{0j}b^j\;,\; b^i-\frac{1}{2}h_{ij}b^j\right)\;,\\
E^\alpha_{\;(3)}&=&\left(h_{0j}k^j\;,\; k^i-\frac{1}{2}h_{ij}k^j\right)\;,
\end{eqnarray}
which is written down for the case of weak gravitational field, $g_{\alpha\beta}=\eta_{\alpha\beta}+h_{\alpha\beta}$. Here
the unit spatial vectors ${\bm a}=(a^1,a^2,a^3)$, ${\bm b}=(b^1,b^2,b^3)$, and
${\bm k}=(k^1,k^2,k^3)$ are orthonormal in the Euclidean sense ($\d_{ij}a^ib^j=\d_{ij}a^ik^j=\d_{ij}b^ik^j=0$ and $\d_{ij}a^ia^j=\d_{ij}b^ib^j=\d_{ij}k^ik^j=1$) with vector ${\bm k}$ pointing to the direction of propagation of the light ray at infinity as given in (\ref{ibc}). 
These basis vectors are convenient to track the changes in the parameters of the electromagnetic wave as it travels from the point of emission of light to the point of its observation. 

The local tetrad $e^\alpha_{\;(\beta)}$ of observers moving with four-velocity $u^\a$ with respect to the static observers relates to the tetrad $E^\alpha_{\;(\beta)}$ by means of the Lorentz transformation
\begin{equation}
\label{lort}
e^\alpha_{\;(\beta)}=\Lambda^{\gamma}_{\;\beta}E^\alpha_{\;(\gamma)}\;,\qquad\qquad
E^\alpha_{\;(\beta)}=\lambda^{\gamma}_{\;\beta}e^\alpha_{\;(\gamma)}\;,
\end{equation}
where the matrix of the Lorentz transformation is \cite{mtw}
\begin{eqnarray}
\label{matr}
\Lambda^{0}_{\;0}&=&u^0\equiv\gamma\\\label{matr1}
\Lambda^{i}_{\;0}&=&\Lambda^{0}_{\;i}=u^{0}\;,\\\label{matr2}
\Lambda^{i}_{\;j}&=&\delta^{ij}+\frac{u^i u^j}{1+\gamma}\;,
\end{eqnarray}
and the inverse matrix of the Lorentz transformation $\lambda^{\alpha}_{\;\beta}$ is obtained from $\Lambda^{\alpha}_{\;\beta}$ by replacing $u^i\rightarrow -u^i$ that complies with the definition of the inverse matrix $\Lambda^{\alpha}_{\;\beta}\lambda^{\b}_{\;\g}=\d^\a_\g$.

The connection between the null tetrad \index{null tetrad} and the observer's tetrad frame, $e^\alpha_{\;(\beta)}$, is given by equations
\begin{eqnarray}
\label{0i}
l^\alpha&=&-(l_\gamma u^\gamma)\left[e^\alpha_{\;(0)}+e^\alpha_{\;(3)}\right]\;,\\\label{1i}
n^\alpha&=&-\frac{1}{2}(l_\gamma u^\gamma)\left[e^\alpha_{\;(0)}-e^\alpha_{\;(3)}\right]\;,\\\label{2i}
m^\alpha&=&\frac{1}{\sqrt{2}}\left[e^\alpha_{\;(1)}+ie^\alpha_{\;(2)}\right]\;,\\\label{3i}
\bar{m}^\alpha&=&\frac{1}{\sqrt{2}}\left[e^\alpha_{\;(1)}-ie^\alpha_{\;(2)}\right]\;.
\end{eqnarray}
Vector pairs $e^\alpha_{\;(0)}$, $e^\alpha_{\;(3)}$ and  $e^\alpha_{\;(1)}$, $e^\alpha_{\;(2)}$ split spacetime at the point in two sub-spaces. In particular, vectors $e^\alpha_{\;(1)}$, $e^\alpha_{\;(2)}$ defines the plane of polarization\index{polarization} in spacetime. 
If vectors $e^\alpha_{\;(0)}$ and
$e^\alpha_{\;(3)}$  are fixed, then,
vectors $e^\alpha_{\;(1)}$ and
$e^\alpha_{\;(2)}$ are defined up to an arbitrary rotation in the plane of polarization. Transformations \eqref{abg7}, \eqref{abg8} with $B=0$ yield
\begin{eqnarray}
\label{abg3}
e'^\alpha_{(1)}&=&\cos\Theta\; e^\alpha_{\;(1)}+\sin\Theta\; e^\alpha_{\;(2)}\;,
\\\label{abg4}
e'^\alpha_{(2)}&=&-\sin\Theta\; e^\alpha_{\;(1)}+\cos\Theta\; e^\alpha_{\;(2)}\;,
\end{eqnarray}
where $\Theta$ is the rotation angle of the vectors in the plane of polarization. We notice that, since vectors $e^\alpha_{\;(1)}$, $e^\alpha_{\;(2)}$ are orthogonal to $e^\alpha_{\;(0)}=u^\a$, the two null vectors, $m^\a$ and $\bar{m}^\a$, are also orthogonal to the four-velocity,
\be\label{nu2d6}
m_\a u^\a=0\qquad,\qquad\bar{m}_\a u^\a=0\;.
\ee
The null vector $n^\a$ is also orthogonal to $u^\a$, $n_\a u^\a=0$. On the other hand, the scalar product of $l^\a$ with four-velocity yields the angular frequency of electromagnetic wave, $\omega\equiv -l_\a u^\a$.

\subsubsection*{Propagation laws for the reference tetrad}\label{plrt}

Discussion of the rotation of the polarization plane and the
change of the Stokes parameters of electromagnetic radiation is
inconceivable without understanding of how the local
reference frame propagates along the light-ray geodesic from the point of emission of light to the point of its observation.
To this end we construct the reference tetrad frame of observer at the point of observation of light
and render a parallel transport of it backward in time along the light-ray geodesic. By
definition, vectors of the tetrad frame of the observer, $e^\alpha_{\;(\beta)}$, and those of the null tetrad
$(l^\alpha, n^\alpha, m^\alpha, \bar{m}^\alpha)$ do not change in a covariant sense as they are parallel
transported along the light ray. The propagation equation for
the tetrad vectors are, thus, obtained by applying the operator
$D\equiv l^\alpha\nabla_\alpha$ of the parallel transport along the null vector $l^\a$. Explicit form of the parallel transport of the reference tetrad is
\begin{equation}
\label{iuk}
\frac{de^\alpha_{\;(\mu)}}{d\s}+\Gamma^\alpha_{\beta\gamma}\,l^\beta\, e^\gamma_{\;(\mu)}=0\;,
\end{equation}
where $\s$ is an affine parameter along the light ray. Using definition of the Christoffel symbols (\ref{bb1}) and changing parameter $\s$ to the proper time $\t$ of the observer we recast (\ref{iuk}) to
\begin{equation}
\label{i9}
\frac{d}{d\tau}\left[e^\alpha_{\;(\mu)}+\frac{1}{2}h^\alpha_{\;\beta}e^\beta_{\;(\mu)}\right]=\frac{1}{2}\eta^{\alpha\nu}\left(\partial_\nu h_{\gamma\beta}-\partial_\gamma h_{\nu\beta}\right)k^\beta e^\gamma_{\;(\mu)}\;.
\end{equation}
The propagation of the null vectors $m^\alpha$ and  $\bar{m}^\alpha$ along the direction of the null vector $l^\a$ is given by 
\begin{eqnarray}
\label{qj}
\frac{dm^\alpha}{d\lambda}+\Gamma^\alpha_{\beta\gamma}\,l^\beta\, m^\gamma=0\;,\\\label{am}
\frac{d\bar{m}^\alpha}{d\lambda}+\Gamma^\alpha_{\beta\gamma}\,l^\beta\, \bar{m}^\gamma=0\;,
\end{eqnarray}
and the same laws are valid for $n^\alpha$ and $l^\alpha$ (see, for example, \eqref{ff2}).
Equations (\ref{i9})--\eqref{am} are the main equations for the discussion of the rotation of the plane of polarization and variation of the Stokes parameters.

\subsubsection*{Relativistic description of polarized electromagnetic radiation}

We consider propagation of a bundle of plane electromagnetic waves from the point of emission to the point of observation. Each of these waves have an electromagnetic tensor $F_{\alpha\beta}={\cal F}_{\alpha\beta}+O(\varepsilon)$ defined in the first approximation by equation \cite{ehlers,frolov}
\begin{eqnarray}
\label{p2}
{\cal F}_{\alpha\beta}&=&a_{\alpha\beta}\exp\left(\frac{i\varphi}{\varepsilon}\right)\;,\\\label{p2+}
a_{\alpha\beta}&=&\Phi\;{m}_{[\alpha} l_{\beta]}+
{\bar\Phi}\;\bar{m}_{[\alpha} {l}_{\beta]}
\end{eqnarray}
where $\Phi$ is a complex scalar amplitude of the wave with a real and imaginary components which are independent of each other in the most general case of incoherent radiation. In the proper frame of the observer with 4-velocity
$u^\alpha$ the components of the electric and magnetic field vectors are defined respectively as
$E^\alpha=-F^{\alpha\beta}u_\beta$ and $H^\alpha=(-1/2\sqrt{-g}) \epsilon^{\alpha\beta\gamma\delta}F_{\gamma\delta}u_\beta$ \citep{wald}. The electric field is a product of slowly-changing amplitude ${\cal E}_\alpha=-a_{\alpha\beta}u^\beta$ and fast-oscillating phase exponent
\begin{equation}
\label{rx}
E_\alpha={\cal E}_\alpha\exp\left(\frac{i\varphi}{\varepsilon}\right)\;.
\end{equation}
The polarization properties of electromagnetic radiation consisting of an ensemble of the waves with equal frequencies but different phases are defined by the components of the electric field measured by observer.
In the rest frame of the observer with 4-velocity $u^\alpha$,
the intensity and polarization properties of the electromagnetic radiation are described in terms of
the {\it polarization} tensor\index{polarization tensor}  \cite{LL, jackson}
\begin{equation}
\label{z2}
J_{\alpha\beta}=
<{E}_\alpha \bar{E}_\beta>=<{\cal E}_\alpha \bar{\cal E}_\beta> \;,
\end{equation}
where the angular brackets represent an average with respect to an ensemble of the electromagnetic waves with randomly distributed phases. This averaging eliminates all fast-oscillating terms from $J_{\alpha\beta}$. One has to notice \cite{LL} that the polarization tensor $J_{\alpha\beta}$ is symmetric only for a linearly polarized radiation. In all other cases, the polarization tensor is not symmetric. The polarization tensor
$J_{\alpha\beta}$ is purely spatial at the point of observation which means it is orthogonal to the four-velocity of observer,
$J_{\alpha\beta}u^\beta$. Furthermore, because the polarization tensor is defined in the sub-space of the polarization plane, it is orthogonal to the wave vector $J_{\alpha\beta}{l}^\beta=0$. This equality follows directly from its definition (\ref{z2}) and \eqref{m5a}.

The vector amplitude ${\cal E}_\alpha$ of the electric field can be decomposed in two independent components in the plane of polarization. Two vectors of the null tetrad, $m^\alpha,\bar{m}^\alpha$, form the circular-polarization basis, and vectors $e^\alpha_{\;(1)}, e^\alpha_{\;(2)}$ form a linear polarization basis. The decomposition reads
\begin{eqnarray}
\label{n5a}
{\cal E}^\alpha&=&{\cal E}_L m^\alpha+{\cal E}_R \bar{m}^\alpha\;,\\
\label{n5b}
{\cal E}^\alpha&=&{\cal E}_{(1)} e^\alpha_{\;(1)} +{\cal E}_{(2)} e^\alpha_{\;(2)}\;,
\end{eqnarray}
where
\begin{equation}
\label{pot}
{\cal E}_L=\frac{1}{2}\omega\Phi\qquad,\qquad {\cal E}_R=\frac{1}{2}\omega{\bar\Phi}\;,
\end{equation}
 are left and right circularly-polarized components of the electric field, 
\begin{equation}
\label{pot+}
{\cal E}_{(1)}=\frac{1}{\sqrt{2}}\left({\cal E}_L+{\cal E}_R\right)\qquad,\qquad {\cal E}_{(2)}=\frac{i}{\sqrt{2}}\left({\cal E}_L-{\cal E}_R\right)\;,
\end{equation}
are linearly polarized components, $\omega=-l_\alpha u^\alpha$ is the angular frequency of the electromagnetic wave, and we have taken into account the condition of orthogonality (\ref{nu2d6}).

There are for electromagnetic Stokes parameters $S_\a=(S_0,S_1,S_2,S_3)$.\index{Stokes parameters} They are defined by projecting the polarization tensor $J_{\a\b}$ on four independent combination of the tensor products of the two vectors, $e^\alpha_{\;(1)}$, $e^\alpha_{\;(2)}$ making up the polarization plane. More specifically, \cite{jackson,anil}
\begin{eqnarray}
\label{z4}
S_0&=&\phantom{i}J_{\alpha\beta}\left[e^\alpha_{\;(1)}e^\beta_{\;(1)}+
e^\alpha_{\;(2)}e^\beta_{\;(2)}\right]\;,\\
\label{z5}
S_1&=&\phantom{i}J_{\alpha\beta}\left[e^\alpha_{\;(1)}e^\beta_{\;(1)}-
e^\alpha_{\;(2)}e^\beta_{\;(2)}\right]\;,\\
\label{z6}
S_2&=&\phantom{i}J_{\alpha\beta}\left[e^\alpha_{\;(1)}e^\beta_{\;(2)}+
e^\alpha_{\;(2)}e^\beta_{\;(1)}\right]\;,\\
\label{z7}
S_3&=&iJ_{\alpha\beta}\left[e^\alpha_{\;(1)}e^\beta_{\;(2)}-
e^\alpha_{\;(2)}e^\beta_{\;(1)}\right]\;,
\end{eqnarray}
where $S_0\equiv I$ is the intensity\index{intensity!electromagnetic radiation}, $S_1\equiv Q$ and $S_2\equiv U$ characterize the degree of a linear polarization, and $S_3\equiv V$ is the degree of a circular polarization\index{degree of polarization} of the electromagnetic wave.

Making use of equation (\ref{z2}) in (\ref{z4})--(\ref{z7}) allows us to represent the Stokes parameters in the linear polarization basis as follows \citep{jackson}
\begin{eqnarray}
\label{z8}
S_0&=&\phantom{i}<|{\cal E}_{(1)}|^2+|{\cal E}_{(2)}|^2>\;,\\
\label{z9}
S_1&=&\phantom{i}<|{\cal E}_{(1)}|^2-|{\cal E}_{(2)}|^2>\;,\\
\label{z10}
S_2&=&\phantom{i}<{\cal E}_{(1)}\bar{\cal E}_{(2)}+\bar{\cal E}_{(1)}{\cal E}_{(2)}>\;,\\
\label{z11}
S_3&=&i<{\cal E}_{(1)}\bar{\cal E}_{(2)}-\bar{\cal E}_{(1)}{\cal E}_{(2)}>\;,
\end{eqnarray}
where ${\cal E}_{(n)}={\cal E}_\alpha e^\alpha_{\;(n)}$ for $n=1,2$.   

It is important to emphasize that though the Stokes parameters have four components, they do not form a 4-dimensional spacetime vector because they do not behave like a vector under transformations of the Lorentz group \cite{LL,jackson}. Indeed, if we perform a pure Lorentz boost all four Stokes parameters remain invariant \cite{LL}. However, for a constant rotation of angle $\Theta$ in the polarization plane, the Stokes parameters transform as \cite{LL}
\begin{eqnarray}
\label{jp4}
S_0'&=&S_0\;,\\\label{jp5}
S_1'&=&S_1\cos 2\Theta+S_2\sin 2\Theta\;,
\\\label{jp6}
S_2'&=&S_1\cos 2\Theta-S_2\sin 2\Theta\;,
\\\label{jp7}
S_3'&=&S_3\;.
\end{eqnarray}
This is what would be expected for a spin-1/2 field. That is, under a duality rotation of $\Theta=\pi/4$, one linear polarization state turns into the other, while the circular polarization state remains the same. The transformation properties \eqref{jp4}--\eqref{jp7} of the Stokes parameters point out that the Stokes parameters $S_1, S_2$ represent a linearly polarized components, and $S_3$ represents a circularly polarized component.

The polarization vector ${\bm P}=(P_1,P_2,P_3)$ \index{polarization vector} and the degree of polarization\index{degree of polarization} $P=|{\bm P}|$ of the electromagnetic radiation can be defined in terms of the normalized Stokes parameters by ${\bm P}=(S_1/I,S_2/I,S_3/I)$. Any partially polarized wave may be thought of as an incoherent superposition of a completely polarized wave with the degree of polarization $P$ and the polarization vector ${\bm P}$, and a completely unpolarized wave with the degree of polarization $1-P$ and nil polarization vector, ${\bm P}=0$, so that for an arbitrary polarized radiation one has: $(S_0,S_1,S_2,S_3)=I(P,P_1,P_2,P_3)+I(1-P,0,0,0)$. For completely polarized waves, vector ${\bm P}$ describes the surface of the unit sphere introduced by Poincar\'e \cite{LL}. The center of the Poincar\'e sphere \index{Poincar\'e sphere} corresponds to an unpolarized radiation and the interior to a partially polarized radiation. Orthogonally polarized waves represent any two conjugate points on the Poincar\'e sphere. In particular, $(P_1=\pm 1,P_2=0,P_3=0)$, and $(P_1=0, P_2=0,P_3=\pm 1)$ represent orthogonally polarized waves corresponding to the linear and circular polarization bases, respectively.

\subsubsection*{Propagation law of the Stokes parameters}

Taking definition (\ref{p2+}) of the electromagnetic tensor and accounting for the parallel transport of the null vectors $l^\alpha$, $m^\alpha$, $\bar{m}^\alpha$ along the light ray and the laws of propagation of the electromagnetic tensor given by equations \eqref{m6}, yield the law of propagation of the complex scalar functions $\Phi$ and ${\bar\Phi}$
\begin{eqnarray}
\label{b12}
\frac{d\Phi}{d\s}+\vartheta\Phi&=&0\;,\\
\label{b19}
\frac{d{\bar\Phi}}{d\s}+\vartheta{\bar\Phi}&=&0\;,
\end{eqnarray}
where $\s$ is an affine parameter along the ray and $\vartheta$ is the expansion of the light-ray congruence defined in \eqref{dzp}. 

Let us consider a sufficiently small, two-dimensional area ${\cal A}$ in the cross-section of the congruence of light rays lying on a null hypersurface of constant phase $\varphi$ that is in the polarization plane. The law of transportation of the cross-sectional area is \cite{mtw,gl1,frolov}
\begin{equation}
\label{p41}
\frac{d{\cal A}}{d\s}-2\vartheta{\cal A}=0\;.
\end{equation}
Thus, the product, ${\cal A}|\Phi|^2={\cal A}\Phi\bar\Phi$, remains constant along the congruence of light rays:
\begin{equation}\label{pi1}
l^\a\nabla_\a\left({\cal A}|\Phi|^2\right)=0\;.
\end{equation}
This law of propagation for the product ${\cal A}|\Phi|^2$ corresponds to the conservation of photon's flux \cite{mtw,gl1}.

The law of conservation of the number of photons propagating along the light ray, corresponds to the propagation law of vector $|\Phi|^2 l^\alpha$. Indeed, taking covariant divergence of this quantity and making use of the equations \eqref{b12}, \eqref{b19} along with definition \eqref{dzp} for the expansion $\vartheta$ of the bundle of light rays, yields \cite{mtw,gl1}
\begin{equation}\label{pid}
\nabla_\alpha\left(|\Phi|^2 l^\alpha\right)=0\;.
\end{equation}
This equation assumes that the scalar amplitude $\Phi$ of the electromagnetic wave can be interpreted in terms of the number density, ${\cal N}$, of photons in phase space and the energy of one photon, $E_{\omega}=\hbar\omega$, as follows 
\begin{equation}\label{bqpo}
|\Phi|=\sqrt{8\pi}\hbar\left(\frac{{\cal N}}{E_{\omega}}\right)^{1/2}\;,
\end{equation}
where the reduced Planck constant\index{Planck constant} $\hbar=h/2\pi$ and the normalizing factor were introduced for consistency between classical and quantum definitions of the energy of an electromagnetic wave \citep{mtw}.

Each of the Stokes parameter is proportional to the square of frequency of light, $\omega=-u^\alpha l_\alpha$, as directly follows from equations \eqref{z8}--\eqref{z11} and \eqref{pot+}. Therefore, the variation of the Stokes parameters along the light ray can be obtained directly from their definitions (\ref{z4})--(\ref{z7}) along the light ray and making use of the laws of propagation \eqref{b12}, \eqref{b19}. However, the set of the Stokes parameters $S_\a$ $(\a=0,1,2,3)$ is not directly observed in astronomy and we do not discuss their laws of propagation. Instead the set of four other polarization parameters $(F_\omega,P_1,P_2,P_3)$ is practically measured \cite{gl1,plwa} and we focus on the discussion of the laws of propagation for these parameters. Here ${\bm P}=(P_1,P_2,P_3)$ is the polarization vector as defined at the end of the preceding section, and $F_\omega$ is the {\it specific flux} \index{monochromatic flux}of radiation (also known as the {\it monochromatic flux} of a light source \cite{plwa}) entering a telescope from a given source. The specific flux is defined as an integral of the {\it specific intensity} (also known as the {\it surface brightness} \cite{plwa}) index{surface brightness} of the radiation, $I_\omega\equiv I_\omega(\omega,{\bm l})$, over the total solid angle (assumed $\ll 4\pi$) subtended by the source on the observer's sky:
\begin{equation}\label{a34s}
F_\omega=\int I_\omega(\omega,{\bm l}) d\Omega(\hat{\bm l})\;,
\end{equation}
where $\hat{\bm l}=(\sin\hat\theta\cos\hat\phi,\sin\hat\theta\sin\hat\phi,\cos\hat\theta)$ is the unit vector in the direction of the radiation flow and $d\hat\Omega({\bm l})=\sin\hat\theta d\hat\theta d\hat\phi$ is the element of the solid angle formed by light rays from the source and measured in the observer's local Lorentz frame.

The {\it specific intensity} $I_\omega$ of radiation at a given frequency $\omega=2\pi\nu$, flowing in a given direction, $\hat{\bm l}$, as measured in a specific local Lorentz frame, is defined by
\begin{equation}\label{aovl}
I_\omega=\frac{d({\rm energy})}{d({\rm time})d({\rm area})d({\rm frequency})d({\rm solid\;angle})}\;.
\end{equation}
A simple calculation (see, for instance, the problem 5.10 in \cite{prbook}), reveals that
\begin{equation}\label{hjui}
{\cal N}=\frac{8\pi^3}{h^4}\frac{I_\omega}{\omega^3}\;,
\end{equation}
where $h$ is the Planck's constant. The number density ${\cal N}$ is invariant along the light ray and does not change under the Lorentz transformation. Invariance of ${\cal N}$ is a consequence of the kinetic equation for photons (radiative transfer equation) which in the case of gravitational field and without any other scattering processes, assumes the following form \cite{mtw}
\begin{equation}\label{a23q}
\frac{d{\cal N}}{d\lambda}=0\;.
\end{equation}
Equations \eqref{hjui}, \eqref{a23q} tell us that the ratio $I_\omega/\omega^3$ is invariant along the light-ray trajectory, that is
\begin{equation}
\label{gtv3}
\frac{I_{\omega}}{I_{\omega_0}}=\left(\frac{\omega}{\omega_0}\right)^3\;,
\end{equation}
where $\omega_0$, $\omega$ are frequency of light at the point of emission and observation respectively, $I_{\omega_0}$ is the {\it surface brightness} of the source of light at the point of emission, and $I_{\omega}$ is the {\it surface brightness} of the source of light at the point of observation.

Equations \eqref{b12}--\eqref{gtv3} make it evident that in the geometric optics approximation the gravitational field does not mix up the linear and circular polarizations of the electromagnetic radiation but can change its {\it surface brightness} $I_\omega$ due to gravitational (and Doppler) shift of the light frequency caused by the time-dependent part of the gravitational field of the isolated system emitting gravitational waves. Furthermore, the {\it monochromatic flux} from the source of radiation changes due to the distortion of the domain of integration in equation \eqref{a34s} caused by the gravitational light-bending effect. Taking into account that the gravitationally-unperturbed solid angle $d\Omega({\bm k})=\sin\theta d\theta d\phi$, and introducing the Jacobian, $J(\theta,\phi)$, of transformation between the spherical coordinates ($\hat\theta$,$\hat\phi$) and $(\theta$,$\phi)$ at the point of observation, one obtains that the measured monochromatic flux is
\begin{equation}\label{qw45}
F_\omega=\int d\phi\int I_{\omega_0}(\theta,\phi)J(\theta,\phi)(\omega/\omega_0)^3\sin\theta d\theta  \;.
\end{equation}
Equation \eqref{qw45} tells us that the monochromatic flux of the source of light can vary due to:
\begin{enumerate}
\item the gravitational Doppler shift of the electromagnetic frequency of light when it travels from the point of emission to the point of observation;
\item the change in the solid angle at the point of observation caused by the gravitational light deflection. The "magnification" matrix is the Jacobian of the transformation of the null directions on the celestial sphere generated by the bending of the light-ray trajectories by the gravitational field of the isolated system.
\end{enumerate}

Spatial orientation of two components, $P_1$ and $P_2$, of the polarization vector ${\bm P}$ at the point of observation differs from that taken at the point of emission of the electromagnetic wave due to the parallel transport of the polarization vector. The reference tetrad with resepct to which the orientation of the polarization vector is measured is subject to the same law of the parallel transport. Therefore, in order to understand the change in the orientation of the polarization vector it is sufficient to consider the change in the orientation of the reference tetrad as it propagates from the point of emission of light to the point of observation. The change in the polarization can be easily extracted then from the law of transformation of the Stokes parameters under the change of the reference tetrad. As we will see later in section \ref{grpp}, only two vectors of the tetrad, $e^\a_{\;(1)}$ and $e^\a_{\;(2)}$, will undergo the change in their orientation as they propagate along the light ray. Consequently, only two components of the polarization vector, $P_1$ and $P_2$ will change. This can be seen after taking into account equations \eqref{abg3}, \eqref{abg4} where the rotational angle $\Theta$ is determined as a solution of the parallel transport equation \eqref{i9} for the reference tetrad (see section \ref{grpp}). Thus, the gravitational field changes the tilt angle\index{tilt angle} of the polarization ellipse as light propagates along the light-ray trajectory. This gravitationally-induced rotation of the polarization ellipse is called the Skrotskii effect \cite{skrot, skrot+}. It was also predicted and discussed independently in \cite{balaz,pleb}. Its observation would play a significant role for detection of gravitational waves of cosmological origin by CMBR-radiometry space missions \cite{polnarev_2008,keating_2006}. The third component of the polarization vector, $P_3$, remains the same along the light-ray trajectory because it represents the circularly polarized component of the radiation and is not affected by the rotation of the reference tetrad as it propagates along the light-ray path.

\section{Mathematical Technique for Analytic Integration of Light-Ray Equations}\label{acv5d}
\label{int}
This section provides mathematical technique for performing integration of equations of propagation of various characteristics of electromagnetic wave from the point of emission of light ${\bm x}_0$ to the point of its observation ${\bm x}$. It has been worked out in \citep{smk1,ksh1,kokopol,kopmak_2007}.
The basic function, which is to be integrated, is the metric tensor perturbation, $h_{\alpha\beta}(t,{\bm x})$, generated by the localized astronomical system and taken at the points lying on the light-ray trajectory. The metric tensor depends on the multipole moments and/or their derivatives taken at the retarded instant of time $s=t-r$\index{retarded time} divided by the radial distance from the system, $r$, taken to some power. The most difficult integrals are those taken from the functions like this, $F(s)/r$, with $F(s)$ being a multipole moment of the gravitating system depending on the retarded time $s$. Calculation of the integrals goes slightly differently for stationary and non-stationary components of the metric tensor and these issues are treated in the next two subsections.
\subsection{Monopole and dipole light-ray integrals}\label{ispmt}

The monopole and dipole parts of the metric tensor perturbation \eqref{bm}--\eqref{em} are
formed by terms being proportional to the mass ${\cal M}$ and spin ${\cal S}^i$ of the
isolated system. These terms are given in the solution of the light-ray equations by integrals $[\Mc(s)/r]^{[-1]}$, $[\Mc(s)/r]^{[-2]}$ and $[\mathcal{S}^i(s)/r]^{[-1]}$, $[\mathcal{S}^i(s)/r]^{[-2]}$. In this section we shall assume for simplicity that mass $\Mc$ and spin $\mathcal{S}^i$ of the isolated system are constant during the time of propagation of light from the source of light to observer. The assumption of constancy of the mass $\Mc$ and spin $\mathcal{S}^i$ of the isolated system is valid as long as one neglects the energy emitted in the form of the gravitational waves by the isolated system under consideration. For light ray propagating from the sources at the edge of our visible universe (quasars) the characteristic interval of time of emission of gravitational waves by an isolated system is comparable with the interval of time the light takes to travel from the point of emission to observer. In this case the time evolution of the mass-monopole and spin-dipole is to be taken into account for correct calculation of the perturbations of the trajectory of light ray (see section \ref{ki4c6} for more detail). Mass and spin of the isolated system can also change due to a catastrophic disruption of the isolated system resulting in supernova explosion. Specific details of how this process affects the light-ray propagation are not considered in the present chapter but, perhaps, are worthwhile to study.

In the case when mass and spin are constant, the integrals that we need to carry out are reduced to
$[1/r]^{[-1]}$ and $[1/r]^{[-2]}$ which
are formally divergent at the lower limit of integration at the past null infinity when time $t\rightarrow -\infty$. However, one must bear in mind that
these integrals do not enter equations of light-ray geodesics
\eqref{eq2+} alone but appear in these equations after taking at least one partial derivative with
respect to either $\xi^i$ or $\tau$ parameters. This differentiation effectively
eliminates divergent parts of the integrals from the final result. Hence, in
what follows, we drop out the formally divergent terms so that the
integrals under discussion assume the following form
\begin{eqnarray}
\label{is1}
  \left[\frac{1}r\right]^{[-1]}&\equiv&
                    \int^t_{-\infty}
                      \frac{d\tau}r=
                       \int
                         \frac{d\tau}{\sqrt{d^2+\tau^2}}=
                          -\ln\left[\frac{r(\tau)-\tau}{r_{\scriptscriptstyle\rm E}}\right]\;,
\\\nonumber\\
\label{is2}
  \left [\frac{1}r\right]^{[-2]}&\equiv&
                   \int^t_{-\infty}
                    \left[\frac{1}r\right]^{[-1]}
                      d\tau=
                         -\tau\ln\left[\frac{r(\tau)-\tau}{r_{\scriptscriptstyle\rm E}}\right]- r(\tau)\;,
\end{eqnarray}
where $\tau=t-t^*$ and we used equation \eqref{r_N} explicitly while expressing the distance $r=r_N(\tau)$ as a function of time $\tau$. The constant distance $r_{\scriptscriptstyle\rm E}$ was introduced to make the argument of the logarithmic function dimensionless. This constant is not important for calculations as it always cancel out in final formulas. However, in the case of gravitational lensing it is convenient to identify the scale constant $r_{\scriptscriptstyle\rm E}$ with the radius of the Einstein ring \cite{gl1,gl2,gl3,sazhin_1998SvPhU}
\begin{equation}
\label{ring9}
r_{\scriptscriptstyle\rm E}=\left(\frac{4G\Mc}{c^2}\frac{rr_0 }{R}\right)^{1/2}\;,
\end{equation}
where $r$, $r_0$ are distances from the isolated system (the deflector of light) to observer and to the source of light respectively, and $R=\tau+\tau_0$. The radius of the Einstein ring is a characteristic distance separating naturally the case of weak gravitational lensing ($d> r_{\scriptscriptstyle\rm E}$) from the strong lensing ($d< r_{\scriptscriptstyle\rm E}$). The Einstein ring at the observer's point has the angular size given by
\begin{equation}
\label{ring7}
\theta_{\scriptscriptstyle\rm E}=\frac{r_{\scriptscriptstyle\rm E}}{r}=\left(\frac{4G\Mc}{c^2}\frac{r_0 }{Rr}\right)^{1/2}\;.
\end{equation}
The angular radius $\theta_{\scriptscriptstyle\rm E}$ defines the angular scale for a lensing situation. For example, in cosmology typical mass of an isolated system is $M\simeq 10^{12} M_\odot$ and distances $r$, $r_0$, $R$ are of the order of 1 Gpc (Gigaparsec\index{Gigaparsec}). Consequently, the angular Einstein radius $\theta_{\scriptscriptstyle\rm E}$ is of the order of one arcsecond, and the linear radius $r_{\scriptscriptstyle\rm E}$ is of the order of 1 Kpc (Kiloparsec\index{Kiloparsec}). A typical star within our galaxy has mass $M\simeq M_\odot$ and distances $r$, $r_0$, $R$ are of the order 10 Kpc. It yields an angular Einstein radius $\theta_{\scriptscriptstyle\rm E}$ of the order of one 1$\mu$as (one milli-arcsecond) and the linear radius $r_{\scriptscriptstyle\rm E}$ is of the order of 10 au (10 astronomical units)\index{astronomical unit}.
\subsection{Light-ray integrals from quadrupole and higher-order multipoles}\label{itd}
\noindent
Time-dependent terms in the metric tensor
\eqref{bm} -- \eqref{em} result from the multipole moments which
can be either periodic (a binary system) or aperiodic (a supernova
explosion) functions of time. The most straightforward way to calculate the impact of the gravitational field of such a source on the propagation of light would be to decompose its multipole moments $F(s):=\left\{{\cal I}_{L}(s),{\cal S}_{L}(s)\ri\}$ in the Fourier series \cite{ksh1}
\begin{equation}
\label{for5}
F(t-r)=\frac{1}{\sqrt{2\pi}}\int^{+\infty}_{-\infty}\tilde F(\tilde\omega)e^{i\tilde\omega(t-r)}d\tilde\omega\;,
\end{equation}
where $\tilde\omega$ is a Fourier frequency, then to substitute this decomposition to the light-ray propagation equations and to integrate them term by term. This method makes an impression that in order to obtain the final result of the integration the (complex-valued) Fourier image $\tilde F(\tilde\omega)$ of the multipole moments of the isolated system must be specified explicitly, otherwise neither explicit integration nor the convolution of the integrated Fourier series will be possible. However, this impression is misleading, at least in general relativity. We shall show below that the explicit structure of $\tilde F(\tilde\omega)$ is irrelevant for general-relativistic calculation of the light-ray integrals. This may be understood better, if one recollects that in general relativity gravitational field propagates with the same speed as light in vacuum. In alternative theories of gravity the speed of gravity and light may be different \citep{kopqcg}, so it can lead to the appearance of terms being proportional to the difference between the speed of gravity and the speed of light that will drastically complicates calculations which will do require to know the explicit form of the Fourier images of gravitational-wave sources as functions of $\tilde\omega$. All such terms, which might depend on the difference between the speed of gravity and the speed of light are cancelled out in general relativity, thus, making integration of the light ray propagation equations manageable and applicable for any source of gravitational waves without particular specification of its temporal behaviour. It may be worth mentioning that the experimental limit on the difference between the speed of gravity and the speed of light in general relativity can be measured in the solar system experiments with major planets as predicted in \cite{kopapjl}. It was experimentally tested with the precision 20\% in the jovian light-ray deflection experiment with VLBI network \cite{fkapj}. No deviation from general relativity was detected.

We notice that the integration of the
light-ray equations is effectively reduced to the calculation of only two
types of integrals along the light ray:  $[F(s)/r]^{[-1]}$ and
$[F(s)/r]^{[-2]}$, where $F(s)=F(t-r)$ denotes any type of the time-dependent multipole moments of the gravitational field of the localized astronomical system. These integrals can be performed after
introducing a new variable \cite{ksh1}
\begin{equation}
\label{y}
  y\equiv s-t^*=\tau-r(\tau)=\tau-\sqrt{d^2+\tau^2}\;,
\end{equation}
which is the interval of time between two spacetime events: position of photon $x^\alpha=(\tau,{\bm x})$ on the light ray and that of the center of mass of the isolated system $z^\alpha=(y,0)$ taken at the retarded time $y$ \footnote{The retarded time $y$ should not be confused with the Cartesian coordinate $y$.}. Equation \eqref{y} is a retarded solution of the null cone equation in flat spacetime
\begin{equation}
\label{ty8}
\eta_{\alpha\beta}(x^\alpha-z^\alpha)(x^\beta-z^\beta)=0\;,
\end{equation}
giving the time of propagation of gravity from the isolated system to the photon along the null cone characteristic of Einstein's equations. This retardation of gravity effect presents in the time argument of the solution \eqref{bm}--\eqref{em} of the linearized Einstein equations. 
The replacement of the time argument $\tau$ with the retarded time $y=\tau-r(\tau)$ allows us to perform integration of equations of light-ray geodesics completely without making specific assumptions about the time dependence of the multipole moments. It is worth noting that while the parameter $\tau$ runs from $-\infty$ to $+\infty$, the retarded time $y$ runs from $-\infty$ to 0, that is, $y$ is always negative, $y\leq 0$.

Equation \eqref{y} leads to other useful transformations
\begin{equation}
\label{dy}
\tau=\frac{y^2-d^2}{2y}\;, \qquad
\sqrt{d^2+\tau^2}=-\frac{1}2\frac{d^2+y^2}y\;, \ee
and
\be\label{bv5x7}
d\tau=\frac{1}2\frac{d^2+y^2}{y^2}dy\;.
\end{equation}
Making use of the new variable $y$ and relationships \eqref{dy}, \eqref{bv5x7} the integrals under discussion can be explicitly displayed as follows:
\begin{eqnarray}
\label{ins1}
\left[\frac{F(t-r)}r\right]^{[-1]}&=& -
\int\limits_{-\infty}^{y}
          \frac{F(t^*+\zeta)}\zeta
  d\zeta\;,
\\\nonumber\\
\label{ins2}
\left[\frac{F(t-r)}r\right]^{[-2]}&=& -\frac{1}2
  \int\limits_{-\infty}^{y}
   \int\limits_{-\infty}^{\eta}\frac{F(t^*+\zeta)}\zeta
    d\zeta
     d\eta
-\frac{d^2}2
      \int\limits_{-\infty}^{y}
    \!\!\!\frac{1}{\eta^2}\!\!\!
        \int\limits_{-\infty}^{\eta}\frac{F(t^*+\zeta)}\zeta
    d\zeta
     d\eta\;,
\end{eqnarray}
where $\zeta$, $\eta$ are the dummy variables of the integration replacing the integration along the light-ray trajectory by that along a null characteristic of the gravitational field, and $t^*$ is the time of the closest approach of photon to the origin of the coordinate chart coinciding with the center of mass of the isolated gravitational system. The time $t^*$ has no physical meaning in the most general case and appears in calculations as an auxiliary (constant) parameter which vanishes in the final result. The reason is simple, by choosing the origin of the coordinate chart at a different point we change the numerical value of $t^*$ but it cannot change physical observables -- the angle of light-ray deflection, frequency shift, etc. 
The time of the closest approach $t^*$ can be used as an approximation of the retarded time $s=t-r$ in the case of a small impact parameter $d$ of the light ray with respect to the isolated astronomical system, that is $s\simeq t^*+O(d^2/cr)$ (see equation \eqref{tl}). This approximation makes an erroneous impression that the physical effect of propagation of gravity is unimportant for calculation of observable effects caused by variable gravitational field of the isolated system. This misinterpretation of the role of gravity's propagation in the interpretation of the observable astronomical effects did happen in some papers \citep{willapj,asada_2002} which used the simplifying approximation $s=t^*$ from the very beginning of calculations but it precludes to spatially disentangle the null cone characteristics of gravity and light. The only case when the time $t^*$ makes physical sense is when observer is located in infinity. Indeed, in this case the radial distance $r=\infty$, and the retarded time $s=t^*$ exactly, because the residual terms $O(d^2/r)=0$ irrespectively of the choice of the coordinate origin \cite{23}.

A remarkable property of the integrals in the right side of equations \eqref{ins1}, \eqref{ins2} is that they depend on the parameters $\xi^i$ and $\tau$ only through either the upper limit of the integration, which is the variable $y=\tau-\sqrt{d^2+\tau^2}$, or the square of the impact parameter, $d^2=({\bm\xi}\cdot{\bm\xi})$ standing in front of the second integral in the right side of equation \eqref{ins2}. For this reason, differentiation of the integrals in the left part of equations \eqref{ins1}, \eqref{ins2} with respect to either $\xi^i$ or $\tau$ will effectively eliminate the integration along the light ray trajectory. For example,
\begin{equation}
\label{d11}
\dksi_i\left\{\left[\frac{F(t-r)}r\right]^{[-1]}\right\}= -\frac{F(t^*+y)}y\;\dksi_iy=
\frac{\xi^i}{yr}\,F(t-r)\;.
\end{equation}
Similar calculation can be easily performed in case of differentiation of integral $[F(t-r)/r]^{[-1]}$ with respect to $\tau$. It results in
\begin{equation}
\label{dt1}
\hat\partial_\tau\left\{\left[\frac{F(t-r)}r\right]^{[-1]}\right\}=-\frac{F(t^*+y)}y\;\hat\partial_\tau
y=-\frac{F(t^*+y)}y\left(1-{\frac{\tau}r}\right)=\frac{F(t-r)}r\;,
\end{equation}
as it could be expected because the partial differentiation with respect to $\tau$ keeps $\xi^i$ and $t^*$ fixed and, hence, is equivalent to taking a total time derivative with respect to $t$ along the light ray.
Since all terms in the solution of the light geodesic equation are represented as partial derivatives from the integrals $[F(t-r)/r]^{[-1]}$ and $[F(t-r)/r]^{[-2]}$ with respect to the parameters $\xi^i$ and/or $\tau$, it is clear from equations \eqref{d11}, \eqref{dt1} that the solution will not contain any single integral like $[F(t-r)/r]^{[-1]}$ at all -- only the derivatives of this integral will appear which are ordinary functions. 

Dealing with the double integrals of type $[F(t-r)/r]^{[-2]}$ is more sophisticated.
We notice that taking the first and second derivatives from the double integral $[F(t-r)/r]^{[-2]}$ do not eliminate the integration along the light ray trajectory
\begin{eqnarray}
\label{d12}
  \dksi_k\left\{
   \left[\frac{F(t-r)}r\right]^{[-2]}\right\}&=
&\xi^k
      \left\{\frac{1}y\left[\frac{F(t-r)}r\right]^{[-1]}-
      \int\limits_{-\infty}^{y}
     \frac{1}{\eta^2}
         \int\limits_{-\infty}^{\eta}\frac{F(t^*+\zeta)}\zeta\,
    d\zeta
     d\eta
     \right\}\,,
\\
\label{d22}
 \dksi_{jk}\left\{\left[\frac{F(t-r)}r\right]^{[-2]}\right\}
 &=&
 \frac{\xi^k\xi^j}{y^2}
\frac{F(t-r)}r\\\notag
&+&P^{jk}
 \left\{\frac{1}y\left[\frac{F(t-r)}r\right]^{[-1]}-
      \int\limits_{-\infty}^{y}
     \frac{1}{\eta^2}
         \int\limits_{-\infty}^{\eta}\frac{F(t^*+\zeta)}\zeta\,
    d\zeta
     d\eta
 \right\}\,,
\end{eqnarray}
However, taking one more (a third) derivative eliminates all integrals from equation (\ref{d22}) completely. More specifically, we have
\begin{equation}
\label{d32}
 \dksi_{ijk}\left\{\left[\frac{F(t-r)}r\right]^{[-2]}\right\}
 =
 \frac{1}y
 \left\{
         \left(
        P^{ij}+\frac{\xi^{ij}}{yr}
            \right)\dksi_k
          +P^{jk}\dksi_i
          +\xi^j\dksi_{ik}
 \right\}
\left[\frac{F(t-r)}r\right]^{[-1]}\,,
 \end{equation}
and making use of equations \eqref{d11}, \eqref{d22} for explicit calculation of partial derivatives in the right part of equation \eqref{d32} proves that all integrations disappear, that is
\begin{eqnarray}
\label{d99}
\dksi_{ijk}\left\{\left[\frac{F(t-r)}r\right]^{[-2]}\right\}
 &=&\frac{P^{ij}\xi^k+P^{jk}\xi^i+P^{ik}\xi^j}{y^2}\frac{F(t-r)}{r}\\\notag&+&\frac{\xi^i\xi^j\xi^k}{y^2r^2}\left[\left(\frac{2}{y}-\frac{1}{r}\right)F(t-r)-\dot F(t-r)\right]\;,
 \end{eqnarray}
 where $\dot F(t-r)\equiv\partial_t F(t-r)$.
The same kind of reasoning works for the third derivatives with respect to the parameter $\tau$ and to the mixed derivatives taken with respect to both $\xi^i$ and $\tau$.

We shall obtain solution of the light geodesic in terms of (STF) partial derivatives \index{STF derivative} taken with respect to the impact parameter vector, $\xi^i$, of the light ray acting on the single and double integrals having a symbolic form $\dksi_{<a_1\ldots a_k>}[F(t-r)/r]^{[-1]}$ and
$\dksi_{<a_1\ldots a_k>}[F(t-r)/r]^{[-2]}$ where the angular brackets around the spatial indices indicate the STF symmetry. Explicit expression for this STF derivative can be obtained directly by applying the differentiation rules shown in equations \eqref{d12} -- \eqref{d32}.
In what follows, we shall see that the solution of the equations of propagation of light rays will always contain three or more derivatives acting on the integrals from the higher-order multipole moments (with the multipole index $l\ge 2$) which have the same structure as that shown in equations  \eqref{ins1}, \eqref{ins2}. It means that the final result of the integration of the light-ray propagation equations depending on the higher-order multipoles of the localized astronomical system can be expressed completely solely in terms of these multipoles taken at the retarded instant of time $s=t-r$. In other words, the observed effects of light deflection, etc. does not depend in {\it general relativity} directly on the past history of photon's propagation that is the effects of temporal variations of the gravitational field (gravitational waves) do not accumulate. On the other hand, the past history of the isolated system can affect the propagation of light ray, at least in principle, through the tails of the gravitational waves which contribute to the present value of multipole moments of the system as shown in equations \eqref{zqm2}, \eqref{zqm3}.

We would like to point out that if an electromagentic wave propagated through a dispersive medium, the physical speed of light would be different from the speed of gravity. In such case the effects of the temporal variations of the gravitational field can do accumulate and the motion of the electromagnetic signal depends on its past history. This effect was studied by Bertotti and Catenacci \cite{berc} in order to set an upper limit on the light scintillations caused by the stochastic gravitational wave background. They concluded that such a limit is not very interesting but it might be a good time now to reconsider this conclusion in application to the more advanced level of observational technologies..

\section{Gravitational Perturbations of the Light Ray}\label{dyb5k}

\subsection{Relativistic perturbation of the electromagnetic eikonal}\label{ki4c6}

Perturbation of eikonal (phase) of electromagnetic signal propagating from the point ${\bm x}_0$ to the point ${\bm x}$ are obtained by solving equation \eqref{7b+}. This solution is found by integrating the metric tensor perturbation along the light-ray trajectory and can be written down in the linearized approximation as an algebraic sum of three separate terms
\begin{equation}
\label{pmr1}
\psi=\psi_{\scriptscriptstyle (G)}+\psi_{\scriptscriptstyle (M)}+\psi_{\scriptscriptstyle (S)}\;,
\end{equation}
where $\psi_{\scriptscriptstyle (G)}$ represents the gauge-dependent part of the eikonal, and $\psi_{\scriptscriptstyle (M)}$, $\psi_{\scriptscriptstyle (S)}$ are eikonal's perturbations caused by the mass and spin multipole moments correspondingly. Their explicit expressions are as follows
\begin{eqnarray}
\label{pmr2}
\psi_{\scriptscriptstyle (G)}&=&(k^i\phi^i-\phi^0)+(k^iw^i-w^0)\;,\\\nonumber\\
\label{pmr3}
\psi_{\scriptscriptstyle (M)}&=&2\left[\frac{{\Mc}(t-r)}r\right]^{[-1]}+
2\sum_{l=2}^\infty\sum_{p=0}^{l-1}
            \frac{(-1)^{l+p}}{l!}
                                  C_l(l-p,p)
\left(
    1-\frac{p}l
  \right)\\&&\times
\left\{
    \left(
        1+\frac{p}{l-1}
      \right)
            k_{<a_1\hdots a_p}\dksi_{a_{p+1}\hdots a_l>}
       \left[\ilrrp\right]^{[-1]}
\notag\right.    \\&&\left.-
    \frac{2p}{l-1}
            k_{<a_1\hdots a_{p-1}}\dksi_{a_p\hdots a_{l-1}>}
                           \left[\frac{k^i{\Ic}^{(p)}_{iA_{l-1}}(t-r)}r\right]^{[-1]}
             \right\}
 \notag    \\\notag\\\label{pmr4}
\psi_{\scriptscriptstyle (S)}&=& 2\epsilon_{abi}k^a{\cal S}^b\dksi_i \ln\left(r-\tau\right)
+4\dksi_a
\sum_{l=2}^{\infty}\sum_{p=0}^{l-1}
        \frac{(-1)^{l+p}l}{(l+1)!}
C_{l-1}\left(l-p-1,p\right)
\\&&\times \left(1-\frac{p}{l-1}\right)
 k_{<a_1\hdots a_p}\dksi_{a_{p+1}\hdots a_{l-1}>}
        \left[\frac{k^i\epsilon_{iba}{\cal S}^{(p)}_{bA_{l-1}}(t-r)}{r}\right]^{[-1]}\;,
    \notag
\end{eqnarray}
where $H(p-q)$ is the Heaviside function defined by the expression \eqref{H}, $C_l(p,q)$ are the polynomial coefficient given by equation \eqref{pco} and ${F}^{(p)}(t-r)$ denotes $\partial^p F(t-r)/\partial t^p$ where $r$ is considered as constant.

We note that the gauge-dependent part of the eikonal, $\psi_{\scriptscriptstyle (G)}$, contains combination of terms
$k^i\phi^i-\phi^0$ defined by equations \eqref{phiphi}--\eqref{phis} which can be, in principle,
eliminated with an appropriate choice of the gravitational field
gauge functions $w^0$ and $w^i$. However, such a procedure
will introduce a reference frame in the sky with a coordinate grid being
dependent on the direction to the source of light rays
that is, to the direcion of the unit
vector $k^i$. The coordinate frame obtained in this way will have the direction-dependent distortions which can change as time goes on because of the proper motion of stars. For this reason the elimination of
functions $\phi^0$ and $\phi^i$ from equation \eqref{pmr2} may be not practically justified. The ADM-harmonic coordinate chart admits a more straightforward treatment of observable relativistic effects. Thus, we leave the gauge functions
$\phi^0$ and $\phi^i$ in equation \eqref{pmr2} along with the gauge
functions $w^0$ and $w^i$ that are defined by formulas (\ref{nw0}), (\ref{nwi}).

Expressions \eqref{pmr3}, \eqref{pmr4} for the eikonal contain derivatives from the retarded integrals of the mass- and spin-multipoles. After taking the derivatives one can prove that the integrals from all high-order multipoles $(l\ge 2)$ are eliminated. Indeed, scrutiny inspection of equations \eqref{pmr3}, \eqref{pmr4} elucidates that all the integrals from the multipole moments enter the equation in combination with at least one derivative with respect to the impact parameter vector $\xi^i$. Hence, the differentiation rule \eqref{d11} is applied which eliminates the integration.

The only integral which must be performed, is that from the mass monopole in equation \eqref{pmr3}. It can be taken by parts as follows
\begin{equation}
\label{xcu7}
   \left[\frac{{\Mc}(t-r)}r\right]^{[-1]}=-\Mc(t-r)\ln(r-\tau)+\int_{-\infty}^y\dot\Mc(t^*+\zeta)\ln\zeta d\zeta\;,
               \end{equation}
where $\dot{\cal M}\equiv\partial{\cal M}/\partial t^*$.
If one assumes that the mass $\Mc$ is constant, then the second term in the right side of equation \eqref{xcu7} vanishes and the eikonal does not contain any integral dependence on the past history of the light propagation. This assumption is usually implied, for example, in the theory of gravitational lensing \cite{gl1,gl2,gl3} and other applications of the relativistic theory of light propagation. Here, we extend our approach to take into account the case of time-dependent mass of the isolated system, $\dot\Mc\not=0$. The mass may change because stars are losing mass in the form of the stellar wind. We shall not consider this case but focus on another process of the lost of mass by the isolated system due to the emission of gravitational radiation. The rate of the mass loss is, then, given by \cite{mtw,LL}
\begin{equation}
\label{jq8}
\dot{\cal M}(t)=-\frac{1}{c^7}{\cal I}^{(3)}_{ij}(t){\cal I}^{(3)}_{ij}(t)+O\left(c^{-9}\right)\;,
\end{equation}
where ${\cal I}^{(3)}_{ij}$ represents a third time derivative from the mass quadrupole moment and terms of order $O\left(c^{-9}\right)$ describe contribution of higher-order mass and spin multipoles which we neglect here. Making use of equation \eqref{jq8} in (\ref{xcu7}) yields
\begin{eqnarray}
\label{pos4}
     \left[\frac{{\Mc}(t-r)}r\right]^{[-1]}&=&-\Mc(t-r)\ln(r-\tau)\\\notag&-&\frac{1}{c^7}\int_{-\infty}^y{\cal I}^{(3)}_{ij}(t^*+\zeta){\cal I}^{(3)}_{ij}(t^*+\zeta)
               \ln\zeta d\zeta+O\left(c^{-9}\right)\;.
               \end{eqnarray}

It is instructive to evaluate contribution of the second term in the right side of equation \eqref{xcu7} in the case of light propagating in gravitational field of a binary system consisting of two stars with masses $m_1$ and $m_2$ orbiting each other on a circular orbit. The total mass $\Mc$ of the system is defined in accordance with equation \eqref{zqm} (for $l=0$) which takes into account the first post-Newtonian correction
\begin{equation}
\label{hq6}
\Mc(t)=m_1+m_2-\frac{1}{2c^2}\frac{\mu\Mc}{a(t)}+O\left(c^{-4}\right)\;,
\end{equation}
where $\mu=m_1m_2/\Mc$ is the reduced mass of the system and $a=a(t)$ is the orbital radius of the system. Assuming that the lost of the orbital energy is due to the emission of gravitational waves from the system in accordance with the quadrupole formula approximation \eqref{jq8}, the time evolution of the orbital radius is defined by equation \cite{LL,mtw}
\begin{equation}
\label{tnz5}
\dot a=-\frac{64}{5c^5}\frac{\mu\Mc^2}{a^3}\;.
\end{equation}
It has a simple solution \cite{mtw}
\begin{eqnarray}
\label{pcq8}
a(t)&=&a_0\left(1-\frac{\tau}{T}\right)^{1/4}\;,\\\notag\\\label{pcq9}
T&=&\frac{5c^5}{256}\frac{a_0^4}{\mu\Mc^2}\;,
\end{eqnarray}
where $\tau=t-t^*$, $t^*$ is the time of the closest approach of light ray to the binary system, $a_0=a(t^*)$ is the orbital radius of the binary system at the time of the closest approach, and $T$ is the spiral time of the binary. The lost of the total mass due to the emission of the gravitational waves is
\begin{equation}
\label{sh7}
\dot\Mc=-\frac{1}{8c^2}\frac{\mu\Mc}{a_0T}\left(1-\frac{\tau}{T}\right)^{-5/4}\;.
\end{equation}
Substituting equation \eqref{sh7} to the right side of equation \eqref{xcu7} and performing integration yields
\begin{eqnarray}
\label{sh8}
\int_{-\infty}^y\dot\Mc(t^*+\zeta)\ln\zeta d\zeta&=&-\frac{1}{2}\frac{\mu\Mc}{c^2a(s)}\ln(r-\tau)\\\notag&+&
\frac{1}{2}\frac{m_1m_2}{c^2a_0}\Biggl\{\ln\left[\frac{a(s)-a_0}{a(s)+a_0}\right]+2\arctan\left[\frac{a(s)}{a_0}\right]\Biggr\}\;,
\end{eqnarray}
where $a(s)\equiv a(t-r)$ denotes the orbital radius of the binary system taken at the retarded time $s=t-r$. Putting all terms in equations \eqref{xcu7}, \eqref{hq6}, \eqref{sh8} together yields
\begin{equation}
\label{sh9}
             \left[\frac{{\Mc}(t-r)}r\right]^{[-1]}=-(m_1+m_2)\ln(r-\tau)+
\frac{1}{2}\frac{m_1m_2}{c^2a_0}\Biggl\{\ln\left[\frac{a(s)-a_0}{a(s)+a_0}\right]+2\arctan\left[\frac{a(s)}{a_0}\right]\Biggr\}\;,
\end{equation}
where the first term in the right side of this equation is the standard Shapiro time delay in the gravitational field of the binary system with constant total mass $m_1+m_2$, and the second term represents relativistic correction due to the emission of gravitational waves by the system causing the overall loss of its orbital energy.

Eikonal describes propagation of a wave front of the electromagnetic wave. Light rays are orthogonal to the wave front and their trajectories can be easily calculated as soon as the eikonal is known. In the present chapter we do not use this technique and obtain solution for the light rays directly from the light-ray geodesic equations that gives identical results.

\subsection{Relativistic perturbation of the coordinate velocity of light}

Integration of the light-ray propagation equations \eqref{ddx} -- \eqref{ddxs} is fairly
straightforward. Performing one integration of these equations with respect to time yields
\begin{eqnarray}
\label{tr1}
         \dot x^i(\tau)&=&k^i+\dot\Xi^i(\tau,{\bm\xi})\;,\\\label{truk}
         \dot\Xi^i(\tau,{\bm\xi})&=&\DXGi+\DXiim+\DXiis\;,
\end{eqnarray}
where the relativistic perturbations of photon's trajectory are given
by
\begin{equation}
\label{po23}
\DXGi=\dtau\left[(\phi^i-k^i\phi^0)+(w^i-k^iw^0)\right]\;,
\end{equation}
\begin{align}
\label{xidm}
\DXiim=\;&2(\dksi_i-k_i\dtau)\left[\frac{\Mc}r\right]^{[-1]}\\\notag&+
2\dksi_i
   \sum_{l=2}^\infty\sum_{p=0}^l\sum_{q=0}^p
     \frac{(-1)^{l+p-q}}{l!}
       C_l(l-p,p-q,q)H(2-q)
\times
\\&
\left(
    1-\frac{p-q}l
   \right)
       \left(
               1-\frac{p-q}{l-1}
         \right)
            k_{<a_1\hdots a_p}\dksi_{a_{p+1}\hdots a_l>}
    \dtau^q
         \left[\frac{{\Ic}^{(p-q)}_{A_{l}}(t-r)}r\right]^{[-1]}-
\notag    \\&
            2\sum_{l=2}^\infty\sum_{p=0}^{l-1}
            \frac{(-1)^{l+p}}{l!}C_l(l-p,p)
\left(
    1-\frac{p}l
  \right)
\Biggl\{
    \left(
        1+\frac{p}{l-1}
      \right)
            k_{i<a_1\hdots a_p}\dksi_{a_{p+1}\hdots a_l>}
       \left[\frac{{\Ic}^{(p)}_{A_{l}}(t-r)}r   \right]-
\notag    \\&
    \frac{2p}{l-1}
            k_{<a_1\hdots a_{p-1}}\dksi_{a_p\hdots a_{l-1}>}
        \left[\frac{{\Ic}^{(p)}_{iA_{l-1}}(t-r)}r\right]
             \Biggr\}\;,
  \notag
\end{align}
\begin{align}
\label{xids}
\DXiis=\;&
2k_j\dksi_{ia}\left[\ejsbr\right]^{[-1]}
-2\dksi_a\eisbr-\\&
4k_j\dksi_{ia}
    \sum_{l=2}^{\infty}\sum_{p=0}^{l-1}\sum_{q=0}^{p}
        \frac{(-1)^{l+p-q}l}{(l+1)!}
C_{l-1}\left(l-p-1,p-q,q\right)H(2-q)\times
    \notag
\\&
    \left(1-\frac{p-q}{l-1}\right)
k_{<a_1\hdots a_p}\dksi_{a_{p+1}\hdots a_{l-1}>}
        \dtau^q
        \left[\frac{\epsilon_{jba}{\mathcal {S}}^{(p-q)}_{bA_{l-1}}(t-r)}r\right]^{[-1]}+
    \notag
\\&
4\left(\dksi_a-k_a\dtz\right)
\sum_{l=2}^{\infty}\sum_{p=0}^{l-1}
        \frac{(-1)^{l+p}l}{(l+1)!}
C_{l-1}\left(l-p-1,p\right)\times
    \notag
\\& \left(1-\frac{p}{l-1}\right)
 k_{<a_1\hdots a_p}\dksi_{a_{p+1}\hdots a_{l-1}>}
        \left[\eislrrp\right]+
    \notag
\\&
4k_j
\sum_{l=2}^{\infty}\sum_{p=0}^{l-1}
        \frac{(-1)^{l+p}l}{(l+1)!}
C_{l-1}\left(l-p-1,p\right)
k_{<a_1\hdots a_p}\dksi_{a_{p+1}\hdots a_{l-1}>}
        \left[ \frac {\epsilon_{jba_{l-1}}{\mathcal {S}}^{(p+1)}_{\hat ibA_{l-2}}(t-r)}r\right]\;.
    \notag
\end{align}
Here $H(p-q)$ is a Heaviside function defined by the expression \eqref{H} and $C_l(p,q)$ are the polynomial coefficients (\ref{pco}). The gauge functions are given in equations \eqref{nw0}--\eqref{nwi+}. 

Mass monopole and spin dipole terms are written down in equations \eqref{xidm}, \eqref{xids} in symbolic form and after taking derivatives are simplified
\begin{eqnarray}
\label{plo9}
2(\dksi_i-k_i\dtau)
        \left[
            \frac{\Mc}r\right]^{[-1]}&=&\frac{2\Mc}{r}\left(\frac{\xi^i}{y}-k^i\right)\;,
            \\\notag\\
2k_j\dksi_{ia}\left[\ejsbr\right]^{[-1]}&=&
            2k^j\epsilon_{jba}\dksi_{a}\left(\frac{\mathcal{S}_b\xi^i}{yr}\right)\;.
            \end{eqnarray}
Remaining integrals shown in the right side of equations \eqref{xidm}, \eqref{xids} are convenient for presentation of the result of integration in the symbolic form. The integration is actually not required because the integrals are always appear in combination with, at least, one operator of a partial derivative $\hat\partial_i$ with respect to the impact parameter vector of the light ray. The derivative operator acts on the integrals in accordance with equation \eqref{d11} converting the integrals into functions of the retarded time $y=\tau-r(\tau)$
\begin{eqnarray}
\label{xx5}
\hat\partial_i
     \left[\frac{{\Ic}^{(p-q)}_{A_{l}}(t-r)}{r}\right]^{[-1]}&=&{\Ic}^{(p-q)}_{A_{l}}(t^*+y)\hat\partial_i\ln(-y)\;,\\\notag\\\label{xx6}
\hat\partial_i
        \left[\frac{\epsilon_{jba}{\mathcal {S}}^{(p-q)}_{bA_{l-1}}(t-r)}{r}\right]^{[-1]}&=&
\epsilon_{jba}{\mathcal {S}}^{(p-q)}_{bA_{l-1}}(t^*+y)\hat\partial_i\ln(-y)\;.
\end{eqnarray}
We conclude that at each point of the wave front of the electromagnetic wave the relativistic perturbation of the direction of propagation of light ray (wave vector) caused by the time-dependent gravitational field of the isolated system depends only on the value of its multipole moments taken at the retarded instant of time and it does not depend on the past history of the light propagation.

\subsection{Perturbation of the light-ray trajectory}\label{h6tv7}

Integration of equation \eqref{tr1} with respect to time yields the relativistic perturbation of the trajectory of the light ray
\begin{equation}
\label{tr2}
x^i(\tau)=x_N^i+\Delta\Xii\limits_{\scriptscriptstyle (G)}+\Delta\Xii\limits_{\scriptscriptstyle (M)}+\Delta\Xii\limits_{\scriptscriptstyle (S)}\;,
\end{equation}
where
\begin{equation}
\label{mk4}
\Delta\Xii\limits_{\scriptscriptstyle (G)}\equiv\XG-\XGo\;,\quad
\Delta\Xii\limits_{\scriptscriptstyle (M)}\equiv\Xiim-\Xiimo\;,\quad
\Delta\Xii\limits_{\scriptscriptstyle (S)}\equiv\Xiis-\Xiiso\;.
\end{equation}
Here, the term
\begin{equation}
\label{pogi}
\XG=(\phi^i-k^i\phi^0)+(w^i-k^iw^0)\;,
\end{equation}
is the gauge-dependent part of the trajectory's perturbation, and the physically meaningful perturbations due to the mass and spin multipoles
\begin{align}
\label{xim}
\Xiim=\;&2(\dksi_i-k_i\dtau)
        \left[\frac{\Mc}r\right]^{[-2]}+\\\notag&
2\dksi_i
   \sum_{l=2}^\infty\sum_{p=0}^l\sum_{q=0}^p
     \frac{(-1)^{l+p-q}}{l!}
       C_l(l-p,p-q,q)H(2-q)
\times
\\&
\left(
    1-\frac{p-q}l
   \right)
       \left(
               1-\frac{p-q}{l-1}
         \right)
            k_{<a_1\hdots a_p}\dksi_{a_{p+1}\hdots a_l>}
    \dtau^q
       \left[\frac{{\Ic}^{(p-q)}_{A_{l}}(t-r)}r\right]^{[-2]}-
\notag    \\&
            2\sum_{l=2}^\infty\sum_{p=0}^{l-1}
            \frac{(-1)^{l+p}}{l!}
                                  C_l(l-p,p)
\left(
    1-\frac{p}l
  \right)
\Biggl\{
    \left(
        1+\frac{p}{l-1}
      \right)
            k_{i<a_1\hdots a_p}\dksi_{a_{p+1}\hdots a_l>}
                   \left[\ilrrp\right]^{[-1]}-
\notag    \\&
    \frac{2p}{l-1}
            k_{<a_1\hdots a_{p-1}}\dksi_{a_p\hdots a_{l-1}>}
               \left[\frac{{\Ic}^{(p)}_{iA_{l-1}}(t-r)}r\right]^{[-1]}
             \Biggr\}\;,
  \notag
\end{align}
\begin{align}
\label{xis}
\Xiis=\;&2k_j\dksi_{ia}\left[\ejsbr\right]^{[-2]}
-2\dksi_a\left[\eisbr\right]^{[-1]}-
\\&
4k_j\dksi_{ia}
    \sum_{l=2}^{\infty}\sum_{p=0}^{l-1}\sum_{q=0}^{p}
        \frac{(-1)^{l+p-q}l}{(l+1)!}
C_{l-1}\left(l-p-1,p-q,q\right)H(2-q)\times
    \notag
\\&
    \left(1-\frac{p-q}{l-1}\right)
k_{<a_1\hdots a_p}\dksi_{a_{p+1}\hdots a_{l-1}>}
        \dtau^q
        \left[\frac{\epsilon_{jba}{\mathcal {S}}^{(p-q)}_{bA_{l-1}}(t-r)}r\right]^{[-2]}+
    \notag
\\&
4\left(\dksi_a-k_a\dtz\right)
\sum_{l=2}^{\infty}\sum_{p=0}^{l-1}
        \frac{(-1)^{l+p}l}{(l+1)!}
C_{l-1}\left(l-p-1,p\right)\times
    \notag
\\ \left(1-\frac{p}{l-1}\right)
& k_{<a_1\hdots a_p}\dksi_{a_{p+1}\hdots a_{l-1}>}
        \left[\eislrrp\right]^{[-1]}+
    \notag
\\&
4k_j
\sum_{l=2}^{\infty}\sum_{p=0}^{l-1}
        \frac{(-1)^{l+p}l}{(l+1)!}
C_{l-1}\left(l-p-1,p\right)
k_{<a_1\hdots a_p}\dksi_{a_{p+1}\hdots a_{l-1}>}
                \left[ \frac {\epsilon_{jba_{l-1}} \mathcal {S}^{(p+1)}_{\hat ibA_{l-2}}(t-r)}r\right]^{[-1]}\;.
    \notag
\end{align}
Here, again $H(p-q)$ is a Heaviside function defined by equation
\eqref{H} and $C_l(p,q)$ are the polynomial coefficients (\ref{pco}).

Relativistic perturbations
\eqref{xim}, \eqref{xis} contain two types of integrals along the light ray which symbolic form is
$[F(s)/r]^{[-1]}$ and $[F(s)/r]^{[-2]}$ where $s=t-r$ is the retarded time. In section
\ref{int} we have shown that taking a derivative from an integral $[F(s)/r]^{[-1]}$ reduces the integral to an ordinary function of the retarded time $s$ with no integral dependence on the past history of the propagation of light ray. In order to eliminate the integration in integrals $[F(s)/r]^{[-2]}$ one needs to take three or more derivatives. One can notice that in  \eqref{xim}, \eqref{xis} there are terms with the summation index $p=l-1$ or $p=l$ in which the number of the derivatives is less than three that is not sufficient to eliminate the integrals. However, all such integrals are multiplied with a numerical coefficients like $1-p/l$, etc. which gets nil for those values of the summation index $p$. It effectively eliminates all the integrals in \eqref{xim}, \eqref{xis} in which the integration cannot be eliminated by the partial derivatives.
As an example let us examine equation \eqref{xim} for
$\Xiim$. In this expression the term with the double integral
$\left[\mathcal{I}^{(p-q)}_{A_l}(s)/r\right]^{[-2]}$ has $l-p+1$ derivatives $\dksi_a$ and $q$ derivatives $\hat\partial_\tau$.
After performing the differentiation, the expression $\dksi_{<a_1\ldots
a_{l-p+1}>}\hat\partial^q_\tau\left[\mathcal{I}^{(p-q)}_{A_l}(s)/r\right]^{[-2]}$ can contain the explicit integrals in the following cases: $q=0$ and $p=l-1$; $q=0$ and $p=l$; $q=1$ and $p=l$. However, these integrals are multiplied with $[1-(p-q)/l][1-(p-q)/(l-1)]$ and in any case of $p$ and $q$ values mentioned above either the factor $1-(p-q)/l$ or $1-(p-q)/(l-1)$ will vanish. 
Similar
reasoning is applied to the spin-dependent perturbation $\Xiis$ which shows that there is no need to explicitly integrate the spin multipole moments in order to calculate the perturbation of light ray trajectory.

The only integral dependence of the light-ray trajectory perturbation depending on the past history of the propagating photon remains in equation \eqref{pogi} which contains the
gauge functions $\phi^\alpha$ and $w^\alpha$ used later in this chapter
for interpretation of observable effects caused by the gravitational waves from the isolated system. The past-history dependence of the light-ray trajectory may come into play also through the integrals from mass monopole and spin-dipole terms: $\left[\Mc/r\right]^{[-2]}$ and $\left[{\cal S}/r\right]^{[-2]}$, if mass and/or spin of the isolated system are not conserved and change as time passes on. The non-conservation can be caused, for example, by emission of gravitational waves carrying away the orbital energy and angular momentum of the system. The past-history contribution of these integrals can be calculated similarly to the eikonal perturbation considered in equations \eqref{jq8}--\eqref{sh9}.

Solution of the light-ray equation with the initial-boundary conditions \eqref{ibc} depends on the unit vector ${\bm k}$ defining direction of the light-ray propagation extrapolated backward in time to the past null infinity\index{null infinity}. In real practice the light ray is emitted at the point ${\bm x}_0$ where the source of light is located at time $t_0$, and it arrives to observer at time $t$ to the point ${\bm x}$ separated from the source of light by a finite coordinate distance $R=|{\bm x}-{\bm x}_0|$. Therefore, solution of the light-ray equations must be expressed in this case in terms of the integrals of the boundary-value problem. It is formulated
in terms of the coordinates of the initial, ${\bm x}_0$, and the final, ${\bm x}$, positions of the photon
\begin{equation}\label{bvp}
{\bm x}(t)={\bm x}\;,\quad\quad {\bm x}(t_0)={\bm x}_0\;,
\end{equation}
and a unit vector
\begin{eqnarray}
\label{unitv}
K^i=-\frac{x^i-x_0^i}{|{\bm x}-{\bm x}_0|}\;,
\end{eqnarray}
that defines a coordinate direction from the observer towards the source of light and can be
interpreted as a unit vector in the Euclidean space in the sense that $\d_{ij}K^i K^j=1$.

In what follows it is convenient to make use of the astronomical coordinates ${\bm x}\equiv x^i=(x^1,x^2,x^3)$ based on
a triad of the Euclidean unit vectors $({\bm I}_0,{\bm J}_0,{\bm K}_0)$ as shown in Fig. \ref{bundle}. 
Vector ${\bm K}_0$ points from observer towards the isolated system emitting gravitational waves and deflecting light rays. Vectors ${\bm I}_0$
and ${\bm J}_0$ lie in the plane being orthogonal
to vector ${\bm K}_0$. The unit vector ${\bm I}_0$ is directed to the east, and that
${\bm J}_0$ points towards the north celestial pole \index{celestial pole}. The origin of the
coordinate system is chosen to lie
at the center of mass of the isolated system.

We will need in our discussion an another reference frame based on
a triad of the Euclidean unit vectors $({\bm I},{\bm J},{\bm K})$ that are
turned at some angle with respect to vectors
$({\bm I}_0,{\bm J}_0,{\bm K}_0)$.
Vector ${\bm K}$ points from the observer towards the source of light, and
vectors ${\bm I}$ and ${\bm J}$ lie in the {\it plane of the sky}\index{plane of the sky} defined to
be orthogonal
to vector ${\bm K}$. 
Mutual orientation of the triads is
defined by the orthogonal transformation (rigid rotation)
\begin{eqnarray}
\label{rotation}
{\bm I}_0&=&\phantom{-}{\bm I}\cos\Omega+{\bm J}\sin\Omega\;,\\
{\bm J}_0&=&-{\bm I}\cos\theta\sin\Omega+{\bm J}\cos\theta\cos\Omega
+{\bm K}\sin\theta\;,\\
{\bm K}_0&=&\phantom{-}{\bm I}\sin\theta\sin\Omega-{\bm J}\sin\theta\cos\Omega+{\bm
K}\cos\theta\;,
\end{eqnarray}
where the rotational angles, $\Omega$ and $\theta$, are constant.

It is rather straightforward to obtain solution of the boundary value problem (\ref{bvp}) for light propagation in terms of the unit vector ${\bm K}$ instead of the unit vector ${\bm k}$ of the initial-boundary value problem. All what we need, is to convert the unit vector ${\bm k}$ to ${\bm K}$  written in terms
of spatial coordinates of the
points of emission, ${\bm x_0}$, and observation,
${\bm x}$, of the light ray. From formula (\ref{tr2}) one has
\begin{eqnarray}
\label{uuu}
k^i
&=&-K^i-\beta^i(\tau,{\bm{\xi}})\;,
\end{eqnarray}
where the relativistic correction $\beta^i(\tau,{\bm{\xi}})$
to the vector $K^i$ is derived from the solution of the initial-boundary value problem, It is explicitly defined as follows
\begin{eqnarray}
\label{expli}
\beta^i(\tau,{\bm{\xi}})
&=&\frac{P_{\;j}^i\left[\Delta\Xij\limits_{\scriptscriptstyle (G)}+\Delta\Xij\limits_{\scriptscriptstyle (M)}+\Delta\Xij\limits_{\scriptscriptstyle (S)}\right]}{|{\bm x}-{\bm x}_0|}\;,
\end{eqnarray}
with $P_{\;j}^i=\delta_{\;j}^i-k^i k_j$ being the operator of projection on the plane being orthogonal to vector $k^i$.
Denominator of equation \eqref{expli} contains the
distance $R=|{\bm x}-{\bm x}_0|$
between observer and source of light which can be very large, thus, making an impression that the relativistic correction $\beta^i(\tau,{\bm{\xi}})$ is probably negligibly small. However, the difference $\Delta\Xi^j
(\tau,{\bm{\xi}})$ in the numerator of (\ref{expli}) is
proportional either to the distance $r=|{\bm x}|$ between the observer and the isolated system or to the distance $r_0=|{\bm x}_0|$ between the source of light and the isolated system. Either one of them or both distances can be comparable with $R$ so that the relativistic correction  $\beta^i(\tau,{\bm{\xi}})$ cannot be neglected in general case, and
must be
taken into account for
calculation of relativistic perturbations of light-ray trajectory. Only in a case where observer and source of light reside
at extremely large distances on
opposite
sides of the source of gravitational waves can the relativistic correction $\beta^i$
be neglected.

\section{Observable Relativistic Effects}\label{mq6yh}

In this section we derive general expressions for four observable relativistic effects  -- the time delay, the light bending\index{bending of light}, the frequency shift, and the rotation of the polarization plane.

\subsection{Gravitational time delay of light}

Relativistic time delay \index{time delay of light}for light propagating through time-dependent gravitational field can be obtained either directly from expression \eqref{tr2} or from the electromagnetic eikonal \eqref{7z}, \eqref{pmr1} by observing that the eikonal is constant not only on the null hypersurface of the wave front of electromagnetic wave but also along the light rays citep{frolov}. Both derivations lead, of course, to the same result
\ba
\label{td}
t-t_0&=&
         |{\bm x}-{\bm x}_0|+\Delta(\tau,\tau_0)\;,\\
         \label{t93}
         \Delta(\tau,\tau_0)&=&\DelG+\Delm+\Dels\;,\ea
where $x^\alpha_0=(t_0,{\bm x}_0)$ are four-coordinates of the point of emission of light, $x^\alpha=(t,{\bm x})$ are four-coordinates of the point of observation, $\DelG$, $\Delm$ and $\Dels$ are functions describing correspondingly
the delay of the electromagnetic signal due to the gauge-freedom, mass and spin
multipoles of the gravitational field of the isolated system. These functions
are expressed  as follows
\ba
\label{t78}
\DelG&=&-k_i\left[\XG
                -\XGo
             \right],\\
\label{tdm} \Delm&=&
     -k_i\left[\Xiim
                -\Xiimo
             \right],\\
\label{tds}
\Dels&=&
     -k_i\left[\Xiis
                -\Xiiso
             \right].
\ea
There exists a relation between the relativistic perturbations of the eikonal and the light-ray trajectory
\begin{equation}
\label{pev5}
\psi(\tau,{\bm\xi})=-k_i\Xi^i(\tau,{\bm\xi})\;,
\end{equation}
where the eikonal perturbation, $\psi$, reads off the equations \eqref{pmr1}--\eqref{pmr4}.
From equations \eqref{t78}--\eqref{tds} one can infer that in the
linear approximation the functions describing the time delay are just
the projections of vector functions describing the coordinate perturbation of the light-ray
trajectory onto the unperturbed direction $k^i$ from the source of light to
the observer.

Equation \eqref{td} defines the light time delay effect in the global time $t$ of the ADM-harmonic reference frame. In order to convert it
to observable proper time $T$ of the observer, we
assume for simplicity that the observer is in a state of free fall
and moves with velocity $V^i$ with respect to the reference frame of the isolated system.
Transformation from the
ADM-harmonic coordinate time $t$ to the proper time $T$ is made with
the help of the standard formula \cite{LL,mtw}
\begin{equation}
\label{prtim}
dT^2=-g_{\alpha\beta}dx^\alpha dx^\beta\;.
\end{equation}
Substituting the metric tensor expansion $g_{\alpha\beta}=\eta_{\alpha\beta}+h_{\alpha\beta}$, where $h_{\alpha\beta}$ is given by equations \eqref{adm1}--\eqref{adm4}, to equation \eqref{prtim} yields
\begin{eqnarray}
\label{prop}
\frac{dT}{dt}&=&\sqrt{1-{\bm V}^2-h_{00}(1+{\bm V}^2)-2h_{0i}V^i-h^{TT}_{ij}V^i V^j}\;.
\end{eqnarray}
In the most simple case, when observer is at rest (${\bm V}=0$) with respect to the ADM-harmonic reference frame equation \eqref{prop} is drastically simplified and depends only on $h_{00}$ component of the metric tensor. Implementation of formula (\ref{adm1}) for $h_{00}$ and subsequent
integration of (\ref{prop}) with respect to time  yields then
\vspace{0.3 cm}
\begin{eqnarray}
\label{tra}
T&=&\left(1-\frac{{\cal M}}{r}\right)\left(t-t_i\right)\;,
\end{eqnarray}
where $t_{\rm i}$ is the initial epoch of observation.
Another simple case of equation (\ref{prop}) is obtained for observer located at the distance $r$ so large that one can neglect $h_{00}$ and $h_{0i}$ quasi-static perturbations of the metric tensor. Then, the difference between
the observer's proper
time $T$
 and coordinate time $t$ is
 \begin{eqnarray}
\label{wcu6}
\frac{dT}{dt}&=&\sqrt{1-{\bm V}^2-h^{TT}_{ij}V^i V^j}\;,
\end{eqnarray}
leading in the case of a small velocity to
\begin{eqnarray}
\label{wcu85}
T&=&\sqrt{1-{\bm V}^2}(t-t_i)-\frac{V^i V^j}{\sqrt{1-{\bm V}^2}}\int^t_{t_i}h^{TT}_{ij}dt\;.
\end{eqnarray}
The last term in the right side of this equation describes the effect of the gravitational waves on the rate of the proper time of observer which may be important under some specific circumstances but is enormously small in real practice and, as a rule, can be neglected. Nonetheless, it would be interesting to study this effect in more detail as a possible tool to detect gravitational waves emitted, for example, in galactic supernova explosions or by powerful gamma-ray bursts (GRB) in distant galaxies. \index{gamma-ray burst}\index{GRB}

\subsection{Gravitational deflection of light}

The coordinate direction to a source of light measured at the point
of observation ${\bm x}$ is
defined by the spatial components of four-vector of photon $l^\alpha$ incoming to an observer from the source of light. This vector depends on the energy (frequency) of photon which is irrelevant for the present section. Normalization of $l^\alpha$ to its frequency eliminates the frequency-dependence and brings about a four-vector $p^\alpha=(1,p^i)$ where $p^i=-\dot{x}^i$ and the dot denotes derivative with respect to coordinate time $t$. Vector $\p^a$ is null, $p_\a p^\a=0$, and is only direction-dependent. The spatial part of this vector can be presented as a small deviation from the direction $k^i$ of the unperturbed photon's trajectory
\begin{eqnarray}
\label{coor}
p^i&=&-k^i-\dot{\Xi}^i\;,
\end{eqnarray}
where  the minus sign indicates that the tangent vector $p^i$ is directed from the observer to the source of light.

The coordinate
direction $p^i$ is not a directly observable quantity as it is defined with respect to the chosen coordinate grid on the curved spacetime manifold. A real observable vector
towards the source of light, $s^\alpha=(1,s^i)$, is defined with respect to the local
inertial frame co-moving with the observer \citep{kopeikin_scholarpedia}.
In this frame $s^i=-dX^i/dT$, where $T$ is the
observer's proper time, and $X^i$ are the spatial coordinates of the local inertial
frame with the observer at its origin. We shall assume for simplicity that the observer is at
rest with respect to
the global ADM-harmonic coordinates $(t,x^i)$. The case of an
observer moving with respect to the ADM-harmonic system with velocity $V^i$ can
be treated with the help of the Lorentz transformation \index{Lorantz transformation} which is a straightforward procedure so that we do not discuss it.
In case of a static observer, 
transformation from the global ADM-harmonic coordinates $(t,x^i)$ to the local
coordinates $(T,X^i)$ is given by the formulas
\begin{equation}
\label{trans1}
dT=\Lambda^0_{\;0}\; dt+\Lambda^0_{\;j}\; dx^j\qquad,\qquad
dX^i=\Lambda^i_{\;0}\; dt+\Lambda^i_{\;j}\; dx^j\;,
\end{equation}
where the matrix of transformation $\Lambda^{\alpha}_{\beta}$ is defined by the
requirements of orthonormality
\begin{eqnarray}
\label{ortog}
g_{\alpha\beta}&=&\eta_{\mu\nu}\Lambda^{\mu}_{\;\alpha}\Lambda^{\nu}_{\;\beta}\;.
\end{eqnarray}
In particular, the orthonormality condition (\ref{ortog}) assumes that the spatial
angles and lengths at the point of observations are measured with the
Euclidean metric $\delta_{ij}$. Because the vector $s^\alpha$ is null ($s_\a s^\a=0$) with respect to the Minkowski metric $\eta_{\alpha\beta}$, we conclude that the Euclidean length, $|{\bm s}|$, of vector $s^i$ is equal to 1. Indeed, one has
\begin{eqnarray}
\label{unity}
\eta_{\alpha\beta}s^\alpha s^\beta&=&-1+{\bm s}^2=0\;.
\end{eqnarray}
Hence, $|{\bm s}|=1$.

In the linear approximation with respect to the universal gravitational constant $G$,
the matrix of the transformation is
as follows \cite{60,31}
\vspace{0.3 cm}
\begin{eqnarray}
\label{lambda}
\Lambda^0_{\;0}&=&1-\frac{1}{2}h_{00}(t,{\bm x})\;,\nonumber\\
\mbox{} \Lambda^0_{\;i}&=&-h_{0i}(t,{\bm x})\;,\nonumber\\
\mbox{} \Lambda^i_{\;0}&=&0\;,\nonumber\\\mbox{}
 \Lambda^i_{\;j}&=&\left[1+\frac{1}{2}h_{00}(t,{\bm x})\right]\delta_{ij}+
 \frac{1}{2}h^{TT}_{ij}(t,{\bm x})\;.
\end{eqnarray}
Using transformation (\ref{trans1}) we obtain relationship between the
observable vector $s^i$ and the coordinate direction $p^i$
\begin{eqnarray}
\label{rls}
s^i&=&-\frac{\Lambda^i_{\;0}-\Lambda^i_{\;j}\; p^j}{\Lambda^0_{\;0}-\Lambda^0_{\;j}\; p^j}\;.
\end{eqnarray}
In the linear approximation this takes the form
\begin{eqnarray}
\label{form}
s^i&=&
\left(1+h_{00}-h_{0j}p^j\right)p^i+\frac{1}{2}h^{TT}_{ij}p^j\;.
\end{eqnarray}
Remembering that vector $|{\bm s}|=1$,
we find the Euclidean norm of the
vector $p^i$ from the relationship
\begin{eqnarray}
\label{norma}
|{\bm p}|&=&1-h_{00}+h_{0j}p^j-\frac{1}{2}h^{TT}_{ij}p^i p^j\;,
\end{eqnarray}
which brings equation (\ref{form}) to the form
\begin{eqnarray}
\label{bnm}
s^i&=&m^i+\frac{1}{2}P^{ij}m^q h^{TT}_{jq}(t,{\bm x})\;,
\end{eqnarray}
where $P^{ij}=\delta^{ij}-k^ik^j$ is the operator of projection onto the plane being orthogonal to $k^i$, and the Euclidean unit vector $m^i=p^i/|{\bm p}|$.

Let us now denote
by $\alpha^i$ the dimensionless vector describing the total angle of
deflection of the light ray measured at the point of observation,
and calculated
with respect to
vector $k^i$ given at  past null infinity. It is defined according
to  \cite{32}
\begin{eqnarray}
\alpha^i(\tau,{\bm{\xi}})&=&k^i {\bm k}\cdot
\dot{\bm{\Xi}}(\tau,{\bm{\xi}})-\dot{\Xi}^i(\tau,{\bm{\xi}})\;,
\end{eqnarray}
or
\begin{eqnarray}
\label{ang}
\alpha^i(\tau,{\bm{\xi}})&=&-\;P^i_{\;j}\;\dot{\Xi}^j(\tau,{\bm{\xi}})\;.
\end{eqnarray}
As a consequence of definitions (\ref{coor}) and (\ref{ang}), we conclude
that
\begin{eqnarray}
\label{uio}
m^i&=&-k^i+\alpha^i(\tau,{\bm{\xi}})\;.
\end{eqnarray}
Accounting for expressions (\ref{bnm}), (\ref{uio}), and (\ref{uuu}) we
obtain the observed direction to the source of light
\begin{eqnarray}
\label{dop}
s^i(\tau,{\bm{\xi}})&=&K^i+\alpha^i(\tau,{\bm{\xi}})+\beta^i(\tau,{\bm{\xi}})+\gamma^i(\tau,{\bm{\xi}})\;,
\end{eqnarray}
where the unit vector $K^i$ is given by equation (\ref{unitv}), relativistic correction $\beta^i$ is defined by equation
(\ref{expli}), and the perturbation
\begin{eqnarray}
\label{gamma}
\gamma^i(\tau,{\bm{\xi}})&=&-\frac{1}{2}P^{ij}k^q h^{TT}_{jq}(t,{\bm x})
\end{eqnarray}
describes a deformation of the local coordinates of observer with respect to the global
ADM-harmonic frame caused by the transverse-traceless part of the gravitational field of the isolated system at the point of observation.

Let two sources of light (quasars, stars, etc.) be observed along the directions ${\bm s}_1$ and
${\bm s}_2$ corresponding to two different light rays passing near the isolated gravitating system at the minimal distances corresponding to the (vector) impact parameters, ${\bm\xi}_1$ and ${\bm\xi}_2$.  The angle $\Psi$ between them, measured
in the local inertial frame is
\begin{eqnarray}
\label{lkj}
\cos\Psi&=&{\bm s}_1\cdot{\bm s}_2\;,
\end{eqnarray}
where the dot between the two vectors denotes the
usual Euclidean scalar product. It is worth emphasizing
that the observed direction $s^i$ to each source of light includes relativistic
deflection of the light ray in the form of three perturbations. Two of them, $\alpha^i$ and $\gamma^i$, depend only on the quantities taken
at the point of observation, but $\beta^i$, according to equation \eqref{expli}, is also sensitive to the strength of the gravitational field taken at the point of emission of light. This
remark reveals that according to relation (\ref{dop}) a single
gravitational wave signal may cause
different  angular displacements and/or time delays for
different sources of light located at different distances from the source
of gravitational waves even if the directions to the light sources are the same.

Without going into further
details of the observational procedure we give an explicit
expression for the total angle of the light deflection $\alpha^i$. We have
\begin{equation}
\label{qapo1}
\alpha^i(\tau,{\bm\xi})=\aGi+\aMi+\aSi\;,
\end{equation}
where
\begin{eqnarray}
\label{qapo2}
\aGi&=&-P^{ij}\dtau\left(\phi^j+w^j\right)\;,
\\
\label{qapo3}
\aMi&=&-\frac{2\Mc}{r}\frac{\xi^i}{y} \\\notag &&-
2\dksi_i
   \sum_{l=2}^\infty\sum_{p=0}^l\sum_{q=0}^p
     \frac{(-1)^{l+p-q}}{l!}
       C_l(l-p,p-q,q)H(2-q)
\times
\\
&&\left(
    1-\frac{p-q}l
   \right)
       \left(
               1-\frac{p-q}{l-1}
         \right)
            k_{<a_1\hdots a_p}\dksi_{a_{p+1}\hdots a_l>}
    \dtau^q
         \left[\frac{{\Ic}^{(p-q)}_{A_{l}}(t-r)}r\right]^{[-1]}
\notag    \\\notag
            &&-4P^{ij}\sum_{l=2}^\infty\sum_{p=0}^{l-1}
            \frac{(-1)^{l+p}}{l!}
                                  C_{l-2}(l-p-1,p-1)\times\\&&
            k_{<a_1\hdots a_{p-1}}\dksi_{a_p\hdots a_{l-1}>}
        \left[\frac{{\Ic}^{(p)}_{jA_{l-1}}(t-r)}r\right]
             \;,
  \notag
\\
\label{qapo4}
\aSi&=&
-2k^j\epsilon_{jba}\dksi_{a}\left(\frac{\mathcal{S}_b\xi^i}{yr}\right)
+2P^{ij}\dksi_a\ejsbr\\
&&+4\dksi_{ia}
    \sum_{l=2}^{\infty}\sum_{p=0}^{l-1}\sum_{q=0}^{p}
        \frac{(-1)^{l+p-q}l}{(l+1)!}
C_{l-1}\left(l-p-1,p-q,q\right)H(2-q)\times
    \notag
\\
    &&\left(1-\frac{p-q}{l-1}\right)
k_{<a_1\hdots a_p}\dksi_{a_{p+1}\hdots a_{l-1}>}
        \dtau^q
        \left[\frac{k^j\epsilon_{jba}{\mathcal {S}}^{(p-q)}_{bA_{l-1}}(t-r)}r\right]^{[-1]}
    \notag
\\\notag
&&-4\left(\dksi_a-k_a\dtz\right)
\sum_{l=2}^{\infty}\sum_{p=0}^{l-1}
        \frac{(-1)^{l+p}l}{(l+1)!}
C_{l-2}\left(l-p-2,p\right)\times\\&&
k_{<a_1\hdots a_p}\dksi_{a_{p+1}\hdots a_{l-1}>}
        \left[P^{ij}\ejslrrp\right]
    \notag
\\\notag
&&-4
\sum_{l=2}^{\infty}\sum_{p=0}^{l-1}
        \frac{(-1)^{l+p}l}{(l+1)!}
C_{l-1}\left(l-p-1,p\right)\times\\&&
k_{<a_1\hdots a_p}\dksi_{a_{p+1}\hdots a_{l-1}>}
        \left[ \frac {P^{iq}k^j\epsilon_{jba_{l-1}}{\mathcal {S}}^{(p+1)}_{qbA_{l-2}}(t-r)}r\right]\;.
    \notag
\end{eqnarray}
These expressions do not contain any explicit integration along the light ray trajectory because all explicit integrals are either eliminated after taking partial derivatives with respect to the upper limit of the integrals or they are vanish because the numerical coefficient in front of them become nil.

\subsection{Gravitational shift of frequency}\label{gsf}

Exact calculation of the gravitational shift of 
frequency of electromagnetic wave travelling from the point of emission to observer
plays a crucial role for the adequate
interpretation of spectral astronomical investigations of high resolution including the astronomical measurements of radial velocities of stars \cite{Lindegren_2003}, anisotropy of cosmic microwave background radiation (CMBR),
and others. In the last decade the technique for measuring the radial velocity of stars had reached
an unprecedented precision of 
$1$ m/sec \cite{57,57a}. At this level the post-Newtonian effects in the orbital motion of spectroscopic binary stars can be measured \cite{59,58ff}. Gravitational shift of frequency of light affect the apparent brightness of the observed sources according to equation (\ref{a34s}). Therefore, it can be important in highly accurate photometric measurements of faint radio sources with large radio telescope like SKA \citep{kramer_2010htra}  

Let a source of light move with respect to the ADM-harmonic coordinate frame
$(t,x^i)$ with velocity ${\bm V}_0=\dot{\bm x}_0(t_0)$ and emit continuous 
electromagnetic
radiation at frequency $\nu_0=1/(\delta{ T}_0)$, where $t_0$ and
${T}_0$ are
the coordinate and proper time of the source of light, respectively. We denote by
$\nu=1/(\delta{ T})$ the observed frequency of the electromagnetic radiation
measured at
the point of observation by an observer moving with velocity ${\bm V}
=\dot{\bm x}(t)$
with respect to the ADM-harmonic coordinate frame. In the geometric optics approximation we can consider the
increments of the proper time, $\delta{ T}_0$ and $\delta{ T}$, as infinitesimally small which allows us to operate with them as with differentials \cite{60,31}. Time delay equation \eqref{td} can be considered as an implicit function of the emission time $t_0=t_0(t)$ having the time of observation $t$ as its argument. Because the coordinate and proper time at the points of emission of light and its observation are connected through the metric tensor we conclude that the
observed gravitational shift of frequency $1+z=\nu/\nu_0$ can be defined
through
the consecutive differentiation of the proper time of the source of
light,
${T}_0$, with respect to the proper time of the observer, ${ T}$,
\cite{60,31}
\begin{eqnarray}
\label{58}
1+z&=&\frac{d{T}_0}{d{ T}}=
\frac{d{T}_0}{dt_0}\frac{dt_0}{dt}\frac{dt}{d{ T}}\;.
\end{eqnarray}

Synge calls relation (\ref{58}) the
{\it Doppler effect in terms of frequency} \cite[p. 122]{syngebook}. It is
fully consistent with the definition of the {\it Doppler shift in terms of energy}
\citep[p. 231]{syngebook} when one compares the energy of photon at the
points of emission and observation of light. The {\it Doppler shift in terms
of energy} is given by
\begin{eqnarray}
\label{enrg}
1+z&=&\frac{\nu}{\nu_0}=\frac{l_\a u^\alpha}
{l_{0\a}u_0^\alpha}\;,
\end{eqnarray}
where $u_0^\alpha$, $u^\alpha$ and ${l}_{0\alpha}$, ${l}_\alpha$
are 4-velocities of the source of light and observer and 4-momenta of photon at
the points of emission and observation respectively. It is quite easy
to see that both mentioned formulations of the Doppler shift
effect are equivalent. Indeed, taking into account that
$u^\alpha=dx^\alpha/d{ T}$ and ${l}_\alpha=
{\partial \varphi}/{\partial x^\alpha}$, where $\varphi$ is the phase of the
electromagnetic wave (eikonal), we obtain  $l_\a u^\alpha=d\varphi/d{ T}$, and $l_{0\a} u_0^\alpha=d\varphi_0/d{ T_0}$ respectively.
Thus, equation \eqref{enrg} yields
\begin{eqnarray}
\label{volk}
1+z&=&\frac{d\varphi}{d\varphi_0}\frac{d{ T}_0}{d{ T}}\;.
\end{eqnarray}
The
phase of electromagnetic wave remains constant along the light ray trajectory.
For this reason, $d\varphi/d\varphi_0=1$ and, hence, equation (\ref{volk}) is reduced to equation (\ref{58}) as expected on the ground of physical intuition.
Detailed comparison of the two definitions of the Doppler shift and the proof of their identity in general theory of relativity is thoroughly discussed in \cite{ksprd,kopeikin_book}.

We emphasize that in equation (\ref{58}) the time derivative
\begin{eqnarray}
\label{59}
\frac{dT_0}{dt_0}&=&\left[1-{\bm V}_0^2-(1+{\bm V}_0^2)h_{00}(t_0,{\bm x}_0)-2V_0^ih_{0i}(t_0,{\bm x}_0)-V_0^i V_0^jh^{TT}_{ij}(t_0,{\bm x}_0)\right]^{1/2}\;,
\end{eqnarray}
is taken at the time $t_0$ at the point of emission of light ${\bm x}_0$ along the world line of the emitter of light while the time derivative
\begin{eqnarray}
\label{60}
\frac{dt}{dT}&=&\left[1-{\bm V}^2-(1+{\bm V}^2)h_{00}(t,{\bm x})-2V^ih_{0i}(t,{\bm x})-V^i V^jh^{TT}_{ij}(t,{\bm x})\right]^{-1/2}\;,
\end{eqnarray}
is calculated at the time of observation $t$ at the position of observer ${\bm x}$ along the world line of the observer.

The time derivative $dt_0/dt$ is taken along the
light-ray
trajectory and calculated from the time delay equation (\ref{td}) where we have to take
into
account that function $\Delta(\tau,\tau_0)$ depends on times $t_0$ and $t$ indirectly
through the retarded times $s_0=t_0-r_0$ and $s=t-r$ that are arguments of the multipole moments of the isolated system and wherein $r_0\equiv |{\bm x}_0|=|{\bm x}(t_0)|$, $r=|{\bm x}|=|{\bm x}(t)|$ are functions of time $t_0$ and $t$ respectively. Function $\Delta(\tau,\tau_0)$ also depends on the time of the closest approach of light ray, $t^{\ast}$, through variables $\tau=t-t^*$, $\tau_0=t_0-t^*$, and on the unit
vector ${\bm k}$. Both $t^*$ and ${\bm k}$ should be considered as parameters depending on $t_0$ and $t$ because of the relative motion of the observer with respect to the source of light which
causes variation in the relative position of the source of light and
observer and, consequently, to the corresponding change in the parameters characterizing trajectory of the
light ray, that is in $t^\ast$ and ${\bm k}$. Therefore, the function $\Delta(\tau,\tau_0)$ must be viewed as parametrically-dependent on four variables $\Delta=\Delta(s,s_0,t^*,{\bm k})$. Accounting for these remarks the derivative
along the light ray reads as follows
\begin{eqnarray}
\label{61}
\frac{dt_0}{dt}&=&\frac{1+{\bm K}\cdot {\bm V}-\displaystyle{
\left\{\frac{\partial s}{\partial t}
\frac{\partial}{\partial s}+\frac{\partial s_0}{\partial t}
\frac{\partial}{\partial s_0}+\frac{\partial t^{\ast}}{\partial t}
\frac{\partial}{\partial t^{\ast}}+\frac{\partial k^i}{\partial t}
\frac{\partial}{\partial k^i}\right\}\Delta(s,s_0,t^{\ast},{\bm k})}}
{1+{\bm K}\cdot {\bm V}_0+\displaystyle{
\left\{\frac{\partial s}{\partial t_0}
\frac{\partial}{\partial s}+\frac{\partial s_0}{\partial t_0}
\frac{\partial}{\partial s_0}+\frac{\partial t^{\ast}}{\partial t_0}
\frac{\partial}{\partial t^{\ast}}+\frac{\partial k^i}{\partial t_0}
\frac{\partial}{\partial k^i}\right\}\Delta(s,s_0,t^{\ast},{\bm k})}
}\;,
\end{eqnarray}
where the unit vector ${\bm K}$ is defined in (\ref{unitv}) and where
we explicitly
show the dependence of function $\Delta(\tau,\tau_0)$ on all parameters which implicitly depend on time.

The time derivative of vector ${\bm k}$ is calculated using the
approximation ${\bm
k}=-{\bm K}$ and formula (\ref{unitv}) where the coordinates of the
source of light, ${\bm
x}_0(t_0)$, and of the observer, ${\bm x}(t)$, are considered as functions of time. These derivatives are
\begin{equation}
\label{vark}
\frac{\partial k^i}{\partial t}=\frac{({\bm k}\times({\bm V}\times{\bf
k}))^i}{R}\;,\quad\quad\quad
\frac{\partial k^i}{\partial t_0}=-\frac{({\bm k}\times({\bm V}_0\times{\bf
k}))^i}{R}\;,
\end{equation}
where $R=|{\bm x}-{\bm x}_0|$ is the coordinate distance between the observer and
the source of
light.

Time derivatives of the retarded times $s$ and $s_0$ with respect to $t$ and
$t_0$ are calculated from their definitions,
$s=t-r$ and $s_0=t_0-r_0$, where we have to take into account that
the spatial position of the point of observation is connected to the
point of emission of light by the unperturbed trajectory of light,
${\bm x}={\bm x}_0+{\bm k}\;(t-t_0)$.
More
explicitly, we use for the calculations
the following relations
\begin{eqnarray}
\label{ret}
s&=&t-|{\bm x}_0(t_0)+{\bm k}\;(t-t_0)|\;,\\\label{rtuo}
s_0&=&t_0-|{\bm x}_0(t_0)|\;,
\end{eqnarray}
where the unit vector ${\bm k}={\bm k}(t,t_0)$ must be also considered as
a function of two arguments $t$, $t_0$ with its
derivatives given by equation (\ref{vark}).
It is instructive to notice that relation (\ref{ret}) combines two characteristics of the null cone -- the first one
is related to the propagation of gravitational field from the isolated system to an observer, and the second one is related to
the propagation of light from the source of light to the same observer. Equation \eqref{rtuo} describes a null cone characteristic corresponding to the propagation of gravitational field from the isolated system to the point of emission of light.
Calculation of the infinitesimal variations of equations (\ref{ret}), \eqref{rtuo} immediately gives a set of partial derivatives
\begin{eqnarray}
\label{add1}
\frac{\partial s}{\partial t}&=&1-{\bm k}\cdot{\bm N}
-({\bm k}\times{\bm V})\cdot({\bm k}\times{\bm N})\;,\\\nonumber\\
\label{add2}
\frac{\partial s}{\partial t_0}&=&(1-{\bm k}\cdot{\bm V}_0)({\bm k}\cdot{\bm  N})\;,\\\nonumber\\
\label{add3}
\frac{\partial s_0}{\partial t_0}&=&1-{\bm V}_0\cdot{\bm N}_{0}\;,\\\nonumber\\
\label{add4}
\frac{\partial s_0}{\partial t}&=&0\;,
\end{eqnarray}
where $N^i=x^i/r$ and $N^i_0=x^i_0/r_0$ are the unit vectors directed from the isolated system to the observer and to the source of light respectively..

Time derivatives of the parameter $t^{\ast}=t_0-{\bm k}\cdot{\bm x}_0(t_0)$, where again ${\bm k}={\bm k}(t,t_0)$, read
\begin{equation}
\label{tstar}
\frac{\partial t^\ast}{\partial t_0}=1-{\bm k}\cdot{\bm
V}_0+\frac{{\bm V}_0\cdot{{\bm{\xi}}}}{R}\;,
\quad\quad\quad\quad\frac{\partial t^\ast}{\partial t}=-
\frac{{\bm V}\cdot{{\bm{\xi}}}}{R}\;.
\end{equation}
Terms of the order $|{\bm\xi}|/R$ in both formulas are produced by the
time derivatives of vector ${\bm k}$.

In what follows we shall restrict ourselves with a static case of an observer and a source of light, that is we shall assume velocities ${\bm V}={\bm V}_0=0$. Taking into account this restriction in equations \eqref{vark}--\eqref{tstar} we expand denominator of \eqref{61} leaving only the linear with respect to the universal gravitational constant $G$ terms. Reducing, then, similar terms allows us to simplify \eqref{61} to
\begin{eqnarray}
\label{bumc}
\frac{dt_0}{dt}&=&1-\Bigl\{\frac{\partial}{\partial s}
+\frac{\partial}{\partial s_0}+\frac{\partial}{\partial t^*}\Bigr\}\Delta(s,s_0,t^{\ast},{\bm k})\;.
\end{eqnarray}
Function $\Delta=\Delta(s,s_0,t^*,{\bm k})$ is defined
by (\ref{t93}) which (in the case of the static observer and the source of light) depends on the retarded times $s$, $s_0$ and the time of the closest approach $t^*$ through the arguments (see \eqref{y}) $y=s-t^*$, $y_0=s_0-t^*$, and $t^*$, that is
\begin{equation}
\label{px4}
\Delta(s,s_0,t^*,{\bm k})\equiv\Delta(y,y_0,t^*)=\psi(y,t^*)-\psi(y_0,t^*)\;,
\end{equation}
where $\psi=-k^i\Xi$ is the relativistic perturbation of the eikonal defined in \eqref{pev5}. This particular dependence of $\psi$ on its arguments means that the partial derivative of $\psi(y,t^*)$ with respect to $y$ and and that of $\psi(y_0,t^*)$ with respect to $y_0$ vanish in \eqref{bumc} which is reduced to a simpler form
\begin{eqnarray}
\label{bumc1}
\frac{dt_0}{dt}&=&1-\frac{\partial\psi(y,t^*)}{\partial t^*}+\frac{\partial\psi(y_0,t^*)}{\partial t^*}\;.
\end{eqnarray}
The partial time derivative from $\psi(y,t^*)$ is found by differentiating relations (\ref{pmr2})--(\ref{pmr4})
\begin{equation}
\label{aaa0}
\frac{\partial\psi(y,t^*)}{\partial t^*}=\frac{\partial\psi_{\scriptscriptstyle (G)}(y,t^*)}{\partial t^*}+\frac{\partial\psi_{\scriptscriptstyle (M)}(y,t^*)}{\partial t^*}+\frac{\partial\psi_{\scriptscriptstyle (S)}(y,t^*)}{\partial t^*}\;
\end{equation}
where
\begin{eqnarray}
\label{aaa2}
\frac{\partial\psi_{\scriptscriptstyle (G)}(y,t^*)}{\partial t^*}&=&(k^i\frac{\partial\phi^i}{\partial t^*}-\frac{\partial\phi^0}{\partial t^*})+(k^i\frac{\partial{w}^i}{\partial t^*}-\frac{\partial{w}^0}{\partial t^*})\;,\\\nonumber\\
\label{aaa3}
\frac{\partial\psi_{\scriptscriptstyle (M)}(y,t^*)}{\partial t^*}&=&
2\sum_{l=2}^\infty\sum_{p=0}^{l-1}
            \frac{(-1)^{l+p}}{l!}
                                  C_l(l-p,p)
\left(
    1-\frac{p}l
  \right)\\&&\times
\left\{
    \left(
        1+\frac{p}{l-1}
      \right)
            k_{<a_1\hdots a_p}\dksi_{a_{p+1}\hdots a_l>}
       \left[\frac{{\Ic}^{(p+1)}_{A_{l}}(t-r)}r   \right]^{[-1]}
\notag\right.    \\&&\left.-
    \frac{2p}{l-1}
            k_{<a_1\hdots a_{p-1}}\dksi_{a_p\hdots a_{l-1}>}
              \left[\frac{k^i{\Ic}^{(p+1)}_{iA_{l-1}}(t-r)}r\right]^{[-1]}
             \right\}
 \notag    \\\notag\\\label{aaa4}
\frac{\partial\psi_{\scriptscriptstyle (S)}(y,t^*)}{\partial t^*}&=& 4\dksi_a
\sum_{l=2}^{\infty}\sum_{p=0}^{l-1}
        \frac{(-1)^{l+p}l}{(l+1)!}
C_{l-1}\left(l-p-1,p\right)
\\&&\times \left(1-\frac{p}{l-1}\right)
 k_{<a_1\hdots a_p}\dksi_{a_{p+1}\hdots a_{l-1}>}
        \left[\frac{k^i\epsilon_{iba}{\cal S}^{(p+1)}_{bA_{l-1}}(t-r)}{r}\right]^{[-1]}\;.
    \notag
\end{eqnarray}
When deriving these equations we assumed that the time evolution of both the mass ${\cal M}$ and the angular momentum ${\cal S}^i$ of the isolated system can be neglected. In opposite case, the right side of equations \eqref{aaa3} and \eqref{aaa4} would contain also time derivatives from the mass and the angular momentum. Partial derivative from function $\psi(y_0,t^*)$ is obtained from equations \eqref{aaa0}--\eqref{aaa4} after replacements $y\rightarrow y_0$, $t\rightarrow t_0$, and $r\rightarrow r_0$. Equations (\ref{aaa2})--(\ref{aaa4}) do not contain explicit integrals along the light ray trajectory because all of the explicit integrals vanish after taking the partial derivatives in these equations.

Frequency shift, $z$, is given by equation \eqref{58}. In case when both observer and source of light are at rest with respect to the ADM-harmonic reference frame, the frequency (energy) of a photon propagating through the gravitational potential of the isolated astronomical system changes in accordance with 
\begin{equation}
\label{aaa5}
\frac{\nu}{\nu_0}=\sqrt{\frac{1-h_{00}(t_0,{\bm x}_0)}{1-h_{00}(t,{\bm x})}}\left[1-\frac{\partial\psi(y,t^*)}{\partial t^*}+\frac{\partial\psi(y_0,t^*)}{\partial t^*}\right]\;,
\end{equation}
In the linear with respect to $G$ approximation, equation (\ref{aaa5}) is simplified
\begin{equation}
\label{aaa6}
\frac{\delta\nu}{\nu_0}=\frac{1}{2}h_{00}(t,{\bm x})-\frac12 h_{00}(t_0,{\bm x}_0)-\frac{\partial\psi(y,t^*)}{\partial t^*}+\frac{\partial\psi(y_0,t^*)}{\partial t^*}\;,
\end{equation}
where $\delta\nu\equiv\nu-\nu_0$. The time of the closest approach $t^*$ enters equation \eqref{aaa6} explicitly. At the first glance, the reader may think that it must be known in order to calculate the gravitational shift of frequency. However, the partial derivative with respect to $t^*$ have to be understood in the sense of equation \eqref{rf2} which makes it evident that the time derivative with respect to $t^*$ taken on the light-ray path is, in fact, the time derivative with respect to time $t$ taken before the using of the light-ray trajectory substitution. Thus, the gravitational shift of frequency is actually not sensitive to the time of the closest approach $t^*$, and can be recast to the form which is more suitable for practical applications,
\begin{equation}
\label{arz}
\frac{\delta\nu}{\nu_0}=\frac{1}{2}h_{00}(t,{\bm x})-\frac12 h_{00}(t_0,{\bm x}_0)-\frac{\partial\psi(t,{\bm x})}{\partial t}+\frac{\partial\psi(t_0,{\bm x}_0)}{\partial t_0}\;,
\end{equation}
where $\psi(t,{\bm x})$ and $\psi(t_0,{\bm x}_0)$ are relativistic perturbations of the electromagnetic eikonal taken at the points of observation and emission of light respectively. These time derivatives are given by the same equations \eqref{aaa0}--\eqref{aaa4} after making use of (\ref{rf2}).

Equation \eqref{arz} has been derived by making use of the definition \eqref{58}. It is straightforward to prove that the definition \eqref{enrg} brings about the same result. Indeed, in the case of the static observer and the source of light their 4-velocities are, $u^\alpha=(dt/dT,0,0,0)$ and $u_0^\alpha=(dt_0/dT_0,0,0,0)$, and the wave vector $l_\alpha=\omega(k_\alpha+\partial_\alpha\psi)$ (see equation \eqref{7ez}) where $\omega$ is a constant frequency of light wave which is conserved along the light ray because of the equation of the parallel transport (\ref{ff2}). Substituting these relations to \eqref{enrg} leads immediately to equation \eqref{arz} as expected.

Physical interpretation of the relativistic frequency shift given by equation \eqref{arz} is straightforward although the calculation of different components of the time derivatives from the eikonal are tedious. The first two terms in the left side of  \eqref{arz} reads
\begin{equation}
\label{aaa7}
\frac{1}{2}h_{00}(t,{\bm x})-\frac12h_{00}(t_0,{\bm x}_0)=\frac{G\Mc}{r}-\frac{G\Mc}{r_0}\;,
\end{equation}
and represents the difference between the values of the spherically-symmetric part of the Newtonian gravitational potential of the isolated system taken at the point of observation and emission of light. The time derivatives of the eikonal depending on the mass and spin multipole moments of the isolated system are given in \eqref{aaa3}, \eqref{aaa4}. Scrutiny examination of these equations reveal that these components of the frequency shift depend on the first and higher order time derivatives of the multipole moments which vanish in the stationary case. The gravitational frequency shift contains the gauge-dependent contribution as well. This contribution is given by equation \eqref{aaa2} and can be calculated by making use of equations \eqref{nw0}--\eqref{nwi+}. The result is as follows
\begin{eqnarray}
\label{v+}
k^i\frac{\partial{w}^i}{\partial t^*}-\frac{\partial{w}^0}{\partial t^*}&=&
k^i\nabla_i\sum_{l=2}^{\infty}\ffrac{(-1)^l}{l!}\left[\frac{
            {\Ic}^{(-1)}_{A_l}(t-r)}{r}\right]_{,A_l}-
            \sum_{l=2}^{\infty}\ffrac{(-1)^l}{l!}
            \left[\frac{
            {\Ic}_{A_l}(t-r)}{r}\right]_{,A_l}
\\
        &-&4\sum_{l=2}^{\infty}\ffrac{(-1)^l}{l!}
                 \left[\ffrac{k^i\dot{\Ic}_{iA_{l-1}}(t-r)}{r}
                                    \right]_{,A_{l-1}}\\\notag
                                    &+&
4\sum_{l=2}^{\infty}\ffrac{(-1)^ll}{(l+1)!}
              \left[\ffrac{k^i\epsilon_{iba}
                             {\mathcal
                             S}_{bA_{l-1}}(t-r)}{r}\right]_{,aA_{l-1}}\;,\notag\\\notag\\
\label{v++}
k^i\frac{\partial\phi^i}{\partial t^*}-\frac{\partial\phi^0}{\partial t^*}&=&
     -2\sum_{l=2}^\infty\sum_{p=1}^l\sum_{q=1}^p
           \frac{(-1)^{l+p-q}}{l!}C_l(l
                                     -p,p-q,q)\left(1-\frac{p-q}l\right)\times
\\
      &&\Biggl\{
           \left(1+\frac{p-q}{l-1}\right)
           k_{<a_1\ldots a_p}\dksi_{a_{p+1}\ldots a_l>}
                  \dtau^{q-1}
         \left[\frac{\Ic^{(p-q+1)}_{A_l}(t-r)}{r}\right]-
\notag \\
       &&2\frac{p-q}{l-1}
           k_{<a_1\ldots a_{p-1}}\dksi_{a_p\ldots a_{l-1} >}
                   \dtau^{q-1}
            \left[\frac{k^i\Ic^{(p-q+1)}_{iA_{l-1}}(t-r)}r\right]
          \Biggr\}-
\notag
\\
&&4\dksi_a
\sum_{l=2}^{\infty}\sum_{p=1}^{l-1}\sum_{q=1}^{p}
        \frac{(-1)^{l+p-q}l}{(l+1)!}
C_{l-1}\left(l-p-1,p-q,q\right)\times
    \notag
\\&& \left(1-\frac{p-q}{l-1}\right)
k_{<a_1\hdots a_p}\dksi_{a_{p+1}\hdots a_{l-1}>}
        \dtau^{q-1}\left[\frac{k^i\epsilon_{iab}\Sc^{(p-q+1)}_{bA_{l-1}}(t-r)}{r}\right]\;.
    \notag
\end{eqnarray}
Time derivatives $\hat\partial_{t^*_0}(k^i{w}^i-{w}^0)$ and $\hat\partial_{t^*_0}(k^i{\varphi}^i-{\varphi}^0)$ can be obtained from equations \eqref{v+}, \eqref{v++} after making replacements $t\rightarrow t_0$, $r\rightarrow r_0$, and $\tau\rightarrow\tau_0$.

\subsection{Gravity-induced rotation of the plane of polarization of light}\label{grpp}

Any kind of axisymmetric gravitational field induces a relativistic effect of the rotation of the polarization plane of an electromagnetic wave propagating through this field. To some extent this effect is similar to Faraday's effect in electrodynamics \cite{LLE}. The Faraday effect is caused by the presence of magnetic field along the trajectory of propagation of electromagnetic wave while the gravity-induced rotation of the plane of polarization of light is caused by the presence of the, so-called, gravitomagnetic field associated with the angular momentum and spin-type multipoles of the isolated system \cite{bct77,kop_2006IJMPD}. This gravitomagnetic effect was first discussed by Skrotskii (\cite{skrot,skrot+} and a number of other researches \cite{balaz,pleb,m75,piran,su,ckmr,sereno}. Recently, we have studied the Skrotskii effect caused by a spinning body moving arbitrarily fast and derived the Lorentz-invariant expression for this effect \cite{km}. In the present chapter we further generalize the Skrotskii effect to the case of an isolated system emitting gravitational waves of arbitrary multipolarity.

We consider the parallel transport of the reference polarization tetrad $e^\alpha_{\;(\beta)}$ defined by equations \eqref{lort} along the light ray. We assume for simplicity that at the past null infinity the spatial vectors of the tetrad, $e^0_{\;(i)}$, coincide with the spatial unit vectors of the coordinate tetrad defined by equation (\ref{dqx}). The parallel transport of the tetrad along the light ray is defined by equation \eqref{i9}. We are interested only in solving equation \eqref{i9} for vectors $e^\alpha_{\;(n)}$ ($n=1,2$) that are used in description of polarization of light. The propagation equation for the spatial components $e^i_{\;(n)}$ can be written in the following form
\begin{equation}
\label{rnq}
\frac{d}{d\tau}\left[e^i_{\;(n)}+\frac{1}{2}h_{ij}e^p_{\;(n)}\right]+\epsilon_{ijp}e^j_{\;(n)}\Omega^p=0\;,\qquad\qquad(n=1,2.)
\end{equation}
where the angular velocity vector 
\begin{equation}
\label{avw}
\Omega^i=-\frac{1}2\epsilon_{ijp}\partial_j\left(h_{p\alpha}k^\alpha\right)\;,
\end{equation}
describes the rate of the rotation of the plane of polarization of electromagnetic wave caused by the presence of the gravitomagnetic field.
As soon as equation \eqref{rnq} is solved, the time component $e^0_{\;(n)}$ of the tetrad is obtained from the orthogonality condition, $l_\alpha e^\alpha_{\;(n)}=0$, which implies that
\begin{equation}
\label{ops2}
e^0_{\;(n)}=k_i e^i_{\;(n)}+h_{0i}e^i_{\;(n)}+h_{ij} k^i e^j_{\;(n)}+\delta_{ij}\dot{\Xi}^i e^j_{\;(n)}\;,
\end{equation}
where the relativistic perturbation $\dot{\Xi}^i$ of the light-ray trajectory is given in equation \eqref{tr1}.

Let us decompose vector $\Omega^i$ into three components that are parallel and perpendicular to the unit vector $k^i$. We can use in the first approximation the well-known decomposition of the Kroneker symbol\index{Kroneker symbol!orthonormal basis} in the orthonormal basis
\begin{equation}
\label{ao61}
\delta^{ij}=a^ia^j+b^ib^j+k^ik^j\;,
\end{equation}
where $({\bm a},{\bm b},{\bm k})$ are three orthogonal unit vectors of the reference tetrad at infinity. Decomposition of  $\Omega^i$ is, then, given by
\begin{equation}
\label{i14}
\Omega^i=({\bf a}\cdot{\bm{\Omega}})a^i+({\bf b}\cdot{\bm{\Omega}})b^i+({\bm k}\cdot{\bm{\Omega}})k^i\;.
\end{equation}
Taking into account that at any point on the light-ray trajectory, vectors $e^i_{\;(1)}=a^i+O(h_{\alpha\beta})$, $e^i_{\;(2)}=b^i+O(h_{\alpha\beta})$, one obtains equations of the parallel transport of these vectors (\ref{rnq}) in the following form 
\begin{eqnarray}
\label{r-7}
\frac{d}{d\tau}\left[e^i_{\;(1)}+\frac{1}{2}h_{ij}e^j_{\;(1)}\right]-({\bm k}\cdot{\bm{\Omega}})e^i_{\;(2)}&=&0\;,\\\label{r-8}
\frac{d}{d\tau}\left[e^i_{\;(2)}+\frac{1}{2}h_{ij}e^j_{\;(2)}\right]+({\bm k}\cdot{\bm{\Omega}})e^i_{\;(1)}&=&0\;,
\end{eqnarray}
where equalities $\varepsilon_{ijl}e^i_{\;(1)} k^l=-e^i_{\;(2)}+O(h_{\alpha\beta})$, and $\varepsilon_{ijl}e^i_{\;(2)} k^l=e^i_{\;(1)}+O(h_{\alpha\beta})$ have been used.

Solutions of equations \eqref{r-7}, \eqref{r-8} in the linear approximation with respect to the (post-Newtonian) angular velocity $\Omega^i$ read
\begin{eqnarray}
\label{i15}
e^i_{\;(1)}&=&a^i-\frac{1}{2}h_{ij}a^j+\left(\int^\tau_{-\infty}{\bm k}\cdot{\bm{\Omega}}(\sigma)\;d\sigma\right)b^i\;,\\\nonumber\\\label{i15a}
e^i_{\;(2)}&=&b^i-\frac{1}{2}h_{ij}b^j-\left(\int^\tau_{-\infty}{\bm k}\cdot{\bm{\Omega}}(\sigma)\;d\sigma\right)a^i\;,
\end{eqnarray}
where the second term in the right side of \eqref{i15}, \eqref{i15a} preserve orthogonality of vectors $e^i_{\;(1)}$ and $e^i_{\;(2)}$ in the presence of gravitational field while the last term describes the Skrotskii effect which is a small rotation of each of the vectors at the angle
\begin{equation}
\label{i18}
\Phi(\tau)=\int^\tau_{-\infty}{\bm k}\cdot{\bm{\Omega}}\;d\sigma
\end{equation}
about the direction of light propagation, ${\bm k}$, in the local plane of vectors ${\bm e}_{(1)}$ and ${\bm e}_{(2)}$. It is worth noting that the Euclidean dot product
${\bm k}\cdot{\bm{\Omega}}$ can be expressed in terms of partial differentiation
with respect to the vector $\xi^i$ of the impact parameter. This can be done by making use of equation (\ref{za34}) and noting that $\varepsilon_{ijp}k^jk^p\equiv 0$, so that
\begin{equation}
\label{i20}
{\bm k}\cdot{\bm{\Omega}}=\frac{1}{2}k^\alpha k^j\varepsilon_{j{p}{q}}\hat{\partial}_q h_{\alpha{p}}\;.
\end{equation}
Hence, the transport equation for the angle $\Phi$ assumes the following form
\begin{equation}
\label{bu7} \frac{d\Phi}{d\tau}=\frac{1}2
        k^{\alpha}k^j
            \epsilon_{j{p} q}\dksi_q
                h_{\alpha{p}}\;.
\end{equation}

This equation can be split in three, linearly-independent parts corresponding to the contributions of the gauge, $\fig$, the mass, $\fim$, and the spin, $\fis$, multipoles of the gravitational field to the Skrotskii effect. Specifically, we have
\begin{equation}
\label{dfi} \Phi=\fig+\fim+\fis+\Phi_0\;,
\end{equation}
where $\Phi_0$ is a constant angle characterizing the initial orientation of the polarization ellipse of the electromagnetic wave in the reference plane formed by the ${\bm e}_{(1)}$ and ${\bm e}_{(2)}$ vectors.

The gauge-dependent part of the Skrotskii effect \index{Skrotskii effect} is easily integrated so that we obtain
\begin{equation}
\label{sf79}
\fig=\frac{1}2k^j
\epsilon_{jpq}
        \dksi_{q}\left(w^{p}+\chi^{p}\right)\;,
\end{equation}
where the gauge vector functions $w^i=\wm^{i}+\ws^{i}$ are given in equations \eqref{wi}. The gauge functions $\chi^i$ \index{gauge functions} appear in the process of integration of the equation of the parallel transport. They 
can be linearly decomposed in two parts corresponding to the mass and spin multipoles:
\begin{equation}
\label{cccp}
\chi^i=\chim^i+\chis^i\;,
\end{equation}
where
\begin{align}
\label{chim}
  \chim^i =\;
 &4\sum_{l=2}^\infty\sum_{p=1}^{l-1}\sum_{q=1}^p
      \frac{(-1)^{l+p-q}}{l!}
      C_{l-1}(l-p-1,p-q,q)
 \left(1-\frac{p-q}{l-1}
            \right)
\\
 & \phantom{oooooooooo}\times
    k_{<a_1\hdots a_p}\dksi_{a_{p+1}\hdots a_{l-1}>}
     \dtau^{q-1}
      \left[\frac{
      \Ic^{(p-q+1)}_{iA_{l-1}}(t-r)
                }r\right]\;,
\notag    \\
 \label{chis}
  \chis^{i}=\;
 &-4\sum_{l=1}^\infty\sum_{p=0}^{l-1}\sum_{q=0}^p
    \frac{(-1)^{l+p-q}l}{(l+1)!}
 C_{l-1}(l-p-1,p-q,q)
 \left [ 1-
        \frac{p-q}{l-1}\,H(l-1)
    \right ]
\\
 &
 \times
 \left[
 H(q)(\dksi_a-k_a\dtz)+k_a\dtau
 \right]
 k_{<a_1\hdots a_p}\dksi_{a_{p+1}\hdots a_{l-1}>}
 \dtau^{q-1}\left[
 \frac{\epsilon_{iba}\Sc^{(p-q)}_{bA_{l-1}}(t-r)}r\right]
 \notag \\
 &+4\sum_{l=3}^\infty\sum_{p=1}^{l-1}\sum_{q=1}^p
    \frac{(-1)^{l+p-q}l}{(l+1)!}
 C_{l-1}(l-p-1,p-q,q)
 \left ( 1-\frac {p}l\right )
    \left (1-\frac{q}p
    \right )
 \notag  \\
 &\phantom{ooooooooo}\times
    k_{<a_1\hdots a_p}\dksi_{a_{p+1}\hdots a_{l-1}>a}
 \dtau^{q-1}\left[
 \frac{\epsilon_{baa_{l-1}}\Sc^{(p-q)}_{ibA_{l-2}}(t-r)}r\right]\;.
  \notag
\end{align}

The gauge-dependent equation \eqref{sf79} has the following, more explicit form:
 \begin{align}
\label{g54}
 \frac{1}2k^j
\epsilon_{j{m}n
    }
        \dksi_{n}
        \chim^{{m}}
             =
&2\sum_{l=2}^\infty\sum_{p=1}^l\sum_{q=1}^p
     \frac{(-1)^{l+p-q}}{l!}\frac{l-p+q}{(l-1)}\frac{p-q}{p}
     C_l(l-p,p-q,q)
   \\
&\notag \phantom{oooooo}\times
   k_{<a_1\hdots a_p}\dksi_{a_{p+1}\hdots a_l>j}
    \dtau^{q-1}
     \;\left[\frac{
            \epsilon_{j{b}a_l}\Ic^{(p-q)}_{{b}L-1}(t-r)
}r\right]^{[-1]}\;,
\end{align}
\begin{align}
\label{mk23}
 \frac{1}2k^j
\epsilon_{j{m}n
    }
        \dksi_{n}
        \chis^{{m}}
             =
&2\sum_{l=1}^\infty\sum_{p=0}^l\sum_{q=0}^p
    \frac{(-1)^{l+p-q}l}{(l+1)!}
C_l(l-p,p-q,q)
  \\\notag
&\phantom{oo}\times
\left\{
H(q)
\left ( 1-\frac {p}l\right )
    \left [1+H(l-1)
        \left (\frac{l-p-1}{l-1}-\frac{p-q}{l-1}\right )
    \right ]\dtz^2
\right.
\\\notag
&-
\left [
    \frac{l-p}l
        +2\frac{p-q}l
            -H(l-1)\frac{p-q}l
    \left(\frac{l-p}{l-1}
        +2\frac{p-q-1}{l-1}
    \right)
\right ]\dksi_{t^*\tau}\!
\\\notag
&\phantom{ooooooooooo}+
\left.
    \frac{p-q}l
        \left [1-H(l-1)\frac{p-q-1}{l-1}
        \right ]\dtau^2
\right\}
\\\notag
&\phantom{ooooooooooo}
\times
    k_{<a_1\hdots a_p}\dksi_{a_{p+1}\hdots a_l>}
 \dtau^{q-1}\;\left[
\frac{\Sc^{(p-q-1)}_L(t-r)}r\right]^{[-1]}\;,
\end{align}
\begin{align}
\label{qxr43}
 \frac{1}2k^j
\epsilon_{j{m}n
    }
        \dksi_{n}
        \wm^{{m}}
 =
&-2\sum_{l=2}^\infty\sum_{p=0}^l\sum_{q=0}^p
     \frac{(-1)^{l+p-q}}{l!}\frac{p-q}p
     C_l(l-p,p-q,q)
  \\\notag
& \phantom{oooooo}\times
   k_{<a_1\hdots a_p}\dksi_{a_{p+1}\hdots a_l>j}
    \dtau^{q}
     \left[\frac{
            \epsilon_{j{b}a_l}
\Ic^{(p-q-1)}_{{b}L-1}(t-r)
}r\right]\;,
\end{align}
\begin{align}
\label{wer7}
 \frac{1}2k^j
\epsilon_{j{m}n
      }
        \dksi_{n}
        \ws^{{m}}
             =
&-2\sum_{l=1}^\infty\sum_{p=0}^l\sum_{q=0}^p
    \frac{(-1)^{l+p-q}l}{(l+1)!}
C_l(l-p,p-q,q)
 \\\notag
&\times
\left\{
\left ( 1-\frac {p}l
\right )
    \dtz^2-
\left (
    \frac{l-p}l
        +2\frac{p-q}l
\right )
\dksi_{t^*\tau}\!+
    \frac{p-q}l
    \dtau^2
\right\}
\\\notag
&\phantom{oooooooooo}
\times
    k_{<a_1\hdots a_p}\dksi_{a_{p+1}\hdots a_l>}
 \dtau^{q}\left[
\frac{\Sc^{(p-q-2)}_L(t-r)}r\right]\;.
\end{align}
The reader can notice the presence of integrals in the right side of equations \eqref{g54} and \eqref{mk23}. The integrals are actually not supposed to be calculated explicitly since there is sufficient number of partial derivatives in front of them which cancels the integration in correspondence with the rules of differentiation of such integrals which have been explained in section \ref{int}.

The Skrotskii effect \index{Skrotskii effect} due to the mass-type multipoles of the isolated system is given by
\begin{equation}
\label{fim}
 \fim(\tau)=
2\sum_{l=2}^\infty\sum_{p=0}^l
     \frac{(-1)^{l+p}}{l!}
     C_l(l-p,p)
\frac{l-p}{l-1}
   k_{<a_1\hdots a_p}\dksi_{a_{p+1}\hdots a_l>j}
         \left[\frac{
            \epsilon_{j{b}a_l}\Ic^{(p)}_{{b}A_{l-1}}(t-r)}r
       \right]^{[-1]}\;.
\end{equation}
The gravitational field of the spin-type multipoles rotates the polarization plane of the electromagnetic wave at the following angle
\begin{eqnarray}
\label{dfis}
\fis(\tau) &=&
2\sum_{l=1}^\infty\sum_{p=0}^l
    \frac{(-1)^{l+p}l}{(l+1)!}
C_l(l-p,p) \left ( 1-\frac {p}l\right )\times\\\notag
    &&\left [1+H(l-1)
        \left (1-\frac{2p}{l-1}\right )
    \right ]
    k_{<a_1\hdots a_p}\dksi_{a_{p+1}\hdots a_l>}
         \left[
\frac{\Sc^{(p+1)}_{A_{l}}(t-r)}r\right]^{[-1]}
\end{eqnarray}
Integrals in equations \eqref{fim} and \eqref{dfis} are eliminated after taking at least one partial derivative so that we do not need to integrate.


\section{Light Propagation through the Field of Gravitational Lens}\label{byq6f}
\begin{figure}
\includegraphics[width=\textwidth]{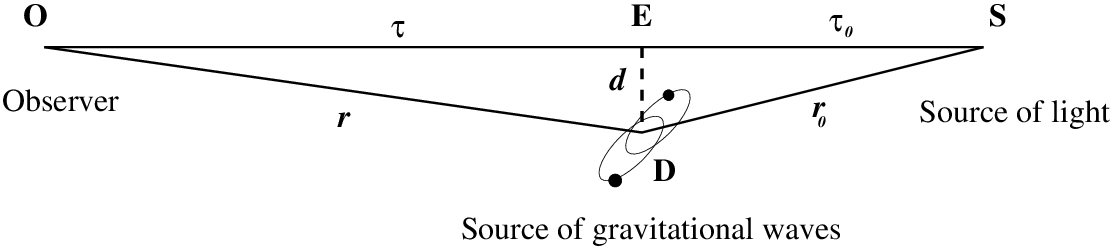}
\caption[Gravitational Lens Approximation]{Relative configuration of an observer (O), a source of light (S),
and a localized system emitting gravitational waves (D). Static part of gravitational field of the localized system and
gravitational waves deflect light
rays which are
emitted at the moment $t_0$ at the point S and received at the moment $t$ at
the point O. The point E on the
line OS corresponds to the moment $t^*$ of the closest approach of a light ray to the
deflector D. We denoted the distances as follows: $OS=R$, $DO=r$, $DS=r_0$, the impact parameter $DE=d$,
$OE=\tau>0$, $ES=\tau_0=\tau-R<0$.
The impact parameter $d$ is much smaller as compared with all other distances.}  
\label{smallimp1}
\end{figure}

\index{gravitational lens}
This section considers propagation of light in a special case of gravitational lens approximation when
the impact parameter $d$ of the light ray with respect to an isolated system, is much smaller than both the distance  $r_0$ from
the isolated system to the source of light and distance $r$ from the
isolated system to observer, as illustrated in Fig. \ref{smallimp1}.

\subsection{Small parameters and asymptotic expansions}

In the case of a small impact parameter of the light ray with respect to the isolated system the near-zone gravitational field of the system strongly affects propagation of the light ray only when the light particle (photon) moves in the close proximity to the system where the effects on the light ray propagation caused by  gravitational waves are suppressed \cite{ksh1,23}. In what follows, we assume that the impact parameter $d$ of the light ray is small as compared with both distances $r$ and $r_0$, that is, $d \ll min[r,r_0]$ (see Fig. \ref{smallimp1}). This assumption allows us to introduce two small
parameters: $\varepsilon \equiv d/r$ and $\varepsilon_0\equiv
d/r_0$. If the light ray is propagated through the near-zone of the isolated system one more small parameter can be introduced, $\varepsilon_{\lambda}\equiv d/\lambda$, where $\lambda$ is the characteristic
wavelength of the gravitational radiation emitted by the system proportional to the product of the speed of gravity ($=c$ in general relativity) and the characteristic period of matter oscillations in the isolated system. Parameters $\varepsilon$, $\varepsilon_0$ and $\varepsilon_{\lambda}$ are not physically correlated. Parameter $\varepsilon_{\lambda}$ is used for the post-Newtonian expansion of the equations describing the observed effects -- deflection of light, time delay, etc. This expansion can be also viewed as a Taylor expansion with respect to the parameter $v/c$, where $v$ is the characteristic speed of motion of matter composing the isolated system and $c$ is the speed of propagation of gravity. If the light ray does not enter the near zone of the system the parameter $\varepsilon_{\lambda}$ is not small and the post-Newtonian expansion can not be performed.

The small-impact-parameter expansions for the retarded time variables $y$ and $y_0$ yield:
\begin{eqnarray}\label{be1}
  y&=&\sqrt{r^2-d^2}-r=-d\left( \frac{\varepsilon}2-\sum_{k=2}^{\infty}\mathcal {C}_k\varepsilon^{2k-1} \right)\;,\\\label{be2}
  y_0&=&-\sqrt{r_0^2-d^2}-r_0=-2r_0+d\left (\frac{\varepsilon_0}2-\sum_{k=2}^{\infty}\mathcal{C}_k\varepsilon_0^{2k-1}\right)\;,\end{eqnarray}
and
\begin{eqnarray}
\label{be3}
  \frac{1}{yr}&=&-\frac{1}{d^2}\left(2+\sum_{k=1}^\infty\mathcal{C}_k\varepsilon^{2k}\right)\;,\\\label{be4}
\frac{1}{y_0r_0}&=&\frac
1{d^2}\sum_{k=1}^{\infty}\mathcal{C}_k\varepsilon_0^{2k}\;,
\end{eqnarray}
where the numerical coefficients entering the expansions are
\begin{equation}\label{be5}
  \mathcal{C}_k=\frac{(-1)^k}{k!}\frac{1}2\left(\frac{1}2-1\right)\cdot\hdots\cdot\left(\frac{1}2-k+1\right)=\frac{(2k-1)!}{(2k)!!}\;.
\end{equation}
Retarded time variables, $s=t-r$ and $s_0=t_0-r_0$, are expanded as follows
\begin{eqnarray}
\label{tl}
s&=&t^*
-d\left(\frac{\varepsilon}2
-\sum_{k=2}^{\infty}\mathcal
{C}_k\varepsilon^{2k-1} \right)\;,
\\
\label{t0l} s_0&=&t^*-2r_0 +d\left(\frac{\varepsilon_0}2
-\sum_{k=2}^{\infty}\mathcal{C}_k\varepsilon_0^{2k-1}\right)\;,
\end{eqnarray}
where $\mathcal{C}_k$ is given by equation \eqref{be5} and $t^*$ is the time of the closest approach of the light ray to the barycenter \index{barycenter} of the isolated system.

Using \eqref{tl} and \eqref{t0l} we can write down the post-Newtonian expansions
for functions of the retarded time $t-r$ as follows
\begin{eqnarray}
\label{Fl}
F(t-r)&=&\sum^\infty_{k=0}\frac{(-1)^kd^k}{k!}\left( \frac{\varepsilon}2-\sum_{k=2}^{\infty}\mathcal {C}_k\varepsilon^{2k-1} \right)^kF^{(k)}(t^*)\\\notag&=&
F(t^*)-\varepsilon\frac{d}2\dot
F(t^*)+O\left(\varepsilon^2\varepsilon^2_\lambda\right)\;,
\\
\label{F0l}
F(t_0-r_0)&=&\sum^\infty_{k=0}\frac{d^k}{k!}\left( \frac{\varepsilon_0}2-\sum_{k=2}^{\infty}\mathcal {C}_k\varepsilon_0^{2k-1} \right)^kF^{(k)}(t^*-2r_0)\\\notag&=&
F(t^*-2r_0)+\varepsilon_0\frac{d}2\dot
F(t^*-2r_0)+O\left(\varepsilon_0^2\varepsilon^2_\lambda\right)\;,
\end{eqnarray}
where the dot above functions denote the total time derivative.
We notice that convergence
of the post-Newtonian
time series for light propagation depends, in fact, not just on a single parameter, $\varepsilon_\lambda\sim v/c$, that is typical in the post-Newtonian celestial mechanics of extended bodies \citep{60,31,kopeikin_book}, but on the product of two parameters. Therefore, the convergence requires satisfaction of two conditions
\begin{eqnarray}
\label{be9}
\varepsilon\varepsilon_\lambda&\ll& 1\;,\\\label{be10}
\varepsilon_0\varepsilon_\lambda&\ll& 1\;,
\end{eqnarray}
where the numerical value of the parameter $\varepsilon_\lambda$ must be taken for the smallest wavelength, $\lambda_{\rm min}$, in the spectrum of the gravitational radiation emitted by the isolated system. Conditions \eqref{be9}, \eqref{be10} ensure convergence of the post-Newtonian series in \eqref{Fl} and \eqref{F0l} respectively.
If the source of light rays and observer are at
infinite distances from the isolated system then $\varepsilon\simeq 0$ and $\varepsilon_0\simeq 0$, and the requirements (\ref{be9}), \eqref{be10} are
satisfied automatically, irrespective of the structure of the
Fourier spectrum \eqref{for5} of the gravitational radiation emitted by the isolated system. In
a real astronomical practice such an assumption may not be always satisfied. In such cases it is more natural to avoid the post-Newtonian expansions of the metric tensor and/or observable effects and operate with the functions of the retarded time. It is important to notice that the retarded time $s=t-r$ which enters the result of the calculation of the is a characteristic of the Einstein equations of gravitational field, not the Maxwell equations. Therefore, measuring the effect of the gravitational deflection of light caused by time-dependent gravitational field allows us to measure the speed of gravity with respect to the speed of light. This type of experiments have been proposed in our paper \citep{kopapjl} and successfully performed in 2003 \citep{fkapj}. There were several publications (see review \citep{will-lrv}) arguing that the speed of gravity is irrelevant in the light-ray deflection experiments by moving bodies. Unfortunately, all the authors of those publications operated with the post-Newtonian expansion of the gravitational field which replaces the retarded time $s=t-r$ of the gravitational field with the time $t^*$ of the closest approach of light to the light-ray deflecting body like in equation (\ref{tl}). This explains why the effect of the retardation of gravity was confused with the retardation of light in \citep{will-lrv}.

If we assume that the mass and angular momentum of the isolated system are conserved the asymptotic expansions of integrals \eqref{is1}, \eqref{is2} of the stationary part of
the metric tensor have the following form
\begin{eqnarray}\label{mnb1}
\left[\frac{1}r\right]^{[-1]}
 &=&-\ln\left(\frac{r}{2r_{\scriptscriptstyle\rm E}}\right)-2\ln\varepsilon-\sum^\infty_{k=1}\frac{(2k-1)!}{2^{2k}(k!)^2}\varepsilon^{2k}\\\notag&=&
-2\ln d+\ln r+\ln(2r_{\scriptscriptstyle\rm E})
 +O(\varepsilon^2),
\\
\label{mnb2}
\left[\frac{1}{r_0}\right]^{[-1]}
&=&-\ln\left(\frac{2r_0}{r_{\scriptscriptstyle\rm E}}\right)-\sum^\infty_{k=1}\frac{(2k-1)!}{2^{2k}(k!)^2}\varepsilon_0^{2k}= -\ln\left(\frac{2r_0}{r_{\scriptscriptstyle\rm E}}\right)  +O(\varepsilon_0^2),
\\
\label{mnb3}
\left[\frac{1}r\right]^{[-2]}&=&-r\left\{1+\left(1+
 \sum_{k=1}^{\infty}\mathcal {C}_k\varepsilon^{2k}\right)\left[\ln\left(\frac{\varepsilon^2r}{2r_{\scriptscriptstyle\rm E}}\right)+\sum^\infty_{k=1}\frac{(2k-1)!}{2^{2k}(k!)^2}\varepsilon^{2k}\right]\right\}\\\nonumber
 &=&
-r-2r\ln d+r\ln\left(2rr_{\scriptscriptstyle\rm E} \right)- \varepsilon\frac{d}2\left[\frac{1}2-\ln\left(
\frac{d^2}{2rr_{\scriptscriptstyle\rm E}}\right) \right] +O(\varepsilon^2),
\\
\left[\frac{1}{r_0}\right]^{[-2]}
&=&-r_0\left\{1-\left(1+
 \sum_{k=1}^{\infty}\mathcal {C}_k\varepsilon_0^{2k}\right)\left[\ln\left(\frac{2r_0}{r_{\scriptscriptstyle\rm E}}\right)+\sum^\infty_{k=1}\frac{(2k-1)!}{2^{2k}(k!)^2}\varepsilon_0^{2k}\right]\right\}\\\nonumber
&=&-r_0+r_0\ln\left(\frac{2r_0}{r_{\scriptscriptstyle\rm E}}\right) -\varepsilon_0\frac{d}2 \left[\frac{1}2+ \ln\left(\frac{2r_0}{r_{\scriptscriptstyle\rm E}}\right)
        \right]
+O(\varepsilon_0^2).
\end{eqnarray}
Next several equations yield the estimates for the partial derivatives with respect to the parameters $\xi^i$ and
 $\tau$ from functions of the retarded time (of gravity). In these estimates the numbers $m$
 and $n$ depend on $l$; we give the estimates from below for $m$ and $n$ for $l \geq 1$.
\begin{align}
\label{el}
\dksi_{<a_1\ldots a_l>}
                    \frac{F(t-r)}r= &
                      O\left(\varepsilon^2\frac{F}{d^{l+1}}\right)\;,
\\
\label{el+}\dksi_{<a_1\ldots a_l>}
                    \frac{F(t_0-r_0)}{r_0}
                      = &
                      O\left(\varepsilon^2\frac{F}{d^{l+1}}\right)\;,\\
\dtau^l\left[\frac{F(t-r)}r
                      \right]= &
                      O\left(\varepsilon^2\frac{F}{d^{l+1}}\right)\;,\\
\dtau^l\left[\frac{F(t_0-r_0)}{r_0}
                      \right]= &
                      O\left(\varepsilon\frac{F}{d^{l+1}}\right)\;,\\
\dksi_{<a_1\ldots a_l>}
                         \left[\frac{F(t_0-r_0)}{r_0}
                             \right]^{[-1]}
                        = &O\left(\varepsilon^2\frac{F}{d^l}\right)\;,\\
\label{elend}\dksi_{<a_1\ldots a_l>}\left[\frac{F(t_0-r_0)}{r_0}
                             \right]^{[-2]}
                         =
                         & O\left(\varepsilon^3\frac{F}{d^{l-1}}\right)\;.
\end{align}
Two asymptotic expansions will be also useful.
\begin{align}\label{d4e5r}
\dksi_{<a_1\ldots a_l>}
                           \left[\frac{F(t-r)}r
                             \right]^{[-1]}
                       = &-2F(t-r)\dksi_{<a_1\ldots a_l>}\ln d+
                       O\left(\varepsilon\varepsilon_\lambda\frac{F}{d^l}\right)\;,\\\label{p3c1za}
\dksi_{<a_1\ldots a_l>}\left[\frac{F(t-r)}r
                             \right]^{[-2]}=
                         &
                         -2rF(t-r)\dksi_{<a_1\ldots a_l>}\ln d+
                         O\left(\varepsilon_\lambda\frac{F}{d^{l-1}}\right)\;,
\end{align}
These estimates and asymptotic expansions will be used for obtaining the observable relativistic effects in the gravitational
lens approximation.

\subsection {Asymptotic expressions for observable effects}
This section provides the reader with the asymptotic expressions for potentially observable relativistic effects by taking into account only the leading terms and neglecting all terms which are proportional to the parameter $\varepsilon$. 

The relativistic time delay is given by
\begin{align}
\label{tdl}
\Delt=
    &\Delm +\Dels,\\
    \intertext{where}
\Delm=
    &-4\Mc\ln d+2\Mc\ln(4rr_0)-
\label{tdml}
    \\
    &4\sum_{l=2}^{\infty}\sum_{p=0}^{l-2}
    \frac{(-1)^{l+p}}{l!} C_l(l-p,p)
    \left(1-\frac{p}l\right) \left(1-\frac{p}{l-1}\right)
    \times                    \notag \\
    &
    \dtz^p\Ic_{A_l}(t-r)
                  k_{<a_1\ldots a_p}\dksi_{a_{p+1}\ldots a_l>}
                     \ln d\;,          \notag
\end{align}
\begin{align}
\label{tdsl} \Dels=&-4\epsilon_{iba}k_i\Sc_b\dksi_a\ln d +
\\
      &8\epsilon_{iba}k_i\dksi_a\sum_{l=2}^{\infty}\sum_{p=0}^{l-1}
                     \frac{(-1)^{l+p}l}{(l+1)!}C_{l-1}(l-p-1,p)
                     \left(1-\frac{p}{l-1}\right)
                     \times
\notag
\\
      &
                        \dtz^p\Sc_{bA_{l-1}}(t-r)
                          k_{<a_1\ldots a_p}\dksi_{a_{p+1}\ldots a_{l-1}>}
                  \ln d \;. \notag
\end{align}

The observable unit vector in the direction from the observer to the source of
light is given by the following expression
\begin{align}
\label{sl}
s^i\left(\tau,{\bm \xi}\right)
            =&K^i+
            \alpha^i\left(\tau,{\bm \xi}\right)+
            \beta^i\left(\tau,{\bm \xi}\right),
\end{align}
where we have dropped off the quantities
$\beta^i\left(\tau_0,{\bm\xi}\right)$ and
$\gamma^i\left(\tau,{\bm\xi}\right)$ as being negligibly small. Vector
$\alpha^i\left(\tau,{\bm \xi}\right)$, characterizing the
deflection of light, is given by expression
\begin{align}\label{m8n3d}
\alpha^i\left(\tau,{\bm \xi}\right)=
&\alpha^i_{{\scriptscriptstyle (M)}}\left(\tau,{\bm \xi}\right)+
 \alpha^i_{{\scriptscriptstyle (S)}}\left(\tau,{\bm \xi}\right)\;,\\
 \intertext{where}
\label{aml} \alpha^i_{{\scriptscriptstyle (M)}}\left(\tau,{\bm
\xi}\right)=
&4\Mc\dksi_i\ln d+\\
&4\dksi_i\sum_{l=2}^{\infty}\sum_{p=0}^{l-2}
    \frac{(-1)^{l+p}}{l!} C_l(l-p,p)
    \left(1-\frac{p}l\right) \left(1-\frac{p}{l-1}\right)
    \times                    \notag \\
&
    \dtz^p\Ic_{A_l}(t-r)
                  k_{<a_1\ldots a_p}\dksi_{a_{p+1}\ldots a_l>}
                     \ln d\;,            \notag  \\
\label{asl} \alpha^i_{{\scriptscriptstyle (S)}}\left(\tau,{\bm
\xi}\right)=
 &4\epsilon_{jba}k_j\Sc_b\dksi_{ia}\ln d -
\\
      &8\epsilon_{iba}k_i\dksi_{ia}\sum_{l=2}^{\infty}\sum_{p=0}^{l-1}
                     \frac{(-1)^{l+p}l}{(l+1)!}C_{l-1}(l-p-1,p)
                     \left(1-\frac{p}{l-1}\right)
                     \times
\notag
\\
      &
                        \dtz^p\Sc_{bA_{l-1}}(t-r)
                          k_{<a_1\ldots a_p}\dksi_{a_{p+1}\ldots a_{l-1}>}
                  \ln d \;. \notag
\end{align}
The corresponding relativistic correction to the light-ray deflection is
\begin{align}
\label{bl} \beta^i\left(\tau,{\bm \xi}\right) =&
\beta^i_{{\scriptscriptstyle (M)}}\left(\tau,{\bm\xi}\right)+
\beta^i_{{\scriptscriptstyle (S)}}\left(\tau,{\bm\xi}\right)\;,
\\
\intertext{where}
\label{bml}
\beta^i_{{\scriptscriptstyle
(M)}}\left(\tau,{\bm\xi}\right)
=&-\frac{r}R\alpha^i_{{\scriptscriptstyle
(M)}}\;,
\\
\label{bsl}
\beta^i_{{\scriptscriptstyle
(S)}}\left(\tau,{\bm\xi}\right)
=&-\frac{r}R\alpha^i_{{\scriptscriptstyle
(S)}}\;.
\end{align}
We can use \eqref{Fl} and \eqref{F0l} to rewrite functions of the retarded time $s=t-r$ in \eqref{tdml}, \eqref{tdsl}, \eqref{aml} and
\eqref{asl} as functions taken at the moment of the closest approach $t^*$ of 
photon to the gravitating system. This is accomplished
by formal replacing $t-r\rightarrow t^*$ because the corrections will be of the higher order
with respect to the parameter $\varepsilon$ (see Eq. \eqref{Fl}). It should not confuse the reader about the physical meaning of the retardation, which is due to the finite speed of propagation of gravity.

The time delay (\ref{tdl}) and the light-ray deflection (\ref{m8n3d}) can be written in a very short and concise form by making use of the gravitational lens potential, $\psi$, which is just the eikonal perturbation. More specifically,
\begin{align}
\Delt=&-4\psi+2\Mc \ln(4rr_0),\\
 \alpha^i\left(\tau,{\bm \xi}\right)=&4\dksi_i\psi,
\end{align}
where 
\be\label{d6h1}
\psi=\psi_{{\scriptscriptstyle(M)}}+\psi_{{\scriptscriptstyle(S)}}\;,
\ee
and
\begin{eqnarray}
\label{psikk}
\psi_{{\scriptscriptstyle(M)}}&=&\Mc\ln d+\\\notag
    &&\sum_{l=2}^{\infty}\sum_{p=0}^{l-2}
    \frac{(-1)^{l+p}}{l!} C_l(l-p,p)
    \left(1-\frac{p}l\right) \left(1-\frac{p}{l-1}\right)\times\\\notag&&
    \dtz^p\Ic_{A_l}(t^*)
                  k_{<a_1\ldots a_p}\dksi_{a_{p+1}\ldots a_l>}\ln d\;,\\\notag\\\label{psikk1}
\psi_{{\scriptscriptstyle(S)}}&=& \epsilon_{jba}k_j\Sc_b\dksi_a\ln d  -\\\notag&&
      2\epsilon_{jba}k_j\dksi_a\sum_{l=2}^{\infty}\sum_{p=0}^{l-1}
                     \frac{(-1)^{l+p}l}{(l+1)!}C_{l-1}(l-p-1,p)
                     \left(1-\frac{p}{l-1}\right)\times\\\notag&&
                        \dtz^p\Sc_{bA_{l-1}}(t^*)
                          k_{<a_1\ldots a_p}\dksi_{a_{p+1}\ldots a_{l-1}>}\ln d\;.
\end{eqnarray}
This expression takes into account the multipole moments of the isolated gravitating system of all orders
and presents a generalization of our previous results obtained in~\cite{smk1} (stationary gravitational field)
and~\cite{ksh1} (the quadrupolar gravitational field).

The angle of rotation of the polarization
plane of electromagnetic wave is simplified if one uses the approximation of gravitational lens. More specifically, the total angle of the rotation is given by 
\be\label{b6gu8}
\Delta\Phi=\Delta\fim+\Delta\fis\;,
\ee
where 
\begin{eqnarray} \label{fiml}
\Delta\fim &=& -4\sum_{l=2}^\infty\sum_{p=0}^l
     \frac{(-1)^{l+p}}{l!}
     C_l(l-p,p)
\frac{l-p}{l-1}\times\\\notag&&
\dtz^{p}
     \epsilon_{j{b}a_l}\Ic_{{b}A_{l-1}}(t^*)
   k_{<a_1\hdots a_p}\dksi_{a_{p+1}\hdots a_l>j}
    \ln d\;),
\\\notag\\
\label{fisl}
\Delta \fis&=&
-4\sum_{l=1}^\infty\sum_{p=0}^l
    \frac{(-1)^{l+p}l}{(l+1)!}
C_l(l-p,p) \left ( 1-\frac {p}l\right ) 
    \left [1+H(l-1)
        \left (1-\frac{2p}{l-1}\right )
    \right ]\times
\notag \\
&& \dtz^{p+1} \Sc_{A_{l}}(t^*)
    k_{<a_1\hdots a_p}\dksi_{a_{p+1}\hdots a_l>}\ln d\;.
\end{eqnarray}

\section{Light Propagation through the Field of Plane Gravitational Waves}\label{dg8j4}\index{plane gravitational wave}
\subsection{Plane-wave asymptotic expansions}
\begin{figure}
\includegraphics[width=\textwidth]{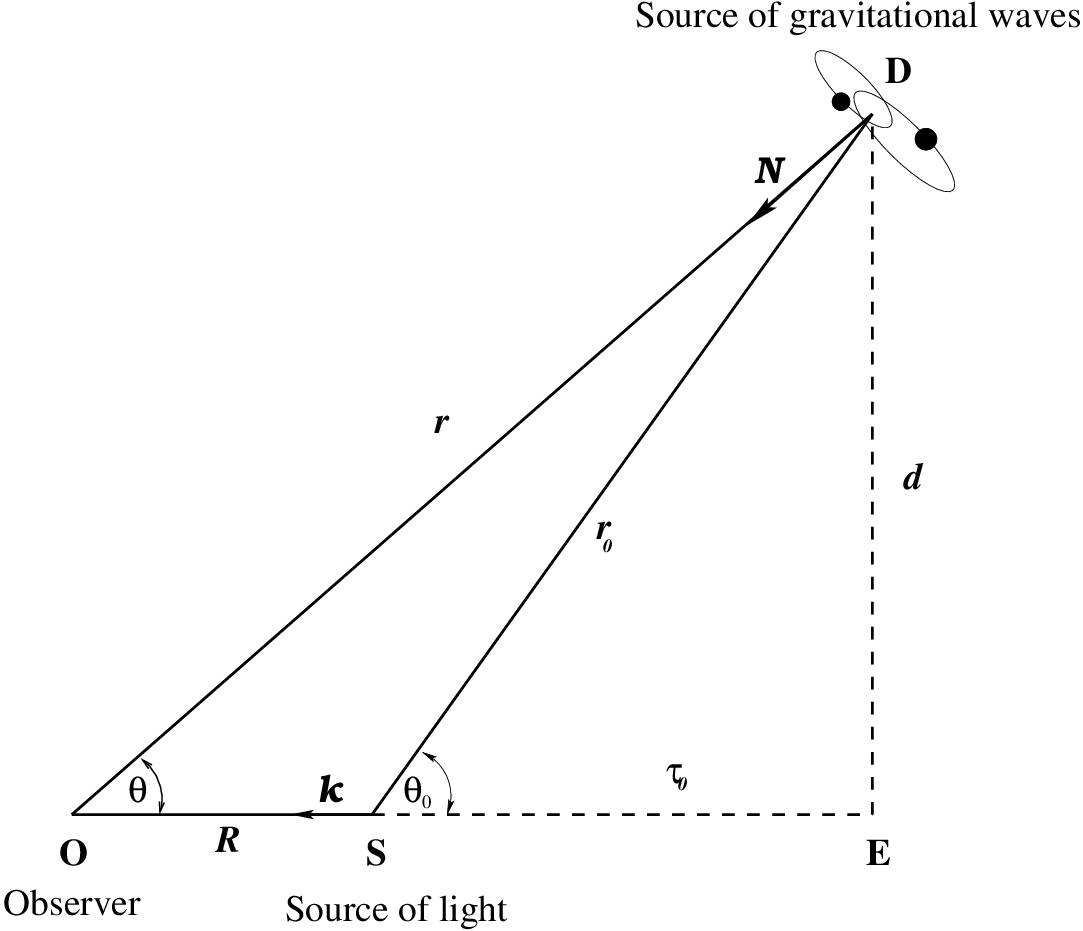}
\caption[Plane-Wave Approximation]{Relative configuration of observer (O), a source of light (S),
and a localized system emitting of gravitational waves (D).
The gravitational waves deflect light
rays which are
emitted at the moment $t_0$ at the point S and received at the moment $t$ at
the point O. The point E on the
line OS corresponds to the moment of the closest approach of light ray to the
system D. Notations for distances are $OS=R$,  $DO=r$, $DS=r_0$, $DE=d$ -  the impact parameter,
$OE=\tau=r\cos\theta$, $ES=\tau_0=\tau-R$.
The distance $R$ is much smaller than both $r$ and $r_0$.
There is no limitation on the impact parameter $d$ which can be or may be not small as
compared to all other distances.} 
\label{largeimp}
\end{figure}

We consider the source of light and observer located respectively at radial distances $r_0=|{\bm r}_0|$ and $r=|{\bm r}|$ from the isolated system emitting gravitational waves. The distance between the source of light and observer is $R=|{\bm r}-{\bm r}_0|$.
In the plane gravitational-wave approximation we assume that 
\begin{itemize}
\item[(a)] the characteristic wavelength of gravitational waves, $\lambda\ll {\rm min}[r,r_0]\;,$ 
\item[(b)] the distance between the source of light and observer, $R~\ll~{\rm min}[r,r_0]\;.$
\end{itemize}
Condition (a) tells us that both the source of light and observer are lying in the wave zone of the isolated system. Condition (b) allows us to consider the gravitational waves emitted by the isolated system as plane waves when they propagate from the source of light to observer. This approximation is vizualized in Fig. \ref{largeimp}.

We introduce small parameters $\delta_\lambda={\rm max}[\lambda/r,\l/r_0]$, $\delta=R/r$ and $\delta_0=R/r_0$
The coordinate relation between $r$, $r_0$ and $R=|{\bm r}-{\bm r}_0|$ can be written as follows
\begin{equation}
\label{ew}
r_0^2=r^2-2rR\cos\theta+R^2=r^2(1-2\delta\cos\theta+\delta^2)\;,
\end{equation}
where $\theta$ -- is the angle between the directions ''observer -- the source of light'' and ''observer -- the gravitating system'' (see Fig. \ref{largeimp}).
From (\ref{ew}) it follows that
\begin{eqnarray}
r_0&=& r(1-\delta\cos\theta)+O(\delta^2)\;,\\
\frac{1}{r_0}&=&\frac{1}{r}(1+\delta\cos\theta)+O(\delta^2)\;.
\end{eqnarray}
For the variables $\tau$ and $\tau_0$ we have exact relations,
\begin{equation}
\label{1q3z}
\tau=r\cos \theta,\qquad \tau_0=\tau-R=r\cos\theta-R.
\end{equation}
Quantities $d$ and $y$ are given in terms of the distance $r$ and the angle $\t$ as follows
\begin{eqnarray}
d&=&r\sin\theta\;,\\
 y&=&\tau-r=-r(1-\cos\theta)\;.
\end{eqnarray}
The retarded instants of time $s=t-r$ and $s_0=t_0-r_0$ are related to each
other as follows
\begin{equation}
\label{t0w}
t_0-r_0=t-r-R(1-\cos\theta)\;.
\end{equation}
Using this expression we can write the Taylor expansion for the
functions of the retarded time $s=t-r$
\begin{equation}
\label{Fw}
F (t_0-r_0)=F(t-r)-R(1-\cos\theta)\dot
F(t-r)+O\left(R^2/\lambda^2\right)\;.
\end{equation}
The impact parameter vector, $\xi^i$, can be decomposed as follows
\begin{equation}
\label{xiw}
  \xi^i=r\left(N^i-k^i\cos\theta\right)\;.
\end{equation}
This form of the impact parameter vector $\xi^i$ is useful in subsequent approximations.

Let us now write out the asymptotic  expressions for the
derivatives with respect to $\xi^i$ and $\tau$ of functions depending on the retarded time $s=t-r$. We have
\begin{eqnarray}\label{g7q8}
\dtau^n\left[\frac{F(t-r)}{r}\right]
       &=&
       (1-\cos\theta)^n\dtz^n\left[\frac{F(t-r)}{r}
\right]+O(\delta^2)\;,
\\
\dksi_{a_1\ldots a_n}\left[\frac{F(t-r)}{r}\right]
       &=&
(-1)^n\frac{\xi_{a_1}\ldots\xi_{a_n}}{r^n}\dtz^n\left[\frac{F(t-r)}{r}
\right] +O(\delta^2)\;,
\\
\dtau^n\left[\frac{F(t-r)}r
                             \right]^{[-1]}
       &=&
       (1-\cos\theta)^{n-1}\dtz^{n-1}\left[\frac{F(t-r)}{r}\right]
       +O(\delta^2)\;,
\\
\dksi_{a_1\ldots a_n}\left[\frac{F(t-r)}r
                             \right]^{[-1]}
       &=&
 \frac{(-1)^n}{1-\cos\theta}
       \frac{\xi_{a_1}\ldots\xi_{a_n}}{r^n}\dtz^{n-1}\left[\frac{F(t-r)}{r}\right]
       +O(\delta^2)\;.
\end{eqnarray}
One more asymptotic expression is given for the integral taken from an STF partial derivative 
\ba
\label{exw1}
\left\{\left[\frac{F(t-r)}r\right]_{,<A_l>}\right\}^{[-1]}&=&
\sum_{p=0}^l\sum_{q=0}^p
     (-1)^{p-q}
     C_l(l-p,p-q,q)\times\\\notag&&
   k_{<a_1\hdots a_p}\dksi_{a_{p+1}\hdots a_l>}
    \dtz^{p-q}\dtau^q\left[\frac{F(t-r)}r
                             \right]^{[-1]}\;.
                             \ea
Taking into account only the leading order terms in (\ref{exw}), we obtain the asymptotic expansion                            
\ba\label{exw}
\left\{\left[\frac{F(t-r)}r\right]_{,<A_l>}\right\}^{[-1]}&=&
\frac{1}{1-\cos\theta} \left[ \frac{\vphantom{F}^{(-1)}F(t-r)}r
\right]_{,<A_l>}-\\\notag&& \frac{(-1)^l}{1-\cos\theta}k_{<A_l>}\dtz^{l-1}
                           \frac{F(t-r)}r+
\\\notag
&&(-1)^lk_{<A_l>}\dtz^l
                             \left[\frac{F(t-r)}r
                             \right]^{[-1]}+O(\delta_\lambda)\;,
\ea
where $N^i=x^i/r$. Equations \eqref{g7q8}--\eqref{exw} can be checked by induction.

Corresponding expressions for the functions taken at the retarded instant of time
$s_0=t_0-r_0$ can be obtained from \eqref{g7q8}--\eqref{exw} by replacing the arguments $t$, $r$ and $\theta$ with $t_0$, $r_0$ and
$\theta_0$ respectively.

\subsection{Asymptotic expressions for observable effects}

In this section we give expressions for the relativistic effects of the time
delay, bending of light and the Skrotskii effect in the gravitational plane-wave approximation. In this approximation we neglect all terms of the order of $\delta^2$,
$\delta_0^2$ and $\delta_\lambda^2$, and higher. For the time delay we have
\begin{equation}
\label{tdw} \Del=\Delm+\Dels,
\end{equation}
 where
\ba
\label{tdmw}
\Delm&=&2\Mc\frac{R}r+\\\notag&&
\frac{2k_ik_j}{1-\cos\theta}
      \sum_{l=2}^{\infty}\frac{(-1)^l}{l!}
\left\{\left[\frac{\dot {\Ic}_{ijA_{l-2}}(t-r)}{r}
           \right]_{,A_{l-2}}^{TT}-
\left[\frac{\dot{\Ic}_{ijA_{l-2}}(t_0-r_0)}{r_0}
           \right]_{,A_{l-2}}^{TT}
           \right\},
\\\notag\\
\label{tdsw}
 \Dels&=&-2\frac{ \epsilon_{iba}k_i\xi_a\Sc_b}{1-\cos\theta}
   \left(\frac{1}{r}-
          \frac{1}{r_0}
          \right)-\frac{ 4k_ik_j}{1-\cos\theta}\times\\\notag&&\sum_{l=2}^{\infty}\frac{(-1)^ll}{(l+1)!}
\left\{\left[\frac{\epsilon_{ba(i}\dot
          {\Sc}_{j)bA_{l-2}}(t-r)}{r}\right]_{,aA_{l-2}}^{TT}-
\left[\frac{\epsilon_{ba(i}\dot
          {\Sc}_{j)bA_{l-2}}(t_0-r_0)}{r_0}\right]_{,aA_{l-2}}^{TT}
          \right\}\;, \notag
\ea
Here the transverse - traceless (TT) part of the tensors depending on the multipole
moments of the isolated astronomical system is taken with respect to the direction $N^i$. Taking into
account expression \eqref{em} for the components of the metric
tensor $h_{ij}$ we can re-write expressions \eqref{tdw} --
\eqref{tdsw} for the time delay as follows
\ba
\label{tdw+} \Del&=&2\Mc \frac{R}r -2\frac{
\epsilon_{iba}k_i\xi_a\Sc_b}{1-\cos\theta}
   \left(\frac{1}{r}-
          \frac{1}{r_0}
          \right)-\\\notag&&\frac{k_ik_j}{2(1-\cos\theta)}
          \left[\int\limits_{-\infty}^t h_{ij}^{TT}(\t,{\bm x})d\t
          -\int\limits_{-\infty}^{t_0}h_{ij}^{TT}(\t,{\bm x}_0)d\t
          \right]\;.
\ea
The observed astrometric direction from observer to the source of light in the plane gravitational-wave approximation is
\be
\label{sw}
 s^i\left(\tau,{\bm \xi}\right)=K^i+
            \alpha^i\left(\tau,{\bm \xi}\right)+
            \gamma^i\left(\tau,{\bm \xi}\right),
\ee
where
\ba
\label{aw+}
 \alpha^i\left(\tau,{\bm \xi}\right)&=&
 \frac{1}2\frac{k_pk_q}{1-\cos\theta}\left[(\cos\theta-2)k^i+N^i\right]h_{pq}^{TT}(t,{\bm x})+
 k^qh_{ip}^{TT}(t,{\bm x}),
\\\notag\\\notag
\gamma^i\left(\tau,{\bm \xi}\right)&= &
   -\ffrac{1}2P^{ij}k^qh_{jq}^{TT}(t, {\bm x}).
\ea
In expression \eqref{sw} we have dropped off the quantities
$\beta^i\left(\tau,{\bm \xi}\right)$ which are negligibly small.
Truncated expressions \eqref{tdw+} -- \eqref{aw+} were obtained
in \cite{ksh1} in the spin-dipole, mass-quadrupole approximation. Current expressions  \eqref{tdw+} -- \eqref{aw+} include all multipole moments of arbitrary order.

In the case when the distance between observer and the source
of light is much smaller than the wavelength of the gravitational
waves $R\ll \lambda$, expression  \eqref{tdw+} is reduced to a
well known result \cite{mtw} for gravitational wave detectors located in a wave-zone of an isolated system
\begin{equation}
\label{tdw++} 
\frac{\Delta
R}R=\frac{1}2k_{ij}h_{ij}^{TT}(t,{\bm x})+O(\delta^2)+O\left(\frac{R}\lambda\right)\;,
\end{equation}
where $\Delta R=c\Delt$.

Relativistic rotation of the polarization plane
of light (the Skrotskii effect) in the gravitational plane-wave approximation assumes the next
form
\begin{equation}
\label{fiw} \Delta\Phi=
            \frac{1}2
            \frac{({\bm k\times N})^i k^j}{1-{\bm k\cdot N}}
                        h^{TT}_{ij}(t,{\bm x})-
            \frac{1}2
            \frac{({\bm k\times N_0})^i k^j}{1-{\bm k\cdot N_0}}
                        h^{TT}_{ij}(t_0,{\bm x}_0)\;.
\end{equation}


\newpage
\bibliographystyle{degruyter-plain}

\begin{thebibliography}{999}
\bibitem[112]{synge} Synge \cite{syngebook} had developed an elegant theory of propagation of null rays in arbitrary transparent medium and gravitational field. This theory makes analysis of the problem under consideration in the present chapter much more sophisticated and will not be considered here.
\end{thebibliography}

\end{document}